\documentclass[twocolumn]{aastex61}
\usepackage{amsmath}
\usepackage{amssymb}

\shorttitle{Twisting by Photospheric Vortices}
\shortauthors{Rappazzo et al.}

\begin{document}

\title{Magnetic field line twisting by photospheric vortices:\\
	   energy storage and release}

\author{A. F. Rappazzo}%
\email{rappazzo@ucla.edu}%
\affil{
	Department of Earth, Planetary and Space Sciences,
	UCLA, Los Angeles, CA 90095, USA
	}%
\author{M. Velli}%
\affil{
	Department of Earth, Planetary and Space Sciences,
	UCLA, Los Angeles, CA 90095, USA
}%
\author{R. B. Dahlburg}%
\affil{
	Laboratory for Computational Physics and Fluid Dynamics, Naval Research Laboratory,
	Washington, DC 20375, USA
}%
\author{G. Einaudi}%
\affil{
	Department of Physics and Astronomy, George Mason University,
	Fairfax, VA 22030, USA
}%

\begin{abstract}
We investigate the dynamics of a closed corona cartesian reduced magnetohydrodynamic (MHD) model where photospheric vortices twist the coronal magnetic field lines. We consider two corotating or counter-rotating vortices localized at the center of the photospheric plate, and additionally more corotating vortices that fill the plate entirely. Our investigation is specifically devoted to study the fully nonlinear stage, after the linear stage during which the vortices create laminar and smoothly twisting flux tubes. Our main goal is to understand the dynamics of photospheric vortices twisting the field lines of a coronal magnetic field permeated by finite amplitude broadband fluctuations. We find that depending on the arrangement  and handedness of the photospheric vortices  an inverse cascade storing a significant amount of magnetic energy  may occur or not. In the first case a reservoir of magnetic energy available to large events such as destabilization of a pre-CME configuration develops, while in the second case the outcome is a turbulent heated corona. Although our geometry is simplified our simulations are shown to have relevant implications for coronal dynamics and CME initiation. 
\end{abstract}

\keywords{magnetohydrodynamics (MHD) --- Sun: corona --- Sun: coronal mass ejections (CMEs) --- 
Sun: magnetic topology --- turbulence}

\section{Introduction}

Solar activity is the manifestation of magnetic energy dissipation, and relevant models involve physical mechanisms able to store and dissipate such energy. In many instances magnetic energy is thought to be stored in flux tubes that are in equilibrium (force-free), and their instability (primarily kink) and/or interaction with neighboring flux tubes could play an important role in flares, corona mass ejections (CMEs) and coronal heating \citep[e.g.,][]{1960MNRAS.120...89G, Lau1996, 1999ApJ...519..884K, Linton2001, 2013ApJ...771...76R, Amari2014, Amari2018, Bareford2016, Hood2016}. Flux tubes can either emerge from sub-photospheric layers where they have been generated by a dynamo mechanism, or they can originate from the twisting of closed coronal magnetic field lines by photospheric motions. In this paper we investigate flux-tube formation as well as energy storage and dissipation due to the effects of photospheric motions.

Numerical simulations of the Parker model for coronal heating \citep{1972ApJ...174..499P, 1988ApJ...330..474P, 1994ISAA....1.....P} have shown that the continuous shuffling of coronal magnetic field line footpoints by random motions uniformly distributed in the photosphere brings about \emph{turbulent} dynamics that transfer the energy injected from the photosphere at large transverse spatial scales toward small spatial scales, thus forming approximately field-aligned current sheets where energy is dissipated impulsively and intermittently in the fashion of \emph{nanoflares} \citep{1996ApJ...457L.113E, 1997ApJ...484L..83D, 1998ApJ...497..957G, 1999ApJ...527L..63D, 1999PhPl....6.4146E, 2003PhPl...10.3584D, 2007ApJ...657L..47R, 2008ApJ...677.1348R, 2011PhRvE..83f5401R}. 

In this case the system is not in equilibrium and flux tubes are not force-free, because the continuous motions at the field lines footpoints keep injecting energy into the corona while the magnetic field strives to relax. On the other hand the dynamics brought about by a single localized photospheric velocity vortex discussed in \cite{2013ApJ...771...76R} show that initially straight field lines are at first twisted in an orderly manner by the smooth photospheric vortex (and the magnetic field is in equilibrium) until internal kink instability sets in. In the subsequent fully nonlinear stage the system does not relax to equilibrium as the footpoints continue to get shuffled by the vortex. Indeed field lines continue to exhibit a certain amount of twist, albeit in a disordered way and with a twist that grows weakly in time and is much lower than the kink instability threshold. Additionally, besides a \emph{direct} energy cascade that forms current sheets where dissipation occurs, an \emph{inverse} cascade is also present, transferring magnetic energy (and twist) from the injection scale toward the large scales, so that the spatial region where field lines are twisted increases in time \citep[see Figures~4-5 in][]{2013ApJ...771...76R}.

We can therefore conclude that a single isolated photospheric velocity vortex can bring about a flux tube with disordered twisted field lines that is out of equilibrium, and consequently is stable to kink modes, that slowly increases its transverse scale in time \citep[doubling its transverse scale in about 500 Alfv\'en crossing times, see][]{2013ApJ...771...76R} well beyond the transverse scale of the photospheric vortex stirring the field line footpoints.

The interaction of two flux tubes has been shown to depend mostly on the curl of the azimuthal component of the magnetic field \citep[e.g.,][]{Lau1996, 1999ApJ...519..884K, Linton2001}. In fact when the curl has the same sign for both flux tubes then magnetic reconnection can occur for the field lines of the azimuthal component, while when the sign is opposite it cannot. Consequently in the first case the magnetic energy of the azimuthal component can be dissipated but not in the second case. This behavior is similar whether the flux tubes have parallel or anti-parallel axial magnetic fields, except that in the anti-parallel case also the axial component can reconnect and more magnetic energy can then be dissipated thus resulting in a much different final magnetic topology.

\cite{1999ApJ...519..884K} and \cite{Linton2001} have both used Gold-Hoyle tubes \citep{1960MNRAS.120...89G} that are in force-free equilibrium, with uniform twist, and a net axial current.
\cite{1999ApJ...519..884K} implemented non-periodic boundary conditions in the axial direction where field line footpoints are either allowed to slip or are line-tied, while \cite{Linton2001} drove one against the other two flux tubes at different contact angles, thus generalizing previous results that considered only parallel or anti-parallel axial magnetic fields. On the other hand \cite{Lau1996} twist at their footpoints the initially straight axially oriented field lines, bringing about different magnetic field topologies depending on the handedness of the photospheric vortices and whether the two resulting flux tubes have parallel or anti-parallel axial fields.

In this paper we investigate only flux tubes with the same axial magnetic fields in the reduced MHD regime,which is maintained by a strong and uniform guide magnetic field \citep{1974JETP...38..283K}, implementing different twists at the footpoints of the magnetic field lines \citep{2013ApJ...771...76R}. While the aforementioned simulations have explored also the dynamics of flux tubes with anti-parallel axial fields, the numerical integration of the reduced MHD equations allows the use of much higher numerical  resolutions and longer durations, thus substantially decreasing the influence of numerical diffusion and most importantly enabling us to explore \textit{and understand} for the first time the behavior of the Poynting flux, i.e., the injection of energy at the boundary by photospheric motions. In fact all previous simulations have studied the coronal dynamics but have not been able to understand how and why the interaction of magnetic field (determined by the internal nonlinear dynamics) and velocity flow at the boundary modifies the Poynting flux into the corona and in turn its back-reaction on the dynamics, that as recently shown by \cite{Rappazzo2018} can be dramatic.

Recently \cite{Reid2018} have investigated the dynamics of three flux tubes within the framework of MHD avalanches and SOC (Self-organized criticality) models for coronal heating \citep{Hood2016}, and \cite{Zhao2015, Knizhnik2015, Knizhnik2017} have simulated the dynamics of multiple flux tubes applied to the so-called helicity ``condensation'' model \citep{Antiochos2013}. We will discuss any relevant impact of our results on these phenomena throughout the paper and in our concluding section. 

The paper is organized as follows. In Section~\ref{sec:ge} we briefly describe the model, equations, and numerical techniques used in the code, while in Section~\ref{sec:bc} we discuss the boundary and initial conditions. The results of our previous work with a single vortex at the boundary are summarized in Section~\ref{sec:sv}, while the results of our numerical simulations with multiple forcing vortices are reported in Section~\ref{sec:ns}, and their impact on coronal physics is finally discussed in Section~\ref{sec:cd}.

\section{Governing Equations} \label{sec:ge}

Flux tubes formation and interaction are studied in a simplified Cartesian geometry, modeling a magnetically closed coronal region with an axially elongated box with orthogonal cross section  of size $\ell$  and axial length $L$,  embedded in a strong, homogeneous and uniform axial magnetic field $\mathbf{B_0} = B_0\, \mathbf{\hat{e}}_z$ aligned to the $z$-direction, thus neglecting any curvature effect. As in previous works \citep[e.g.,][]{2007ApJ...657L..47R, 2013ApJ...771...76R} the dynamics are integrated with the equations of Reduced MHD \citep{1974JETP...38..283K, 1976PhFl...19..134S, 1982PhST....2...83M, 1992JPlPh..48...85Z}, well suited for a plasma embedded in a strong axial magnetic field, with constant and uniform density $\rho$. Introducing the magnetic and velocity field potentials $\psi$ and $\varphi$, the physical fields are then given by $\mathbf{b} = \nabla \psi \times \mathbf{\hat e}_z$, and $\mathbf{u} = \nabla \varphi \times \mathbf{\hat e}_z$, with axial current density $j = -\nabla^2 \psi$, and vorticity $\omega = -\nabla^2 \varphi$. In dimensionless form the equations are then given by:
\begin{equation}
  \partial_t \psi = -\left[ \psi, \varphi \right] + B_0 \partial_z \varphi
  + \eta_n \nabla^{2n}\, \psi, \label{eq:eq1}
\end{equation}
\begin{equation}
  \partial_t \omega = \left[ j, \psi \right] - \left[ \omega, \varphi \right]
  + B_0 \partial_z j + \nu_n \nabla^{2n}\, \omega. \label{eq:eq2}
\end{equation}
Magnetic and velocity fields $\mathbf{b}$ and $\mathbf{u}$ have only components orthogonal to the axial direction z, so as the gradient and Laplacian operators, e.g., 
\begin{equation}
  \nabla = \mathbf{\hat{e}}_x \partial_x + \mathbf{\hat{e}}_y \partial_y.
\end{equation}
The Poisson bracket of generic functions $g$ and $h$ is defined as $[g,h] = \partial_x g \partial_y h - \partial_y g \partial_x h$ (e.g., $[j,\psi] = \mathbf{b} \cdot \nabla j$). The linear terms $\propto \partial_z$ couple the planes along the axial direction through  a wave-like propagation at the Alfv\'en speed $B_0$. Incompressibility in the reduced MHD regime follows from the large value of the axial magnetic field  \citep{1976PhFl...19..134S} and they apply also to low $\beta$ systems such as the corona \citep{1992JPlPh..48...85Z}. Furthermore recent fully compressible simulations of a similar Cartesian coronal loop model have shown that the inclusion of thermal conductivity and radiative losses, that transport away the heat produced by the small scale energy dissipation, keep the dynamics in the reduced MHD regime for strong guide magnetic fields \citep{2012A&A...544L..20D, 2016ApJ...817...47D, Dahlburg2018}.

To render the equations in nondimensional form, we have first expressed the magnetic field as an Alfv\'en velocity [$b \rightarrow b/\sqrt{4\pi \rho}$], where $\rho$ is the density supposed homogeneous and constant, and then all velocities have been normalized to the typical velocity  of photospheric convective motions $u^{\ast} = 1$~km~s$^{-1}$. In order to comply (and for a readily comparison of simulation results) with normalizations used in our previous simulations the aspect ratio of the box axial length L to the photospheric convective length-scale $\ell_c$ is taken equal to $L/\ell_c$=40. Therefore with its typical value $\ell_c \sim$10$^3$~km, we obtain a box axial length $L\sim\,$40,000~km, a typical coronal loop length. Physical lengths are thus normalized with $\ell^{\ast} = 4 \ell_c$ (so that, as in our previous works, the convective length-scale is 1/4 in dimensionless units), and times with the related crossing time $t^{\ast} = \ell^{\ast}/u^{\ast}$. As a result, the linear terms $\propto \partial_z$ in Eqs.~(\ref{eq:eq1})--(\ref{eq:eq2}) are multiplied by the dimensionless Alfv\'en velocity $B_0$ that is the ratio of the Alfv\'en velocity to $u^{\ast}$ (keeping in mind that the magnetic field has been first expressed as an Alfv\'en velocity, i.e., $B_0 \rightarrow B_0/\sqrt{4\pi \rho}$).

In our numerical simulation we have implemented hyperdiffusion \citep[e.g.,][]{2003matu.book.....B}, that effectively limits diffusion to small scales, in all our numerical simulations except run~C (Table~\ref{tbl}). The index $n$ in the diffusive terms in Eqs.~(\ref{eq:eq1})-(\ref{eq:eq2}) is called \emph{dissipativity} and we use n=4, with $\eta_n = \nu_n = (-1)^{n+1}/Re_n$, with $Re_n$ corresponding to standard diffusion for n=1, for which
the kinetic and magnetic Reynolds numbers are given by:
\begin{equation}
Re = \frac{\rho\, \ell^{\ast} u^{\ast}}{\nu}, \qquad 
Re_{_m} = \frac{4\pi\, \ell^{\ast} u^{\ast}}{\eta c^2},
\end{equation}
where $c$ is the speed of light. For numerical stability the dissipative coefficients are given the same value $\eta_n = \nu_n$.
The diffusive timescale $\tau_n$ at the scale $\lambda$ associated with the dissipative terms in Equations~(\ref{eq:eq1})-(\ref{eq:eq2})
\begin{equation}
	\tau_n \sim Re_n \lambda^{2n},
\end{equation}
decreases faster toward the small scales for higher values of $n$. Therefore to have the same diffusive timescale  at the resolution scale $\lambda_\textrm{min} = 1/N$ (with $N$ the number of grid points) for both standard ($n=1$) and hyperdiffusion ($n > 1$), i.e., $\tau_n = \tau_1$, we need to have
\begin{equation}
  Re_n \sim Re_1 N^{2(n-1)}. 
\end{equation}
For instance, assuming a grid with $N=512$ and $Re_1 = 800$, for $n=4$ we obtain $Re_4 \sim 10^{19}$. At the same time the corresponding diffusive timescales at large scales ($\lambda \sim 1$) is much bigger for the hyperdiffusive case, with $\tau_4/\tau_1 \sim Re_4/Re_1 \sim 2 \times 10^{16}$. With the boundary velocity forcings used in the simulations described in this paper (see Sections~\ref{sec:bc} and \ref{sec:sv}) the use of hyperdiffusion with a negligible diffusion at large scale is crucial because, as discussed in \cite{2013ApJ...771...76R}, with the adopted grid resolutions diffusion at the large scales is big enough to impede twisting field lines beyond the threshold of kink instability, a clear numerical artifact.

Our parallel code RMH3D solves numerically Equations~(\ref{eq:eq1})-(\ref{eq:eq2}), by advancing in time the Fourier components in the $x$- and $y$-directions of the scalar potentials. Along the $z$-direction we implement non-periodic boundary conditions (Section~\ref{sec:bc}), and use a central second-order finite-difference scheme, while in the $x$-$y$ plane, a Fourier pseudospectral method  is implemented. Time is discretized with a third-order Runge-Kutta method. More details of the numerical code and methods are discussed in \cite{2007ApJ...657L..47R, 2008ApJ...677.1348R}.

\section{Boundary and Initial Conditions}  \label{sec:bc}

Magnetic field lines are line-tied to the top and bottom photospheric-mimicking plates ($z=0$ and $L$), were we impose as boundary condition a velocity field that convects the footpoints of the magnetic field lines.

\begin{table}
	\begin{center}
		\caption{Simulations summary\label{tbl}}
		\begin{tabular}{lccccc}
			\hline \hline\\[-5pt]
			Run & Forcing & B$_0$ & $n_{\rm x} \times n_{\rm y} \times n_{\rm z}$ &  n & $Re_{_n}$ \\[5pt]
			\hline\\[-5pt]
			A & co2  &  200 & $1024\times512\times208$& 4 & $1\times 10^{19}$ \\[.5em]
			B & cr2  &  200 & $1024\times512\times208$& 4 & $5\times 10^{19}$ \\[.5em]
			C & co4  &  200 & $512^2 \times 208$ & 1 & $800$ \\[.5em]
			D & co16 & 1000 & $512^2 \times 504$ & 4 & $1\times 10^{19}$ \\[.5em]
			E & co16 &  200 & $512^2 \times 504$ & 4 & $1\times 10^{19}$ \\[.5em]
			F & co16 &   50 & $512^2 \times 504$ & 4 & $2\times 10^{20}$ \\[.7em]
			\hline\\
		\end{tabular}
		\tablecomments{The boundary velocity forcing is indicated as `co2' for two corotating photospheric vortices, `cr2' for two counter-rotating vortices, and `co4' and `co16' for respectively four and sixteen corotating vortices (see Section~\ref{sec:bc}). The numerical grid resolution is $n_x \times n_y \times n_z$. The next columns indicate respectively the value of the dissipativity n and the hyperdiffusion coefficient $Re_n$  (Section~\ref{sec:bc}).}
	\end{center}
\end{table}

In \cite{2013ApJ...771...76R} we have used a single velocity vortex at the center of the top plate z=L. Here we use different combinations of the same vortex that is the building block of our boundary velocity forcing. The velocity potential for the single vortex centered in ($x_0$, $y_0$) in the z=L plane and extending in the 2D interval $\mathit{I} = \mathit{I}_x \times \mathit{I}_y$, where $\mathit{I}_x = \left[ x_0 - \Delta/2, x_0+\Delta/2 \right]$, and $\mathit{I}_y = \left[ y_0 - \Delta/2, y_0+\Delta/2 \right]$, both of linear extent $\Delta = 1/4$ (corresponding to the photospheric convective length-scale $\ell_c \sim 10^3$~km in dimensionless units, see Section~\ref{sec:ge}), while vanishing outside, is given by:
\begin{align}
&\varphi\! \left( x, y \right)  =  \frac{1}{2\pi \sqrt{3}}\, \sin^2\! \left[ 4\pi \left( x-x_0+\frac{3}{8} \right) \right] \times \nonumber \\
& \qquad \qquad \qquad \quad \, {} \sin^2\! \left[ 4\pi \left( y-y_0+\frac{3}{8} \right) \right] \label{eq:f0} \\[1em]
&\mathrm{for} \quad (x, y) \in \mathit{I}, \quad \mathrm{and} \quad
\varphi = 0 \quad  \mathrm{for} \quad (x, y) \notin \mathit{I}. \nonumber 
\end{align}
The velocity is linked to the potential by
$\mathbf{u}  = \nabla \varphi \times \mathbf{\hat{e}}_z$
and its components are then:
\begin{align} 
&u_x^L\! \left( x, y \right)  = +\frac{2}{\sqrt{3}}\, \sin^2\! \left[ 4\pi \left( x-x_0+\frac{3}{8} \right) \right] \times \nonumber \\
& \qquad \qquad \qquad \quad \ {} \sin\!   \left[ 8\pi \left( y-y_0+\frac{3}{8} \right) \right]  \label{eq:f01} \\[.5em]
&u_y^L\! \left( x, y \right)  = -\frac{2}{\sqrt{3}}\, \sin\!   \left[ 8\pi \left( x-x_0+\frac{3}{8} \right) \right] \times \nonumber \\
& \qquad \qquad \qquad \quad \ {} \sin^2\! \left[ 4\pi \left( y-y_0+\frac{3}{8} \right) \right] \label{eq:f02}
\end{align}
in the interval \textit{I} and vanish outside.

\begin{figure}
	\begin{centering}
		\includegraphics[scale=.57]{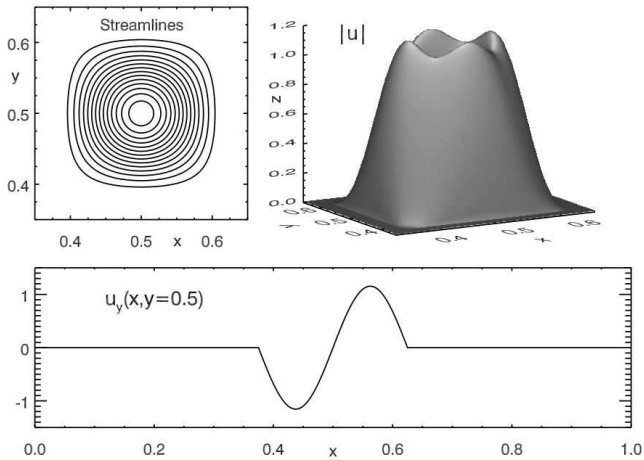}
		\caption{Counter-clockwise rotating quasi-circular vortex of linear length $\Delta$=1/4 and centered at ($x_0$, $y_0$) = (1/2, 1/2) (see Eqs.~(\ref{eq:f0})--(\ref{eq:f02})) employed as the building block to make the boundary velocity forcings in our simulations.
			\emph{Top:} streamlines (\emph{left}), and profile of its absolute velocity value $|u|$ (\emph{right}).
			\emph{Bottom:} plot of the velocity $y$-component as a function of
			$x$ at $y=0.5$.
			\label{fig:fig1}}
	\end{centering}
\end{figure}

An illustration of a single vortex centered in ($x_0$, $y_0$) = (1/2, 1/2) is shown in Figure~\ref{fig:fig1}. This is a counter-clockwise vortex centered in the middle of the plane $z=L$ and has quasi-circular streamlines, with a slight departure from a perfectly circular shape toward the edge of the interval \textit{I}, where all velocity components then vanish. Averaging over the surface \textit{I} the velocity r.m.s. is $\langle (u^L)^2 \rangle_I = 1/2$, the same value of the boundary velocity fields used in our previous works \citep{2007ApJ...657L..47R, 2008ApJ...677.1348R, 2010ApJ...722...65R, 2011PhRvE..83f5401R, 2013ApJ...771...76R}.

In \cite{2013ApJ...771...76R} we considered a single vortex centered at ($x_0$, $y_0$) = (1/2, 1/2), and the domain spanned $0 \le x, y \le 1$. Here we want to investigate the dynamics induced by two nearby vortices. In order for the periodic boundary conditions along $x$ and $y$ not to affect the dynamics through interactions across those boundaries, we double the box length along $x$ and maintain the same length along y: $0 \le x \le 2$, and $0 \le y \le 1$. The two vortices (each of linear extent $\Delta$=1/4) are set side by side and centered respectively in ($x_1$, $y_1$) = (1-$\Delta$/2, 1/2), and ($x_2$, $y_2$) = (1+$\Delta$/2, 1/2). As summarized in Table~\ref{tbl} we performed two simulations with two vortices: in run~A both vortices are the same and therefore corotating (indicated with `co2' in Table~\ref{tbl}), while for run B the velocity of the second vortex is opposite and therefore the two vortices are counter-rotating (`cr2'). 

Additionally we also performed a simulation with four corotating vortices (run~C, Table~\ref{tbl}) of double linear length $\Delta = 1/2$ and centered symmetrically in (x, y) = (1/2 $\pm$ 1/4, 1/2 $\pm$ 1/4), see the top-left panel in Fig.~\ref{fig:fig10} for an illustration. For run~C the orthogonal box size is same as our previous simulations  \citep[and in particular the case with one vortex,][]{2013ApJ...771...76R} with $0 \le x,y \le 1$, so that the four vortices fill entirely the x-y plane. The velocity potential in this case is given by
\begin{equation}
\varphi\! \left( x, y \right)  =  \frac{1}{\pi \sqrt{3}}\,  
\sin^2\! \left( 2\pi x  \right) \sin^2\! \left( 2\pi y \right). \label{eq:co4}
\end{equation}

Finally, in order to better understand the inverse cascade process for corotating vortices we have also performed a parameter study with three additional simulations with sixteen vortices (runs D, E and F, Table~\ref{tbl}) discussed in section~\ref{sec:c16}. In this case the boundary potential is given by:
\begin{equation}
\varphi\! \left( x, y \right)  =  \frac{\sqrt{3}}{2\pi}\,  
\sin^2\! \left( 4\pi x  \right) \sin^2\! \left( 4\pi y \right). \label{eq:co16}
\end{equation}
The velocity fields derived from Equations~(\ref{eq:co4})--(\ref{eq:co16}) yield for the boundary velocity the usual r.m.s. value $\langle (u^L)^2 \rangle = 1/2$ used in our previous simulations.
                                                                
In all simulations a vanishing velocity is imposed at the bottom plate $z=0$:
\begin{equation} 
  \mathbf{u}^0 \left( x, y \right)  =  0. \label{eq:f1}
\end{equation}

At time $t=0$ we start our simulations with a uniform and homogeneous magnetic field along the axial direction $\mathbf{B}_0 = B_0\, \mathbf{\hat{e}}_z$. The orthogonal component of the velocity and magnetic fields are zero inside our computational box $\mathbf{u}=\mathbf{b}=0$, while field lines are line-tied at the bottom plate (z=0) to a motion-less photosphere and at the top-plate footpoints are shuffled by the applied boundary velocity forcing discussed previously (kept constant in time). 

As shown in our previous work \citep{2013ApJ...771...76R}, initially for a time interval smaller than the nonlinear timescale $t < \tau_{nl}$, the time evolution of the magnetic and velocity fields with boundary forcing $\mathbf{u}^L$ in z=L, and $\mathbf{u}^0 = 0$ in z=0, is given by:
\begin{eqnarray}
&&\mathbf{b} (x,y,z,t) \approx 
\frac{t}{\tau_A}\, \mathbf{u}^L, 
\label{eq:lin1} \\[.2em]
&&\mathbf{u} (x,y,z,t) \approx 
\frac{z}{L}\, \mathbf{u}^L.
\label{eq:lin2}
\end{eqnarray}
where $\tau_A = L/B_0$ is the Alfv\'en crossing time along the axial direction $z$. The magnetic field grows linearly in time and is proportional to $\mathbf{u}^L$, while the velocity field is stationary with its value increasing linearly along z from 0 at z=0 up to $\mathbf{u}^L$ at z=L. Both are therefore \emph{mappings} of the boundary velocity field $\mathbf{u}^L$.
We can therefore incidentally use this property to visualize the location of the vortices in the x-y plane by looking at the magnetic field lines of the orthogonal magnetic component $\mathbf{b}$ in the mid-plane z=5 at $t/t_A$=0.61 for simulation~A in Figure~\ref{fig:fig2}, for simulation~B in Figure~\ref{fig:fig6}, and for simulation~C in Figure~\ref{fig:fig10}. The sign of the current, in color, indicates also the vortex rotation, since from Equation~(\ref{eq:lin1}) we can write $j= \mathbf{\hat{e}_z} \cdot \nabla \times \mathbf{b} = t/\tau_A \mathbf{\hat{e}_z} \cdot \nabla \times \mathbf{u}^L = t/\tau_A\ \omega^L$, with $\omega^L$ the axial vorticity of the boundary velocity $\mathbf{u}^L$.

\section{Previous Single Vortex Study Results} \label{sec:sv}

In \citep{2013ApJ...771...76R} we have carried out numerical simulations with a single boundary photospheric velocity vortex (Eqs.~(\ref{eq:f0})-(\ref{eq:f02})) at the center of the top plate z=L as illustrated in Figure~\ref{fig:fig1}.

The originally straight field lines threading the computational box along the z direction ($\mathbf{B_0}=B_0 \mathbf{\hat{e}}_z$), and line-tied to the end-plates z=0 and L, get twisted by the boundary flow, initially following the linear behavior of Eq.~(\ref{eq:lin1}). Our forcing vortex is not perfectly circular, particularly toward the edge (see Figure~\ref{fig:fig1}), so that the resulting orthogonal component of the magnetic field $\mathbf{b}$ is slightly out of equilibrium, but field line tension quickly straightens out the field lines in a round shape (e.g., see Fig.~4 at $t/t_A=80.64$ in \cite{2013ApJ...771...76R}, and Fig.~\ref{fig:fig6} at $t/t_A=33.63$ in this paper), so that $\mathbf{b}$ is in equilibrium at any given time at this stage. The slight difference between the vortex streamlines and the field lines of the orthogonal magnetic field component $\mathbf{b}$ introduces a very small perturbation that makes the system unstable to the internal kink mode, leading during the nonlinear stage of the instability to dissipate about 90\% of the accumulated magnetic energy, with $\Delta E \sim 10^{25}$~erg that is a typical value for a micro-flare.
These results are in full agreement with previous theoretical analysis and numerical simulations of kink instabilities for the closed corona with field lines line-tied both  to a motionless photosphere \citep{1996A&A...308..935B, 1997SoPh..172..257V, 1998ApJ...494..840L, 2008A&A...485..837B, 2009A&A...506..913H} or with an applied non-vanishing velocity at the boundary \citep{1990ApJ...361..690M, 2002A&A...387..687G}.

Nevertheless all these previous simulations have been carried out with smooth and laminar initial magnetic and velocity fields, with only an infinitesimal perturbation added to the magnetic and velocity fields that then triggers the development of kink instability. On the other hand it is clear from observations and simulations that magnetic and velocity fields of finite amplitude are continuously injected in the solar corona from the underlying photosphere, chromosphere and also from dynamical processes occurring in the corona. It can therefore be posited that very often, in a more realistic case, twisting motions are not applied to initially straight and laminar magnetic field lines but rather to a magnetic field where finite-amplitude broadband fluctuations with small-scales and current sheets are present. 

For this reason in \cite{2013ApJ...771...76R} we have continued our numerical simulations beyond the stage when kink instability develops, at which point magnetic energy is small (with b/B$_0\sim$~5\%), but magnetic and kinetic energies exhibit a broadband spectrum from large to small dissipative scales. During this (nonlinear) stage the photospheric vortex continues to twist the magnetic field lines, but it occurs in a disordered way as the magnetic field is always out of equilibrium and it never returns to the laminar and smooth field exhibited during the initial linear stage (Equation~(\ref{eq:lin1})). Therefore the kink instability, in its classic linear version, does not occur again, rather the presence of broadband fluctuations leads to a turbulent cascade of the energy injected by the photospheric vortex at large transverse scales. This is not to say that rapid dynamical large scale phenomena do not occur, they do but in a nonlinear disordered fashion. Field lines still exhibit a  twist whose r.m.s. value remains approximately constant during this non-linear stage fluctuating around its mean $( \sim 180^{\circ} )$. The electric current is no longer structured coherently at the large scales as it was during the linear stage described in Equation~(\ref{eq:lin1}), but similarly to reduced MHD turbulence it is organized in thin current sheets elongated in the axial directions.

Current sheets are formed by a \emph{direct} cascade transporting energy from the large to the small scales. This phenomenon has been detected also in our previous simulations with disordered photospheric motions and a shear flow \citep{2008ApJ...677.1348R, 2010ApJ...722...65R}. But for the single vortex, besides the direct cascade also an \emph{inverse cascade} of magnetic energy occurs, transporting energy from the transverse injection scale toward larger scales. In physical space this corresponds to the twisted field lines occupying an increasingly larger volume, with its transverse scale increasing in time. Because at large scales no dissipation occurs the inverse cascade can store a significant amount of magnetic energy that subsequently can be released and dissipated if interactions with nearby magnetic structures occur \citep[for a more detailed discussion, see][]{2013ApJ...771...76R}.

Notice that the inverse cascade of magnetic energy and twist occurring for a single isolated photospheric vortex \citep{2013ApJ...771...76R} is very different from the inverse energy cascade arising from the coalescence of magnetic islands in two dimensions \citep[e.g.,][]{Politano1989, 1992PhFlB...4.3070M, 1996ApJ...457L.113E} that extended in three dimensions with a strong axial field is at the base of the helicity `condensation' model proposed by \citep{Antiochos2013}. 

Indeed for a single isolated flux tube twisted at its footpoint by a single velocity vortex, and characterized in its nonlinear stage by a single isolated distorted magnetic island in the orthogonal plane, no coalescense and merging of magnetic islands can evidently occur. The inverse cascade is in fact caused by the continued injection of magnetic energy from the photospheric vortex that twists the field lines, thus increasing the orthogonal magnetic field intensity and its magnetic pressure. 

It is interesting to notice though that it is the thin current sheets elongated along the axial direction, that in the nonlinear stage are continuously generated by the direct energy cascade, that through the j$\times$b force (with j axial and b orthogonal) push outward the regions with more intense magnetic field thus enlarging and spreading the region where field lines are twisted (e.g., see \cite{2013ApJ...771...76R}, and it is also well visible in the animation associated to Figure~\ref{fig:fig2} in the fully nonlinear stage at t~$\gtrsim$~22.5~t$_A$). Therefore in this case it is the direct energy cascade, that continuously creates thin current sheets, that drives the inverse energy cascade of magnetic energy that enlarges the region where field lines are twisted. 

Further discussion of the results obtained from the simulations with a single photospheric vortex will be given in our concluding Section~\ref{sec:cd}, together with a discussion about the results from simulations with multiple boundary velocity vortices described in the next Section.

\section{Numerical Simulations} \label{sec:ns}
 
In this section we discuss the results of the numerical simulations outlined in  Table~\ref{tbl}. Simulation~A is forced by two equal vortices at the photospheric plate z=L that are therefore \emph{corotating}, while simulation B employs two \emph{counter-rotating} vortices. Simulation~C has four corotating vortices at the boundary and employs standard diffusion. In all simulations the vortical velocity patterns are applied at the top plate z=L, and a vanishing velocity at the bottom plate z=0. Initially at time t=0 the orthogonal magnetic and velocity field components vanish inside the computational box, i.e., $\mathbf{b} = \mathbf{u} = 0$, and only the uniform and constant axial magnetic field $\mathbf{B}_0 = B_0\, \mathbf{\hat{e}}_z$ is present.

\subsection{Run~A} \label{sec:runa}

In this simulation two identical photospheric vortices, of linear extension $\Delta$ = 1/4 (corresponding in conventional units to the length-scale of photospheric granulation $\sim 10^3$~km) as described in Equations~(\ref{eq:f0})-(\ref{eq:f02}) and illustrated in Fig.~\ref{fig:fig1}, are set side by side in the boundary plane z=L centered respectively in $(x_0, y_0) = (1\pm \Delta/2, 1/2)$. The numerical grid has n$_x \times$ n$_y \times$ n$_z$ = 1024 $\times$ 512 $\times$ 208 points spanning $0 \le x \le 2$, $0 \le y \le 1$, and $0 \le z \le 10$. The hyperdiffusion coefficient is $Re_4 = 10^{19}$, with diffusivity $n=4$. The Alfv\'en velocity is $B_0 = 200$, corresponding to 200~km/s in conventional units. The total duration is  $\sim$~600 axial Alfv\'en crossing times $\tau_A = L/B_0$, where L=10 is the axial box length. 

\begin{figure*}
	\begin{centering}
		\includegraphics[scale=.31]{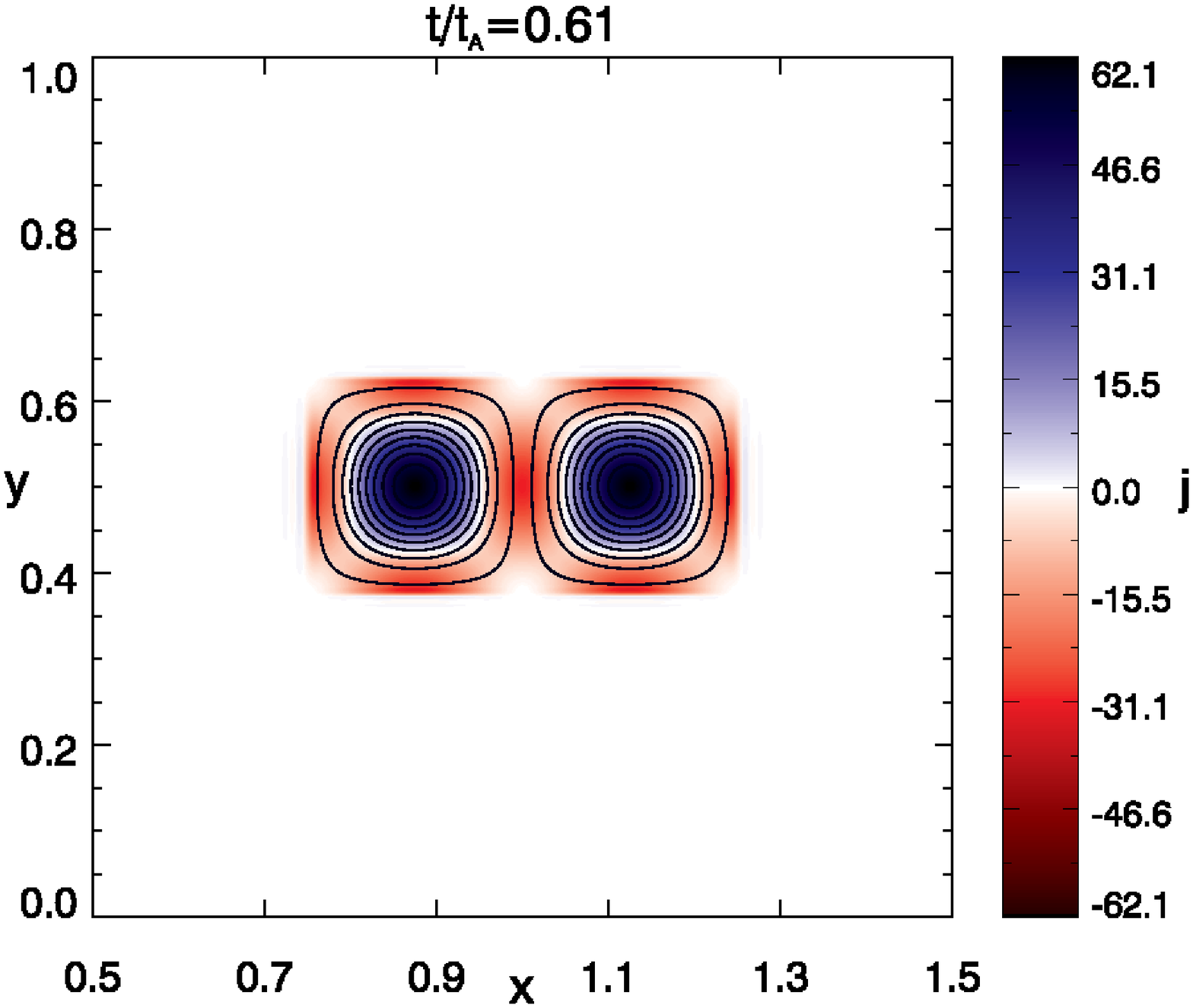}
		\includegraphics[scale=.31]{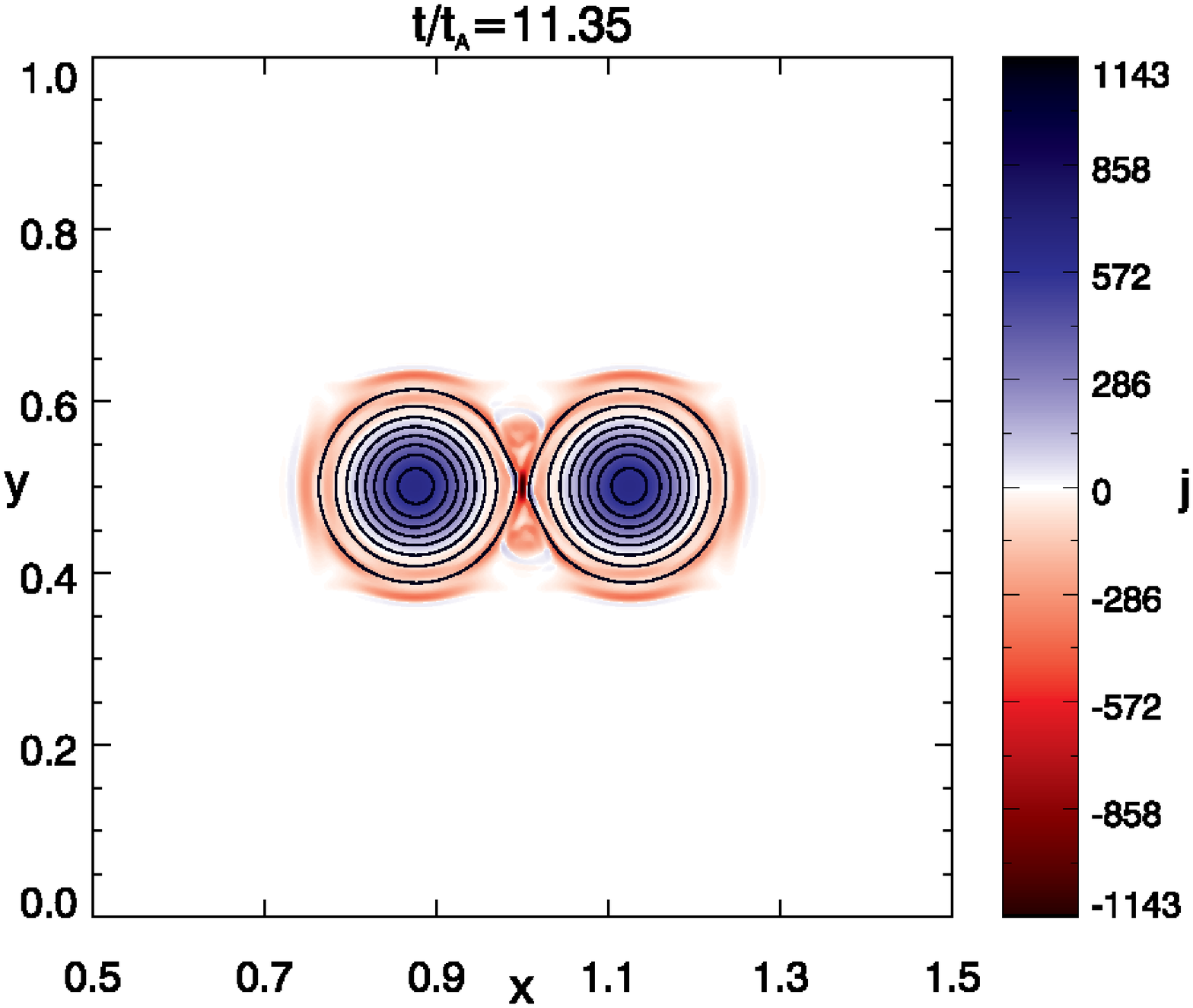}\\[1em]
		\includegraphics[scale=.31]{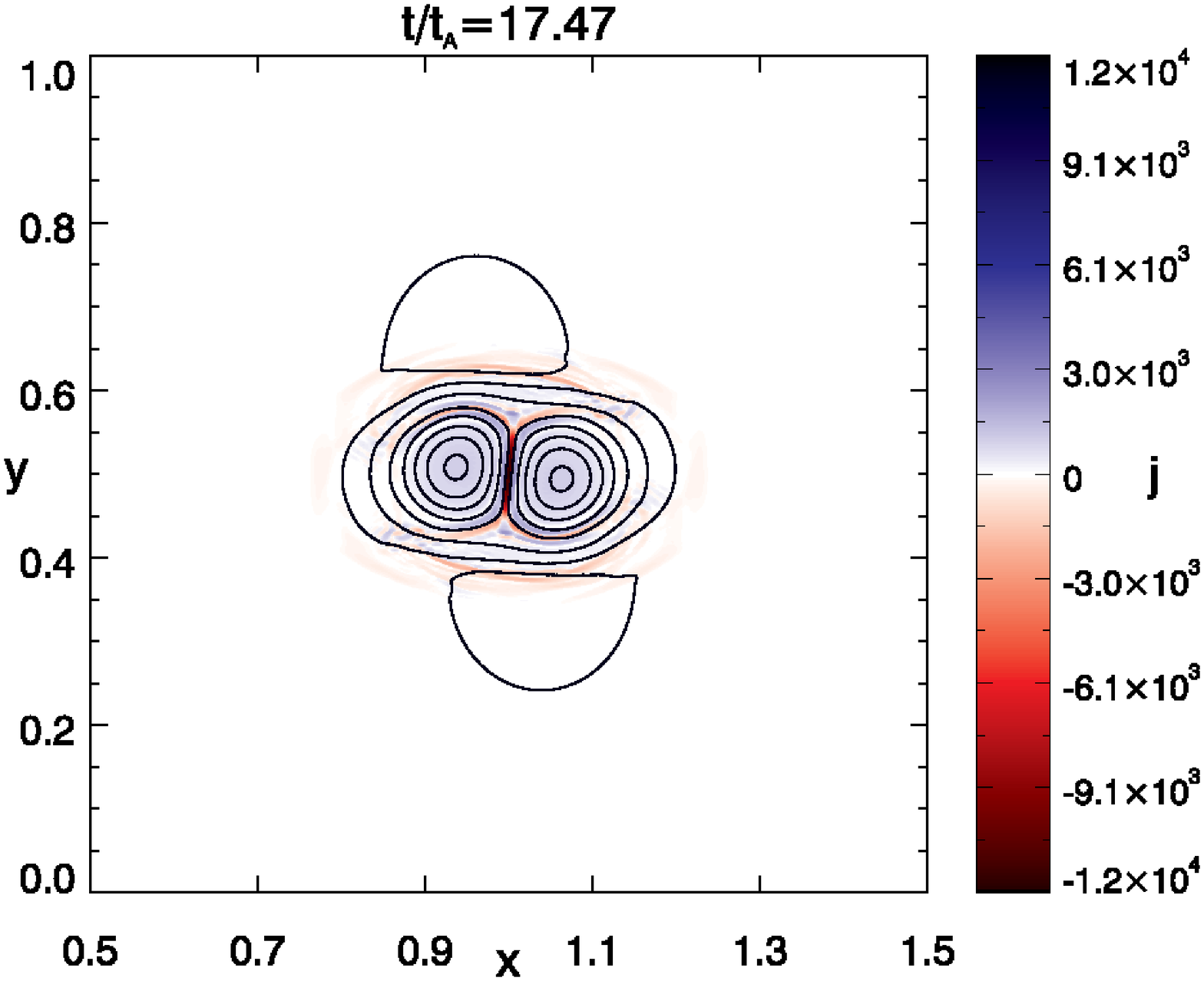}
		\includegraphics[scale=.31]{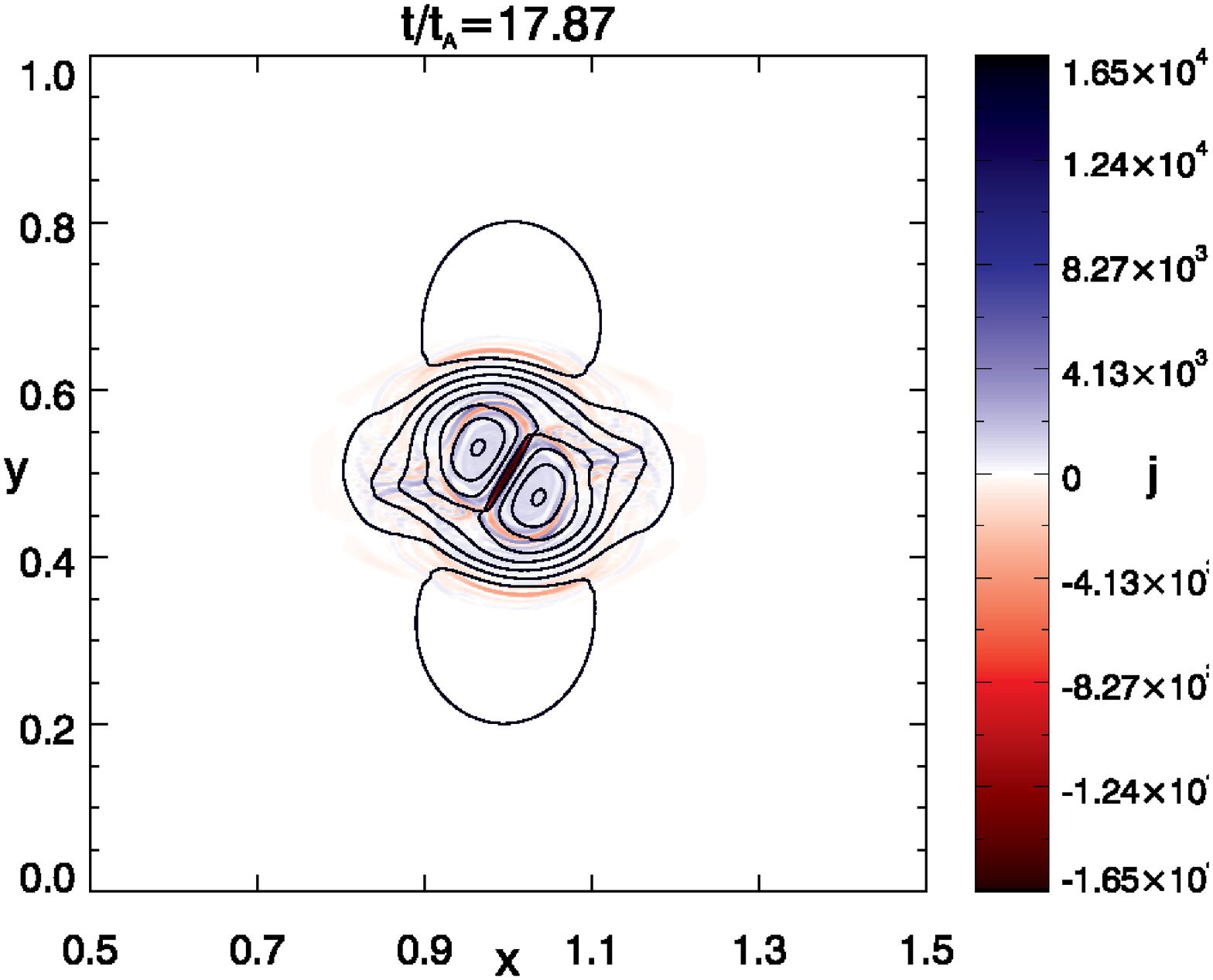}\\[1em]
		\includegraphics[scale=.31]{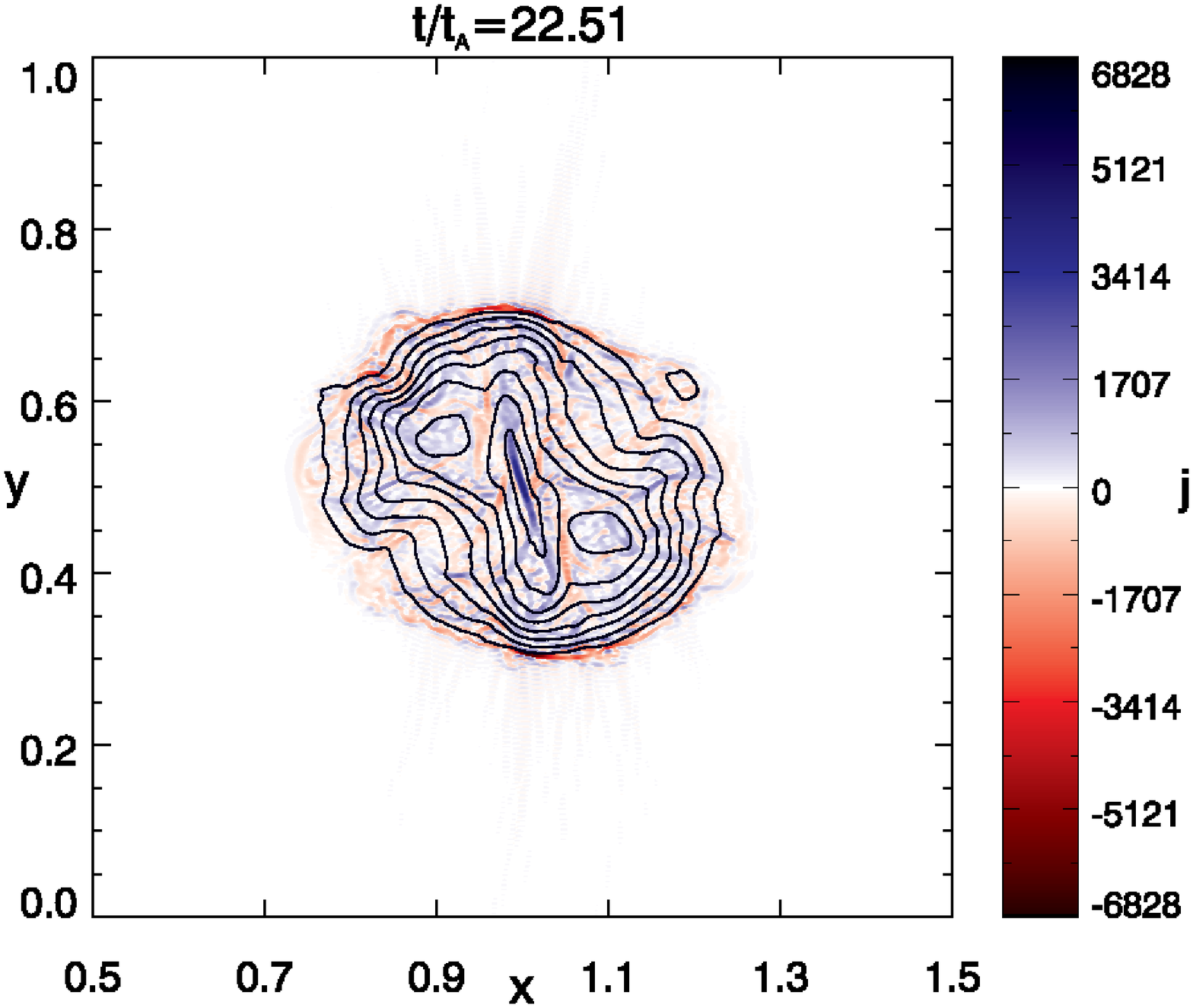}
		\includegraphics[scale=.31]{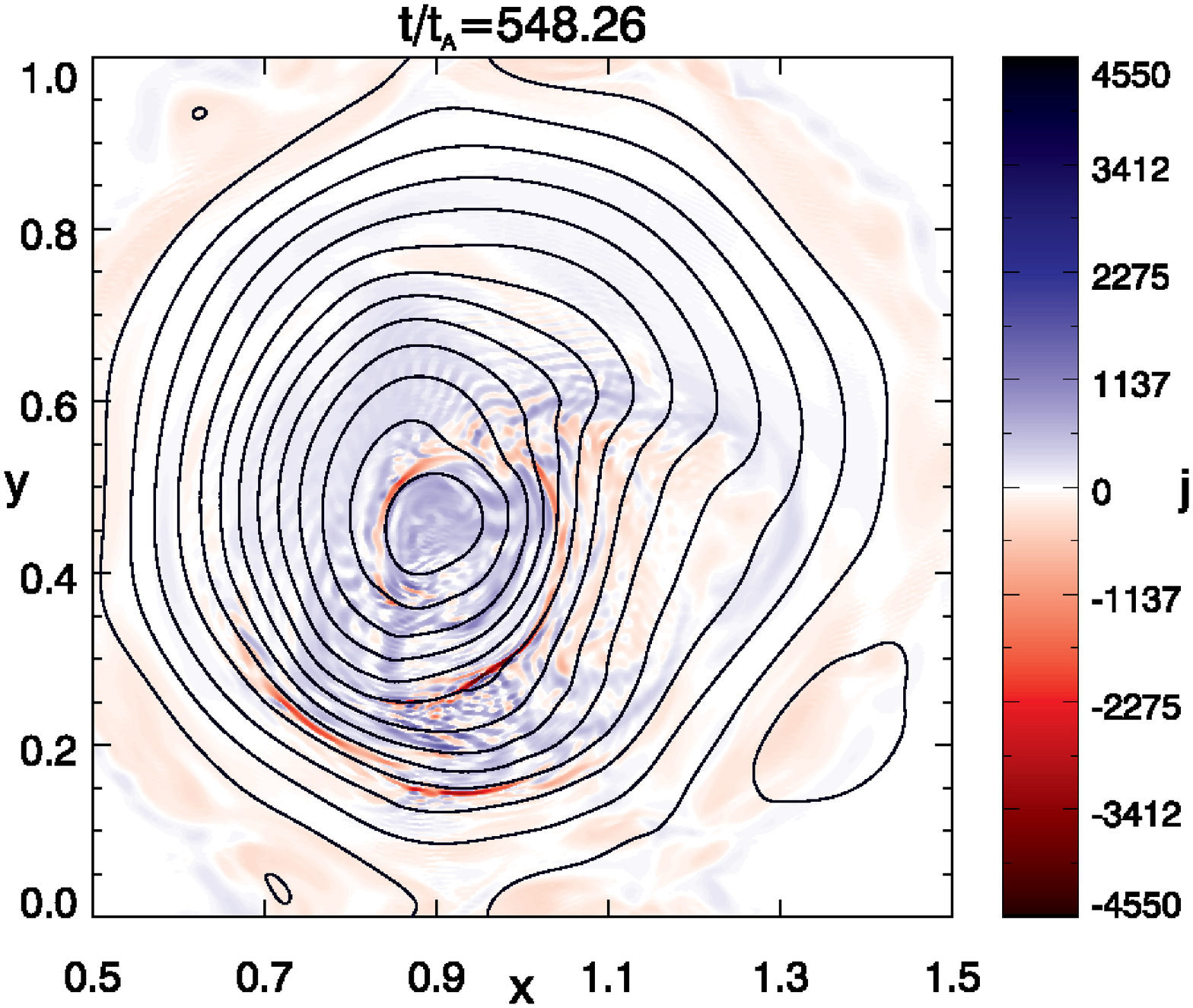}
		\caption{\emph{Run~A -- Corotating vortices}: Axial component of the current $j$ (in color) and field lines of the orthogonal magnetic field component \textbf{b} in the mid-plane $z=5$ at selected times. At the beginning of the linear stage (t=0.61~t$_A$) the orthogonal magnetic field is a mapping of the two corotating boundary vortices (see linear analysis in Section~\ref{sec:bc}, Equation~(\ref{eq:lin1})). Later on the field line tension straightens out in a circular shape the vortices mapping (e.g., t=11.35~t$_A$). Field lines of \textbf{b} are oppositely directed at the boundary between the two flux ropes brought about by the two corotating vortices. Magnetic reconnection therefore starts to occur in the plane z=5 around t=11.35~t$_A$, pushing together the two original flux tubes (t=17.47 and 17.87~t$_A$) until they completely merge (t=22.51~t$_A$). From this point on the dynamics are similar to the simulations with one single vortex centered in the top plate z=10 \citep{2013ApJ...771...76R}, with an inverse cascade of energy increasing the transverse scale where a disorderly twisted magnetic field is present, up to reach the y-direction box size at t$\sim$548.26~t$_A$. \\
(An animation of this figure is available.) \label{fig:fig2}}
	\end{centering}
\end{figure*}
\begin{figure*}
	\begin{centering}
		\includegraphics[scale=.35]{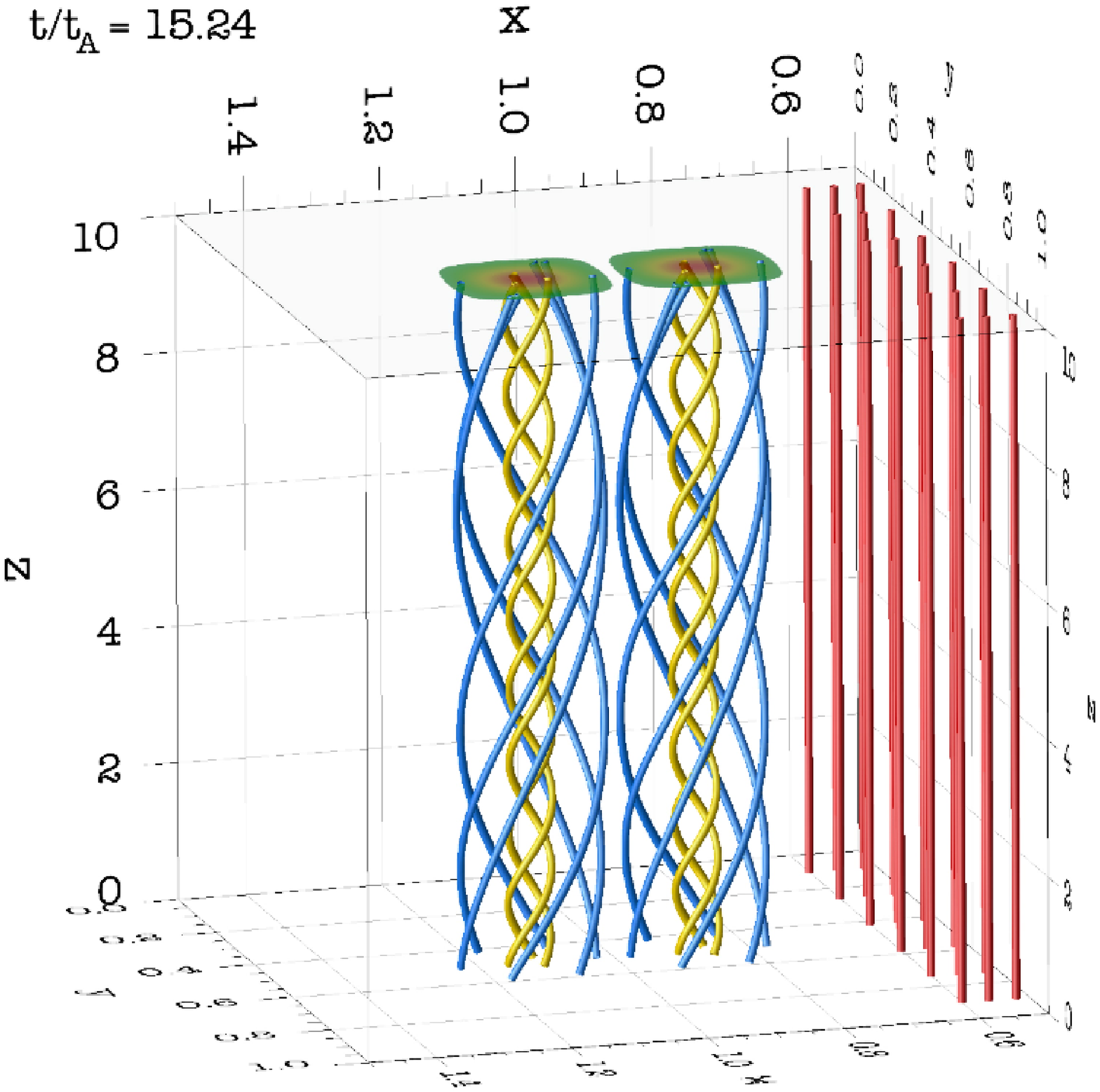}\hspace{1em}
		\includegraphics[scale=.35]{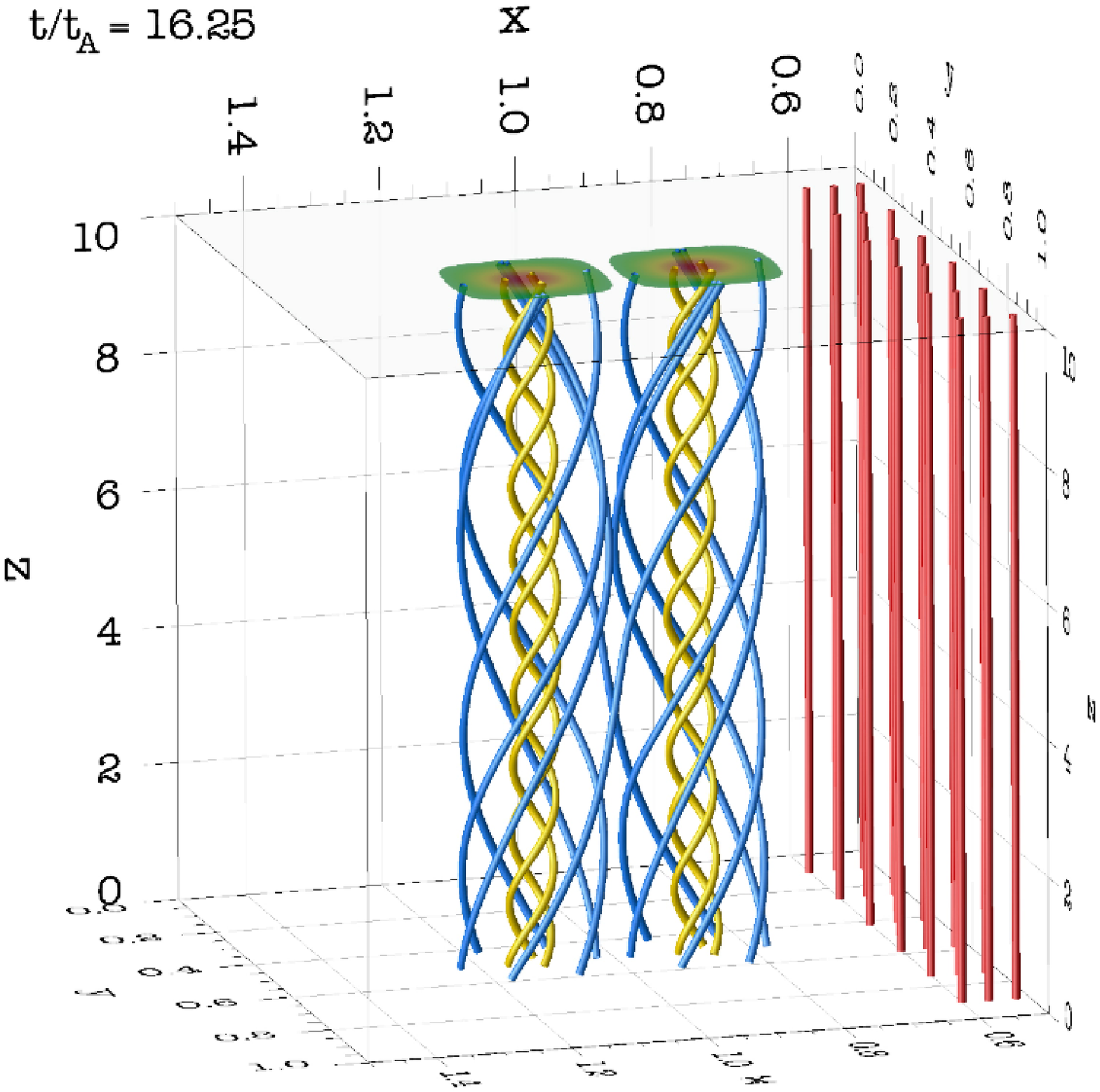}\\[.5em]
		\includegraphics[scale=.35]{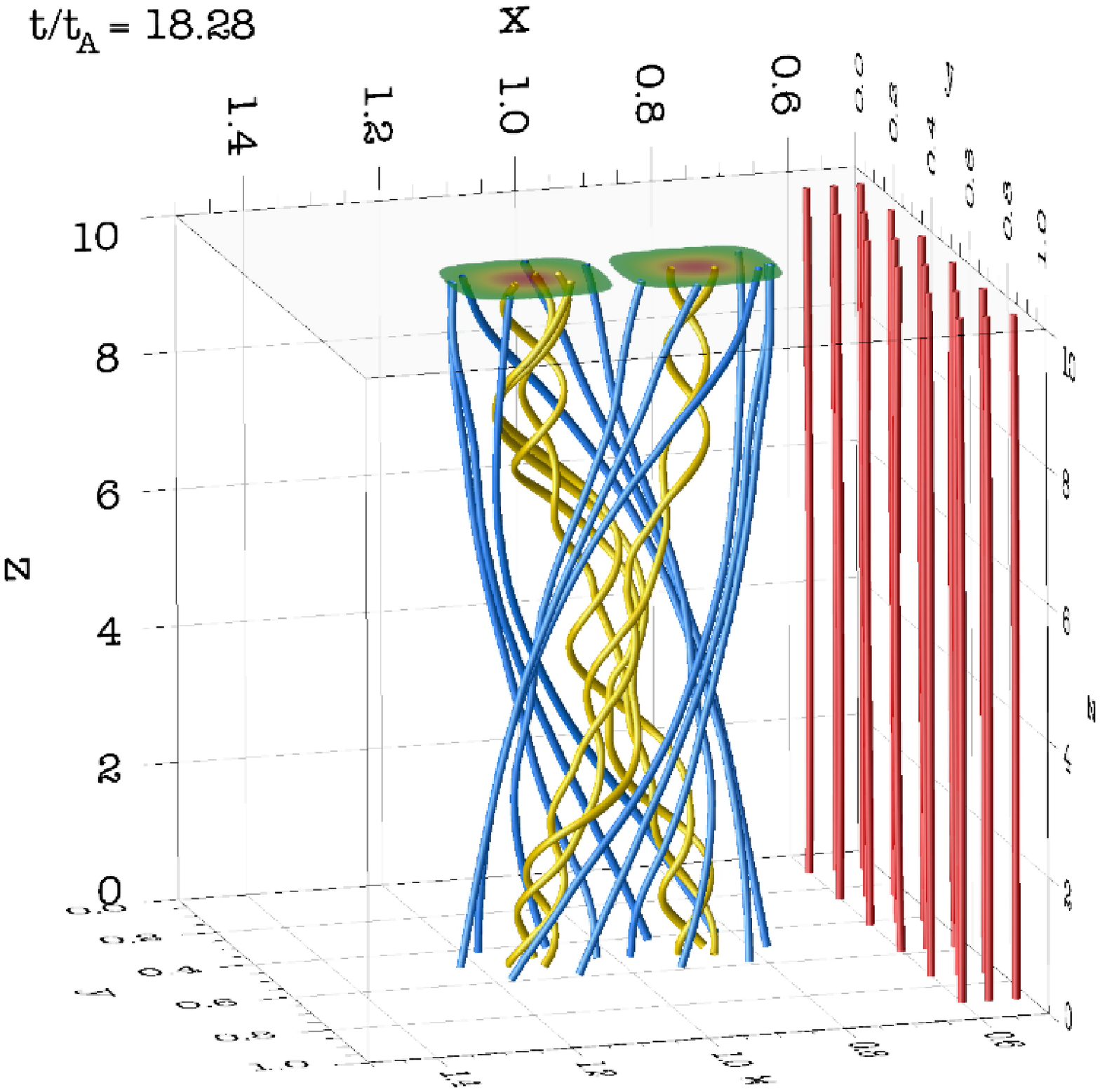}\hspace{1em}
		\includegraphics[scale=.35]{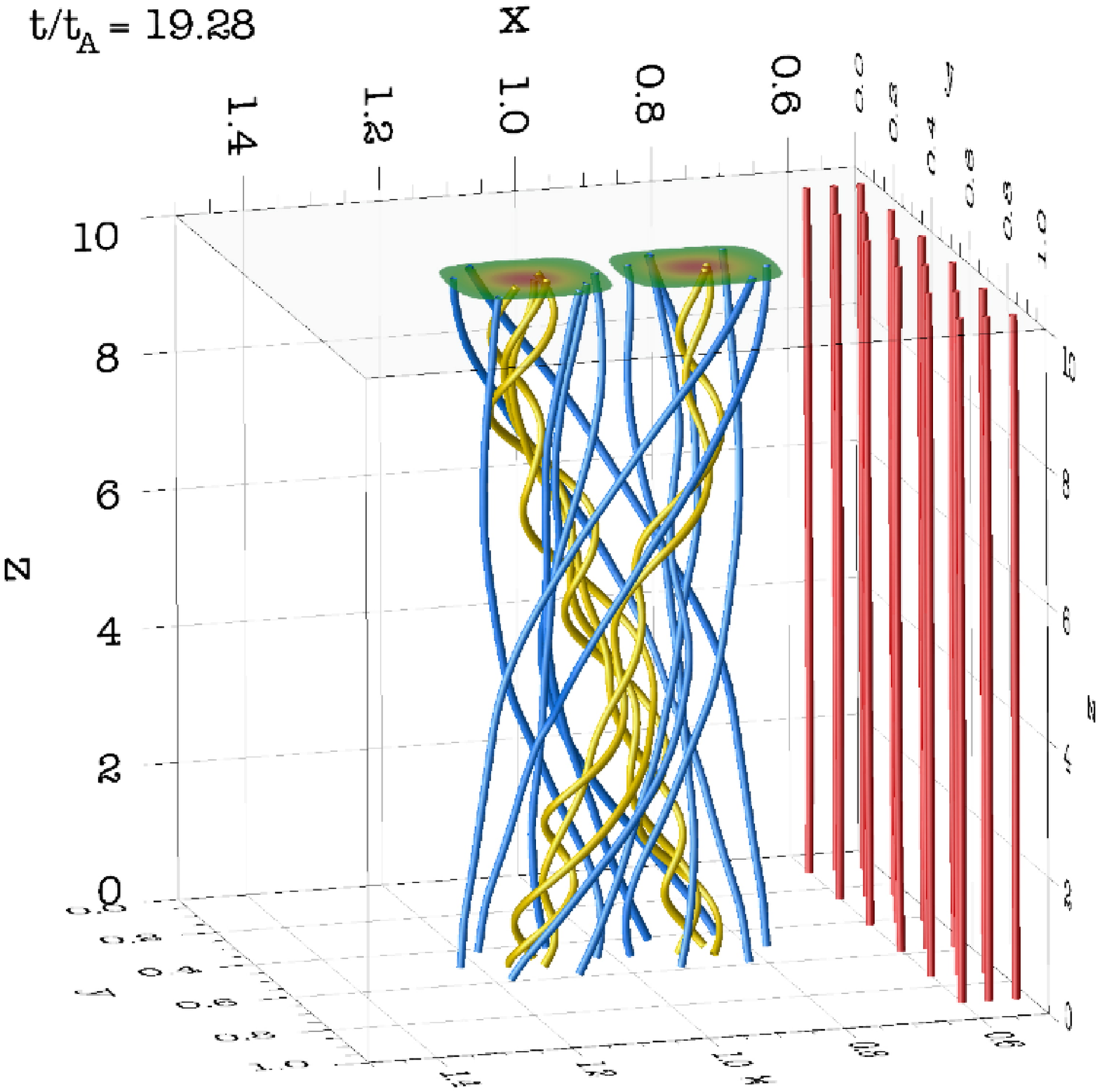}\\[.5em]
		\includegraphics[scale=.35]{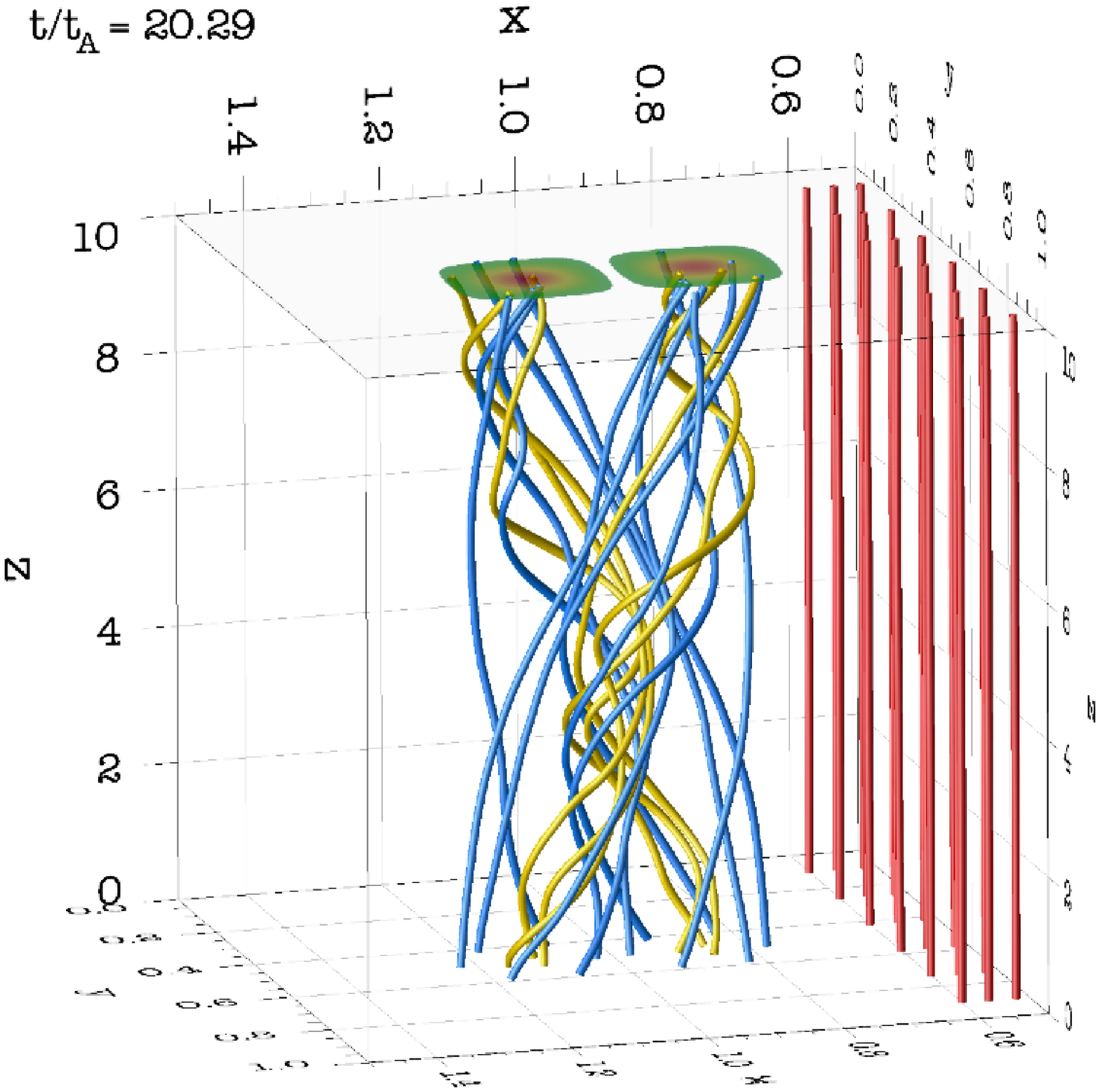}\hspace{1em}
		\includegraphics[scale=.35]{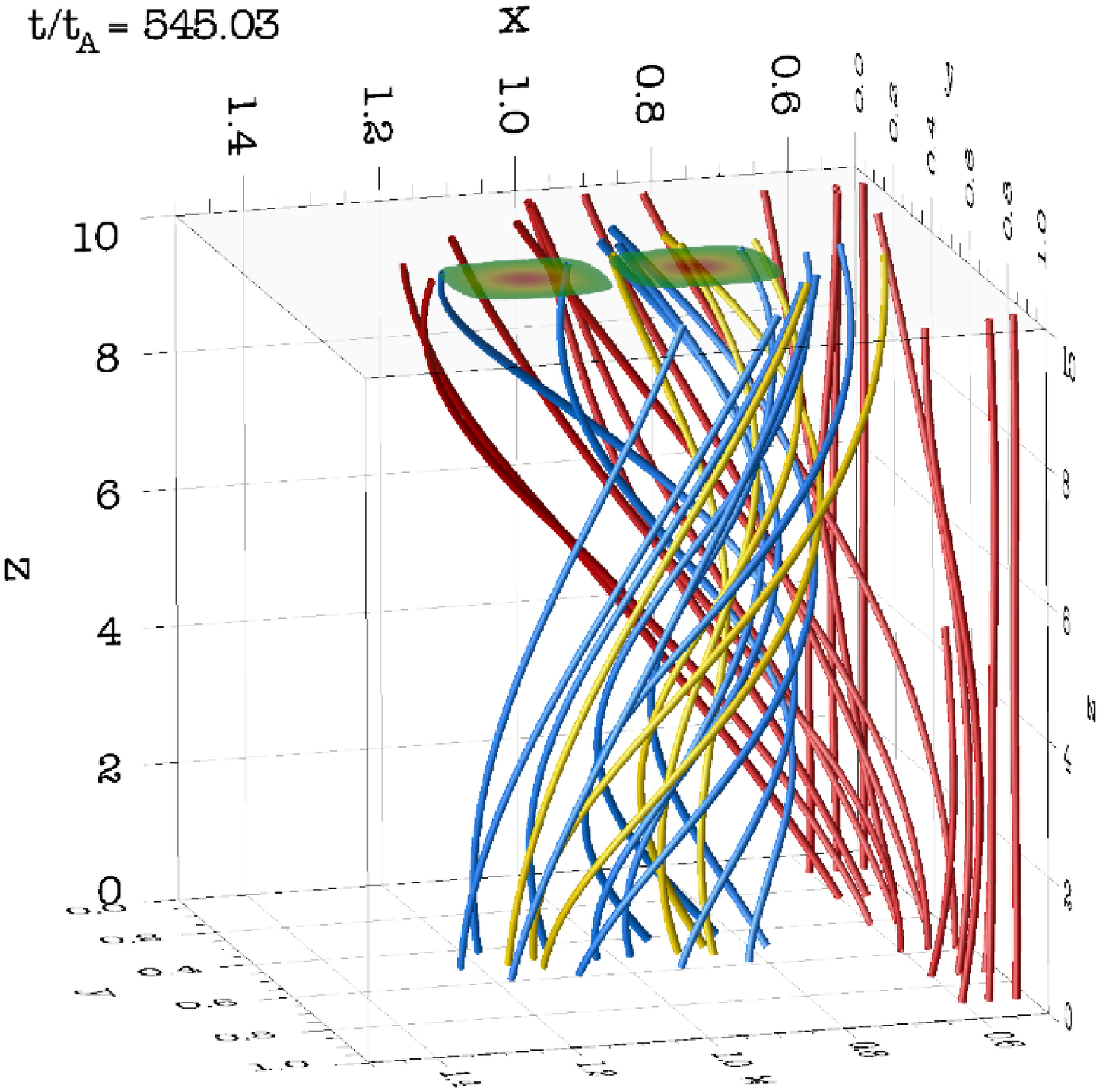}\\[.3em]
		\caption{\emph{Run A}: Three-dimensional views of magnetic field lines at selected times (they are always traced from the same locations in the motionless photospheric boundary z=0, as visible in the panels). In the linear stage ($t < 10\, \tau_A$) the corotating boundary vortices (shown in color in the plane $z=10$) twist into helices the magnetic field lines in the corresponding underlying regions. Those outside these two regions remain straight at first (a sample of is shown in red). As magnetic reconnection starts to develop in the central region (around z=5) field lines start changing connectivity (already visible at t=16.25~t$_A$). As it progresses more field lines change connectivity in and between the two original flux tubes (t=18.28, 19.28, and 20.29~t$_A$). Later on, as the magnetic energy inverse cascades progresses, the region where field lines are twisted increases its transverse scale including also field lines that were not twisted initially by the boundary vortices during the linear stage (confront t=15.24~t$_A$ with t=545.03~t$_A$). 
The box has been rescaled for an improved visualization, the axial length (along $z$) is ten times the length of the  orthogonal cross section (along $x$-$y$). \label{fig:fig3}}
	\end{centering}
\end{figure*}

Figure~\ref{fig:fig2} shows the field lines of the orthogonal magnetic field component $\mathbf{b}$ and the current density $j$ (in color) in the mid-plane z=5 at selected times. Initially the fields evolve linearly according to Equation~(\ref{eq:lin1}). At time t=0.61~t$_A$ magnetic field and current density are then a mapping of the boundary velocity and vorticity fields, with the field lines displaying their slight departure from a circular shape toward the vortices' edge. At later times the field line tension straightens them out into a circular shape (t=11.35~t$_A$). 

For a single vortex kink instability transitions to the nonlinear stage around t~$\sim$~84~t$_A$ \citep{2013ApJ...771...76R}. In the present double vortex case though there is not enough time for the internal kink mode to develop fully. Because the vortices are corotating, then the field lines of the orthogonal magnetic field \textbf{b} are oppositely directed between the two flux-tubes around x=1, and they can therefore reconnect. In fact around time t~$\sim$~15~t$_A$ the more external field lines start to reconnect in the region around x=1. The corresponding removal of magnetic pressure between the flux tubes further contributes to push the two flux tubes toward each other and consequently more flux is reconnected (see Figure~\ref{fig:fig2} at times t/t$_A$=17.47, 17.87, and 22.51).

\begin{figure}
	\begin{centering}
		\includegraphics[scale=.57]{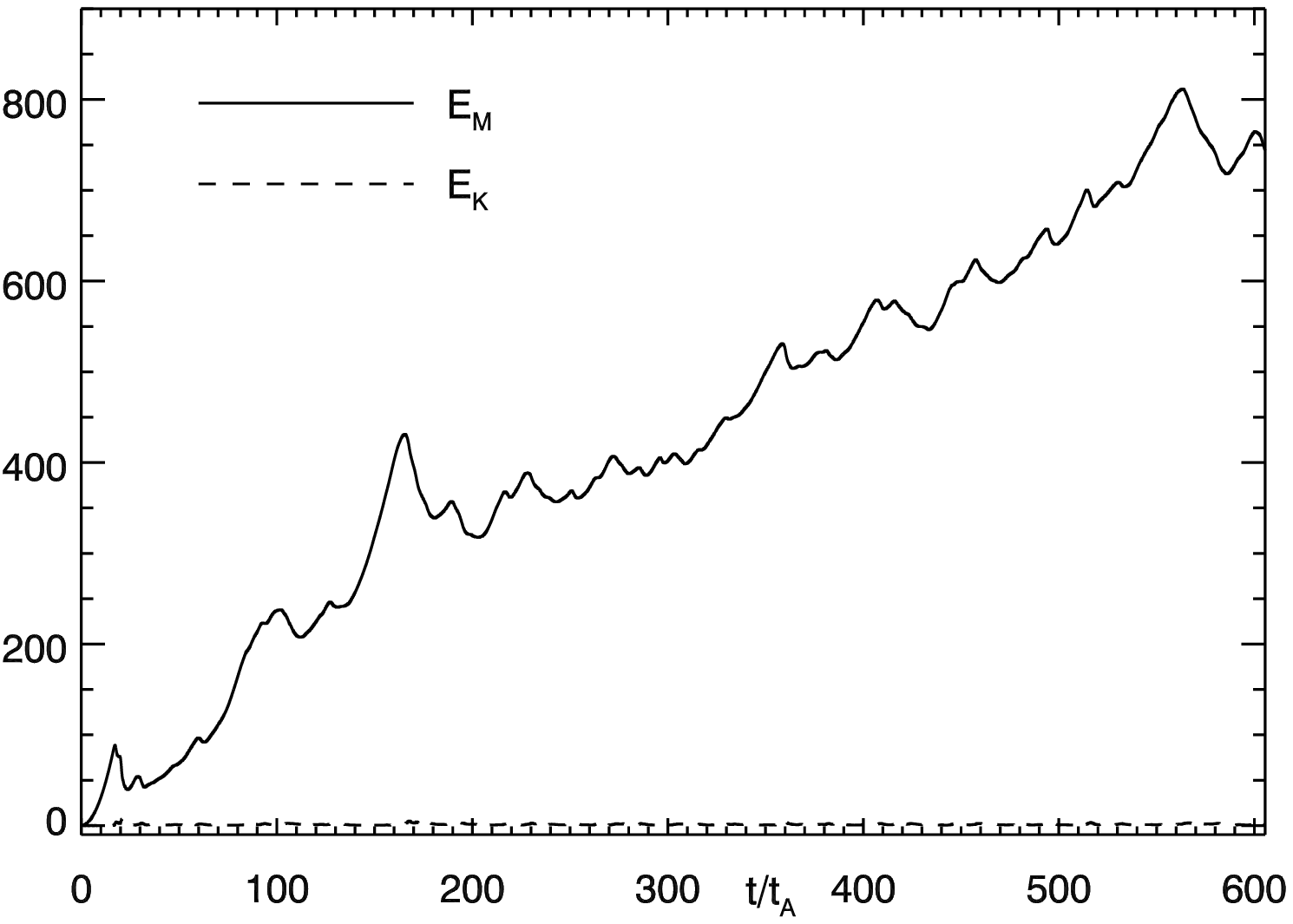}
		\caption{\emph{Run~A:} Magnetic ($E_M$) and kinetic ($E_K$) energies
			as a function of time. $\tau_A = L/B_0$ is the axial Alfv\'en crossing time. \label{fig:fig4}}
	\end{centering}
\end{figure}

As mentioned in Section~\ref{sec:sv} our investigation is aimed at understanding the dynamics in the fully nonlinear stage, when  strong magnetic fluctuations are present in the corona. In particular, as discussed in Section~\ref{sec:sv}, with a single vortex we have seen that after the kink instability develops the system never returns to a laminar state, rather the field lines exhibit a twist fluctuating around a mean value and an inverse energy cascade occurs so that the region of twisted field lines increases its transverse scale in time. 

Therefore we can reasonably suppose that the dynamics we have described so far for the run with two corotating vortices continues to hold also for the case of two separated vortices. In fact even if they are set apart a certain distance, after they transition to the nonlinear stage through kink instability, the inverse cascade will then make them bigger until they get in contact with one another and subsequently they will merge similarly to what already occurs for the simulation presented here. Nevertheless this scenario should be verified with numerical simulations in subsequent work.

Interestingly, once the two flux tubes merge, the subsequent nonlinear dynamics are very similar to the case with a single vortex, i.e., an inverse cascade of magnetic energy (with the mechanism discussed in Section~\ref{sec:sv} and \cite{2013ApJ...771...76R}) occurs and the transverse scale of the region with twisted magnetic field lines increases in time as shown in Figure~\ref{fig:fig2}, where at t=548.26~t$_A$ the transverse scale of the twisted nonlinear flux tube has reached the box size in the y-direction.

A three-dimensional view of selected field lines of the total magnetic field (i.e, $B_0 \mathbf{\hat{e}}_z + \mathbf{b}$) is shown in Figure~\ref{fig:fig3} at different stages of the dynamics. The two photospheric vortices are shown in the plane z=10 in color scale. As expected because of line-tying in z=0 and z=10, magnetic reconnection is more active in the central region of the numerical box (extended around z=5). As magnetic reconnection occurs field lines change connectivity, at first (t=18.28, 19.28, and 20.29~t$_A$) interchanging connectivity between the two original flux tubes, and later on, once the inverse cascade of magnetic energy develops, the change of connectivity includes also field lines external to the two original flux tubes, and giving rise to an overall twisted pattern as can be seen from Figure~\ref{fig:fig3} at t=545.03~t$_A$ and Figure~\ref{fig:fig2} at t=548.26~t$_A$.

Further insight into the dynamics is given by the temporal evolution of the total magnetic and kinetic energies
\begin{equation}
E_M = \frac{1}{2} \int \! \mathrm{d}V\, \mathbf{b}^2, \qquad
E_K = \frac{1}{2} \int \! \mathrm{d}V\, \mathbf{u}^2, \label{eq:en}
\end{equation}
the total ohmic dissipation rate
\begin{equation}
J = \frac{(-1)^{n+1}}{Re_n} \int \! \mathrm{d}V\, \mathbf{b} \cdot \nabla^{2n} \mathbf{b}, \label{eq:ohm}
\end{equation}
and the integrated Poynting flux $S$, i.e., the power injected from the top boundary by the work done by convective motions on the field lines' footpoints
\begin{equation}
	S = B_0 \int\limits_{z=L} \! \mathrm{d}a\, \mathbf{b} \cdot \mathbf{u}^L. \label{eq:poy}
\end{equation}
Because in these simulations viscous dissipation is much smaller than ohmic dissipation the system obeys the energy equation $\mathrm{d} E/\mathrm{d}t = S-J$, where $E$=$E_M+E_K$ \citep[e.g.,][]{2008ApJ...677.1348R}.
\begin{figure}
	\begin{centering}
		\includegraphics[scale=.57]{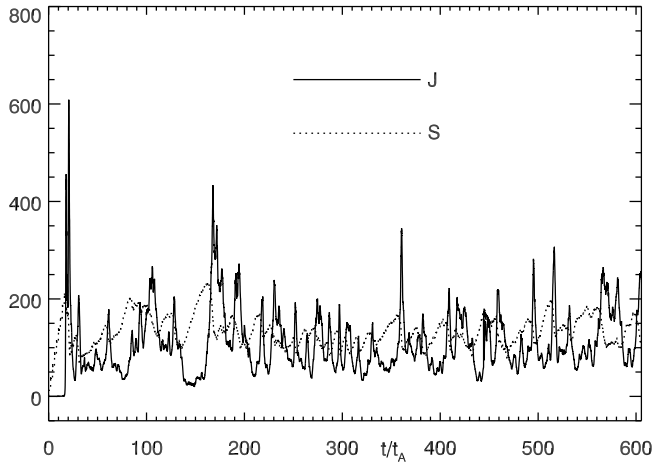}
		\caption{\emph{Run~A:} Ohmic ($J$) dissipation rate and
			the integrated Poynting flux $S$ (the injected power) versus 
			time. Viscous dissipation is negligible respect to the
			ohmic contribution. \label{fig:fig5}}
	\end{centering}
\end{figure}

The ohmic dissipation rate (Eq.~\ref{eq:ohm}) can be derived from the reduced MHD equations (\ref{eq:eq1})-(\ref{eq:eq2}), and is the generalization to the hyperdiffusive case of the dissipative term obtained for standard diffusion with n=1, for which from Eq.~(\ref{eq:ohm}) we can obtain the more familiar form $J=\int \! \mathrm{d}V\, \mathbf{j}^2/Re$ \citep[e.g., see][]{2010PhDT.......193R}. These integrated quantities are shown in Figures~\ref{fig:fig4} and \ref{fig:fig5}. They display a very similar time evolution to the case with a single vortex \citep{2013ApJ...771...76R}, because the increase in magnetic energy (Figure~\ref{fig:fig4}) is determined by the inverse cascade that increases steadily the volume where field lines are twisted in the computational box, while kinetic energy remains approximately constant throughout the nonlinear stage. The main difference is that for the single vortex case there is a strong dissipative event around time t~$\sim$~85~t$_A$, when kink instability transitions to the nonlinear stage, with a dissipation peak for $J$ about 150 larger than the average values at subsequent times. On the other hand in the present case with two corotating vortices ohmic dissipation has statistically steady fluctuations in the nonlinear stage, similarly to the single vortex case (we do not plot the viscuous dissipation rate, because in both cases it is much smaller), but lacks such a strong dissipative peak in the transition from the linear to the nonlinear stage, because as discussed previously magnetic reconnection between the two flux tubes starts earlier around t~$\sim$~12~t$_A$, so that there is less magnetic energy available to dissipate and kink instability has also faster dynamics. Additionally as shown in Figure~\ref{fig:fig5}, the integrated Poynting flux $S$ shows that energy is injected into the computational box in a statistically steady fashion, but with fluctuations that are on average higher than the energy dissipation rate, there is therefore on average a surplus of injected energy to sustain the inverse cascade.

Simulations with two and seven corotating vortices next to each other have been recently carried out by \cite{Zhao2015}. Nevertheless these simulations are stopped when the magnetic islands of the orthogonal magnetic field merge thus forming a single twisted flux tube. Therefore they do not observe the propagation of the magnetic field twist in the surrounding region, that on the opposite is clearly observed in our simulation (e.g., see Figures~\ref{fig:fig2} and \ref{fig:fig3}), thus neglecting this physical mechanism for an inverse cascade of magnetic energy and field line twist. 

More recently \cite{Reid2018} have performed numerical simulations with three corotating vortices. But they are isolated only in the y-direction while along~x because of periodicity they form an infinite chain of adjoining vortices. Although not specifically discussed in the paper, from their figures an inverse cascade not directly linked to the merging of magnetic islands appear to occur along the y-direction in the fully nonlinear stage with a mechanism similar to what described here for two vortices after the flux tubes merge and in \cite{2013ApJ...771...76R} for a single vortex. Their manuscript focuses mostly on MHD avalanches for SOC models and their impact on coronal heating. We will discuss this aspect in Section~\ref{sec:runc} and in our concluding section.

\subsection{Run~B} \label{sec:runb}

\begin{figure*}
	\begin{centering}
		\includegraphics[scale=.31]{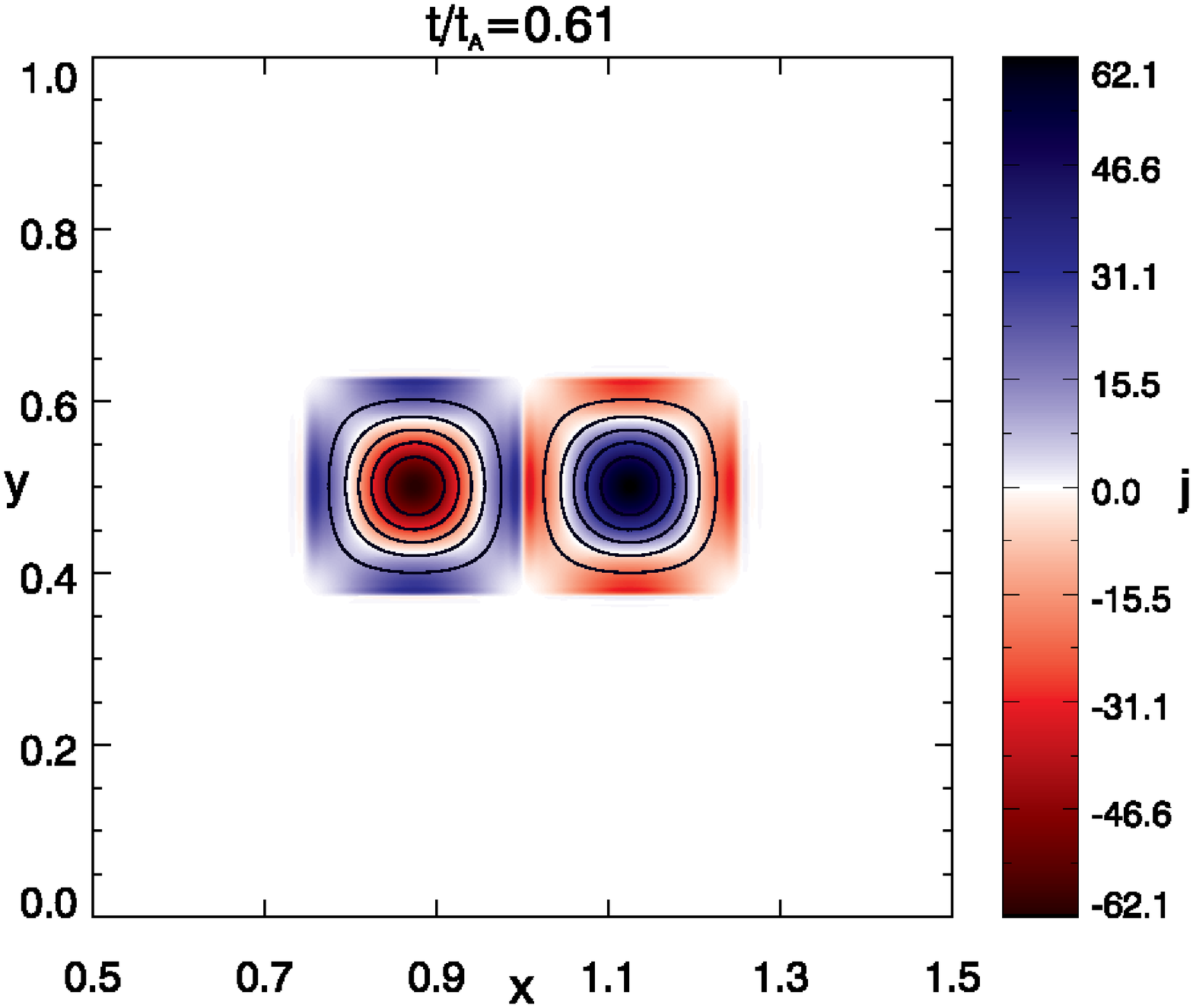}
		\includegraphics[scale=.31]{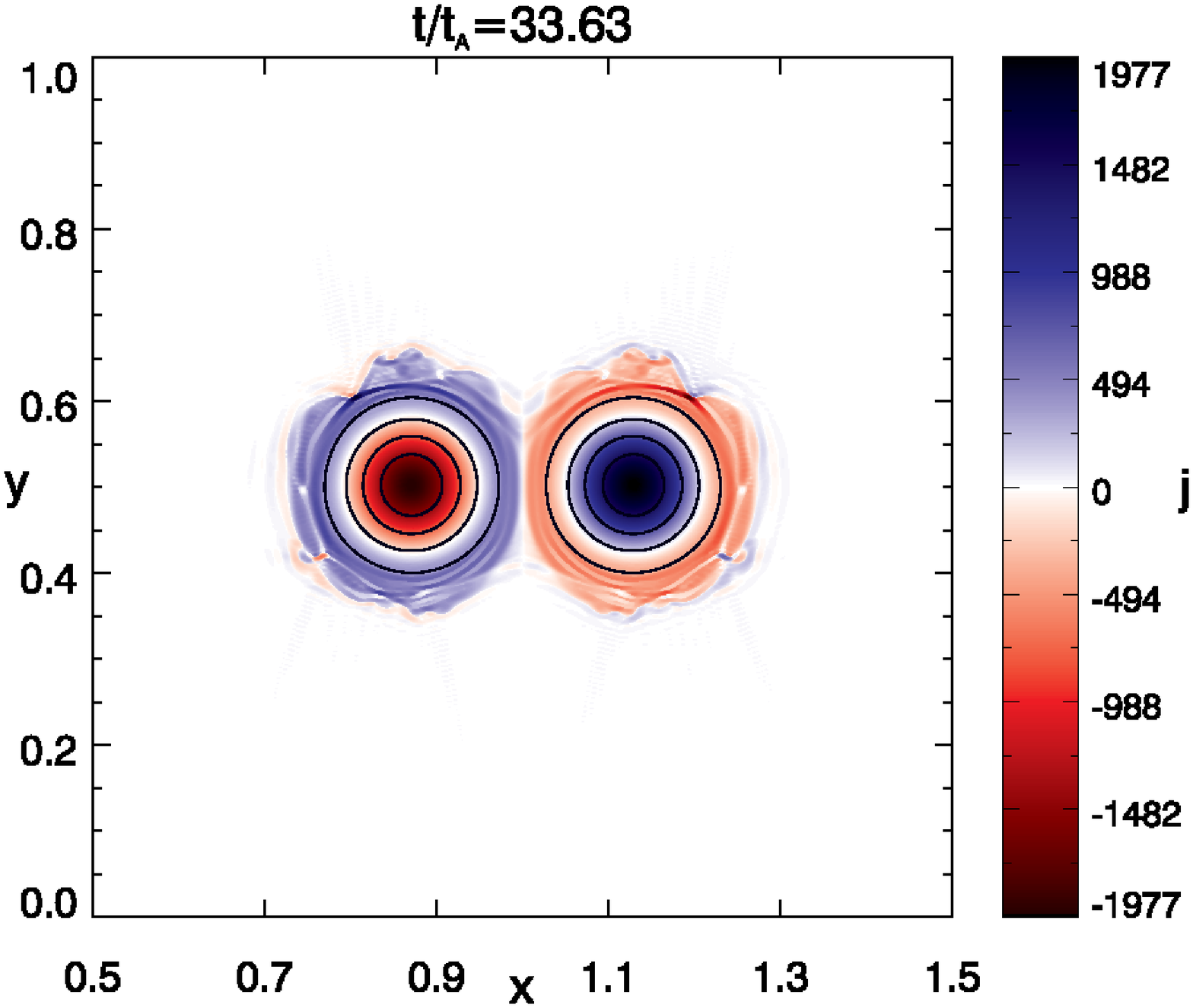}\\[1em]
		\includegraphics[scale=.31]{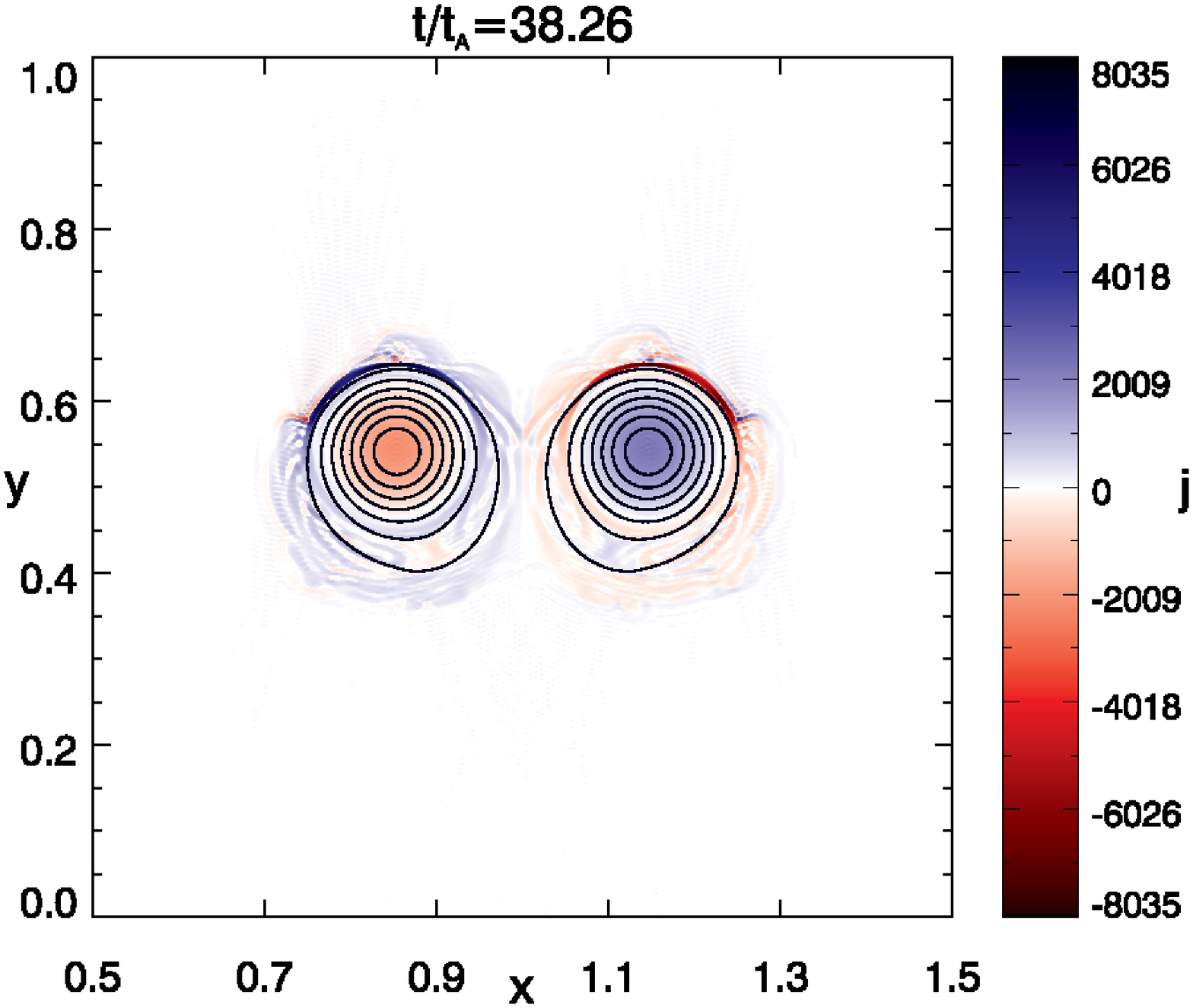}
		\includegraphics[scale=.31]{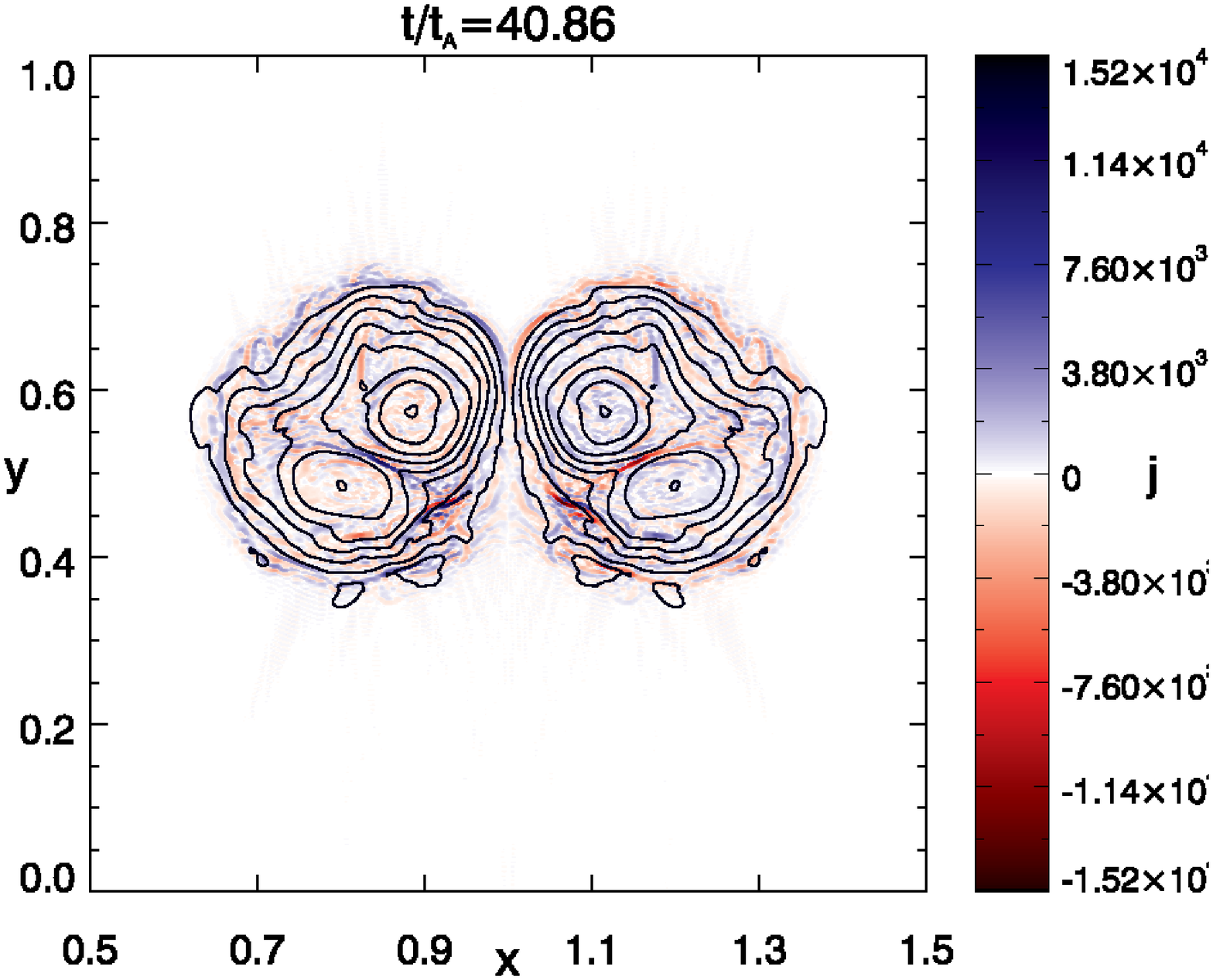}\\[1em]
		\includegraphics[scale=.31]{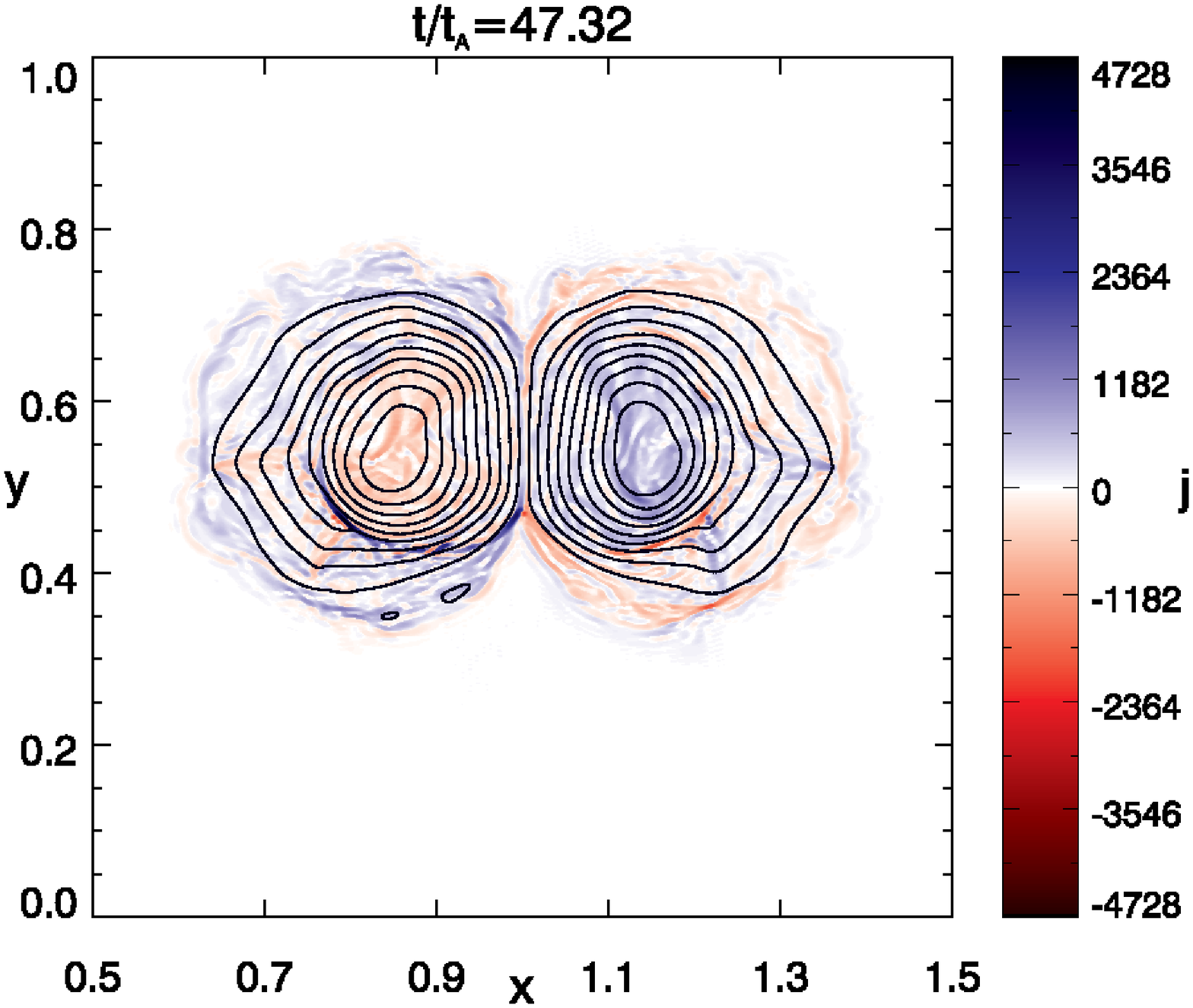}
		\includegraphics[scale=.31]{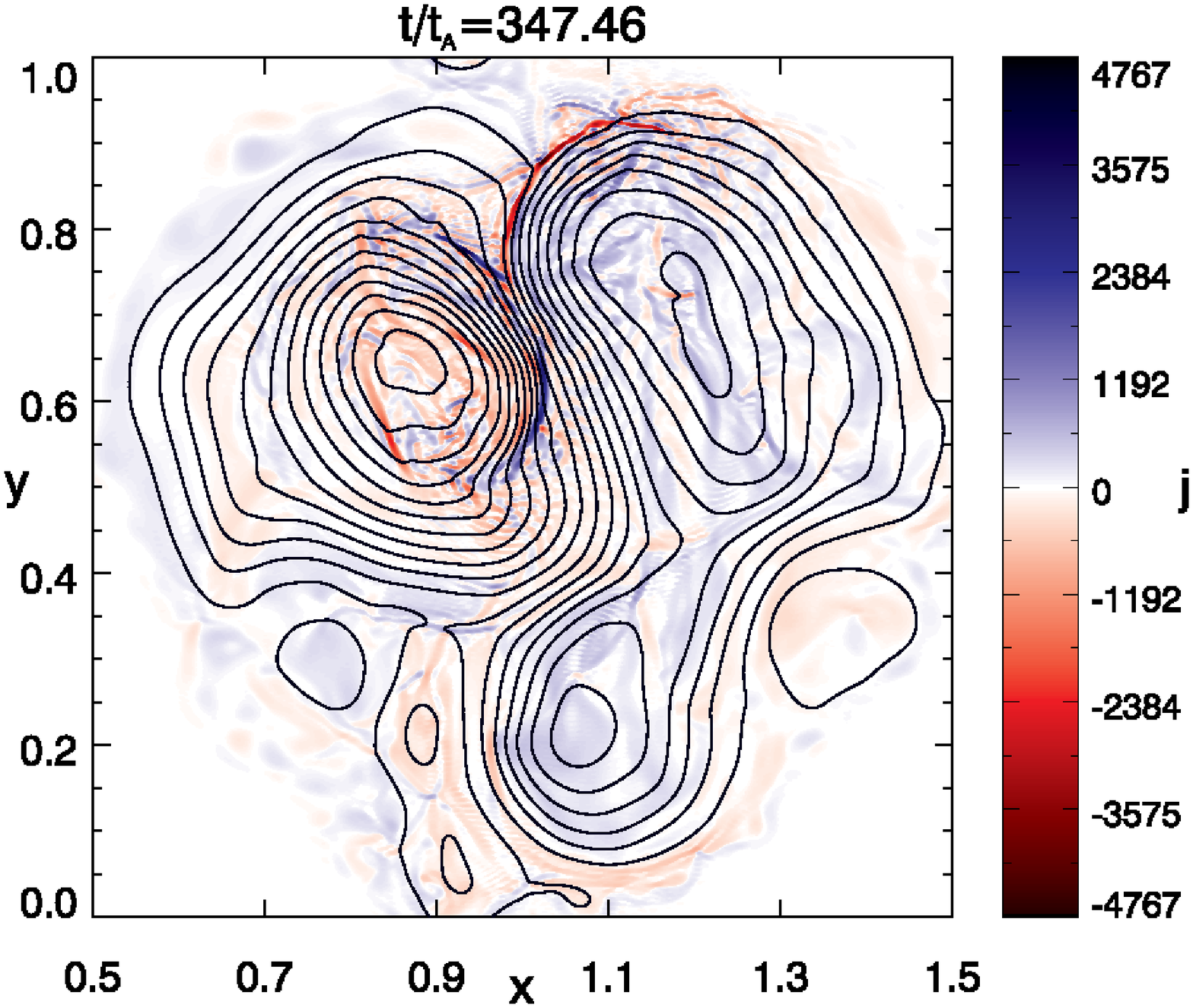}
		\caption{\emph{Run~B -- Counter-rotating vortices}: Axial component of the current $j$ (in color) and field lines 
			of the orthogonal magnetic field in the midplane ($z=5$) at selected times covering 
			the linear and nonlinear regimes up to $t\sim 600\, \tau_A$.
			At the beginning of the linear stage ($t = 0.61\, \tau_A$)
			the orthogonal magnetic field is a mapping of the boundary vortex
			[see linear analysis, Equation~(\ref{eq:lin1})].
			Still in the linear stage but at later times ($t = 80.64\, \tau_A$)
			the field line tension straightens out in a circular shape
			the vortex mapping. An internal kink mode develops ($t \sim 83.85\, \tau_A$)
			and the instability transitions the system to the nonlinear stage.
			In the fully nonlinear stage the field lines are still circular, but in a disordered
			way, exhibit a broad range of scales, including current sheets, and steadily 
			occupy a larger fraction of the computational box. \\
			(An animation of this figure is available.) \label{fig:fig6}}
	\end{centering}
\end{figure*}
\begin{figure*}
	\begin{centering}
		\includegraphics[scale=.35]{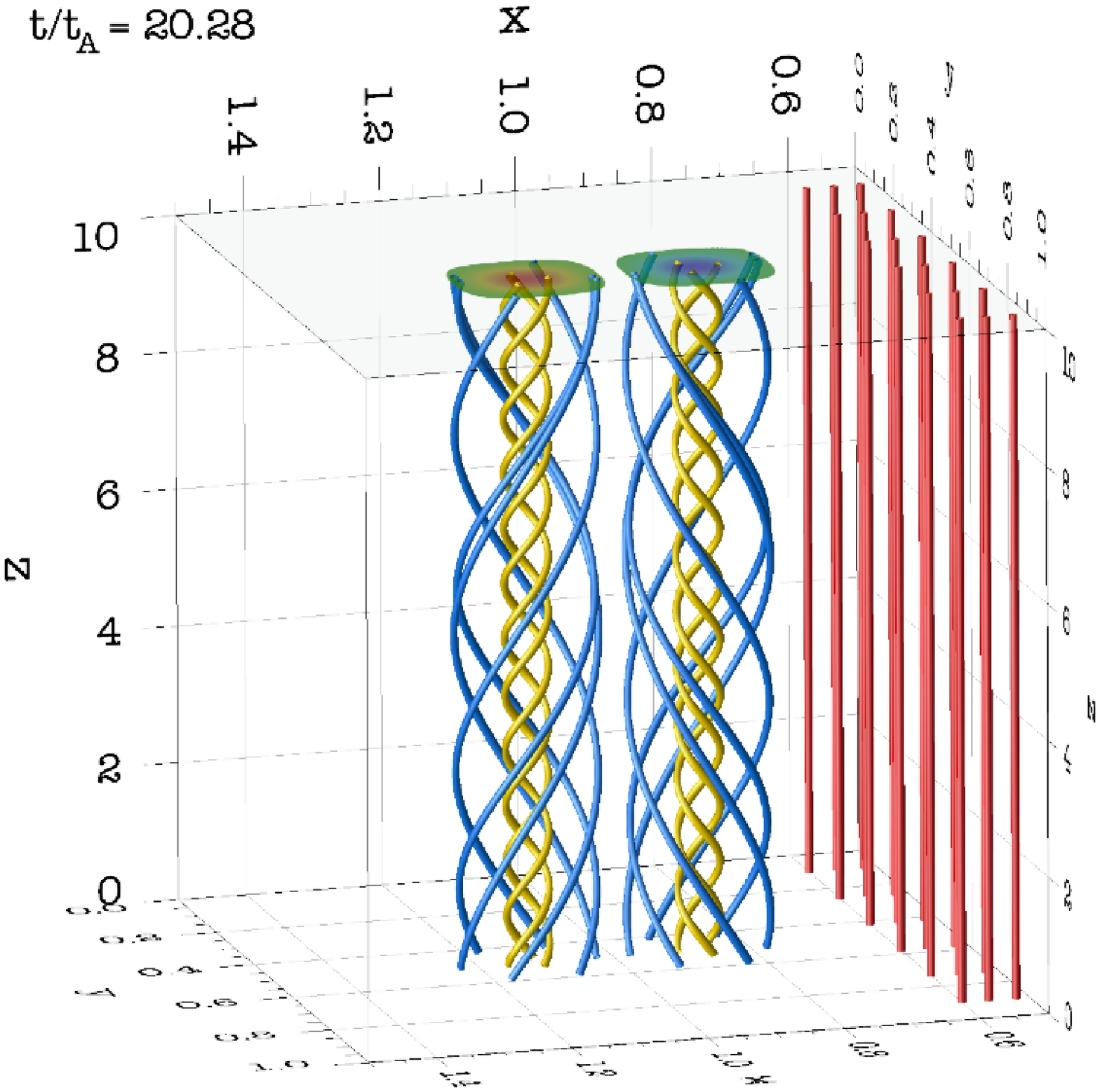}\hspace{1em}
		\includegraphics[scale=.35]{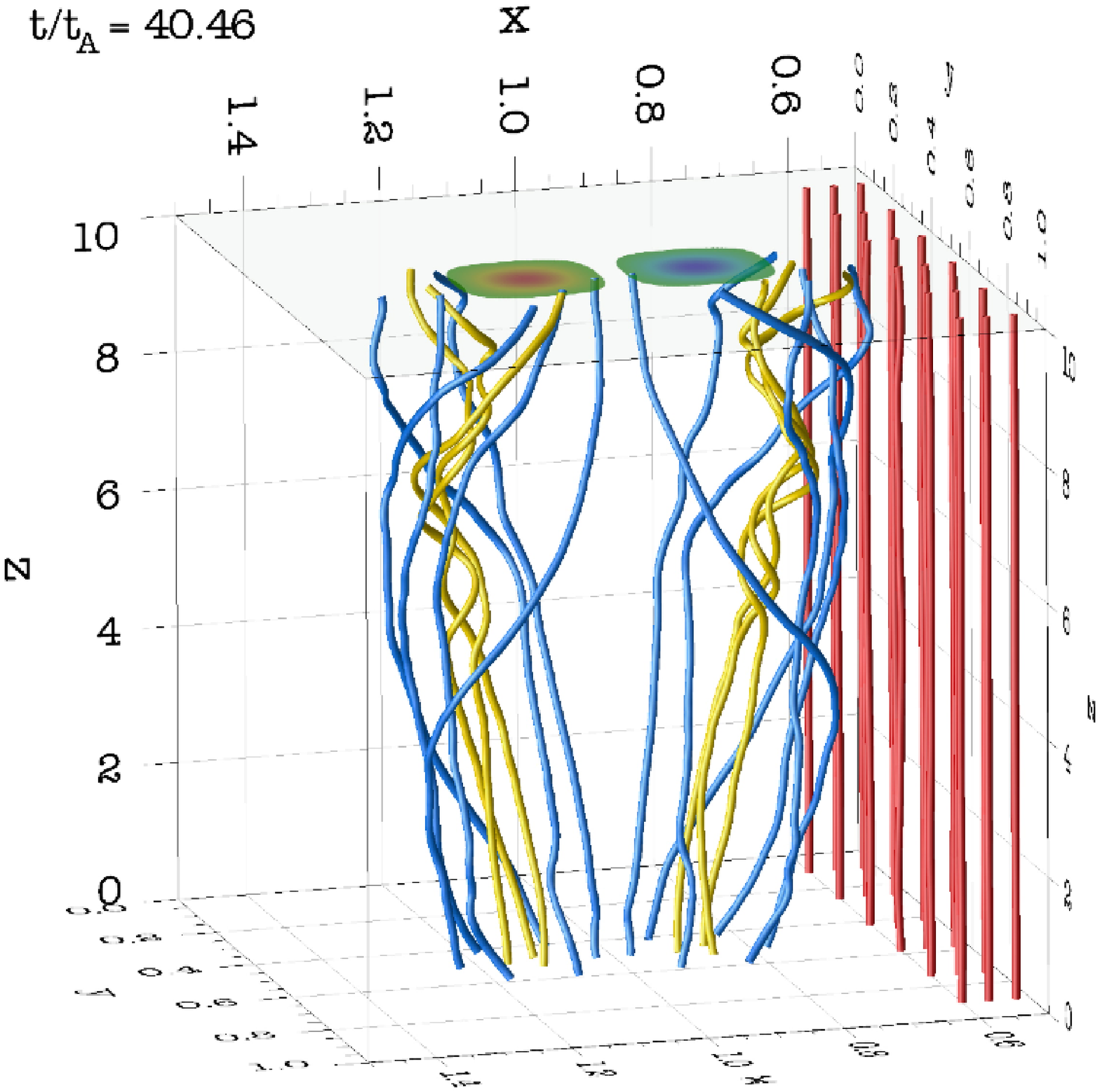}\\[.5em]
		\includegraphics[scale=.35]{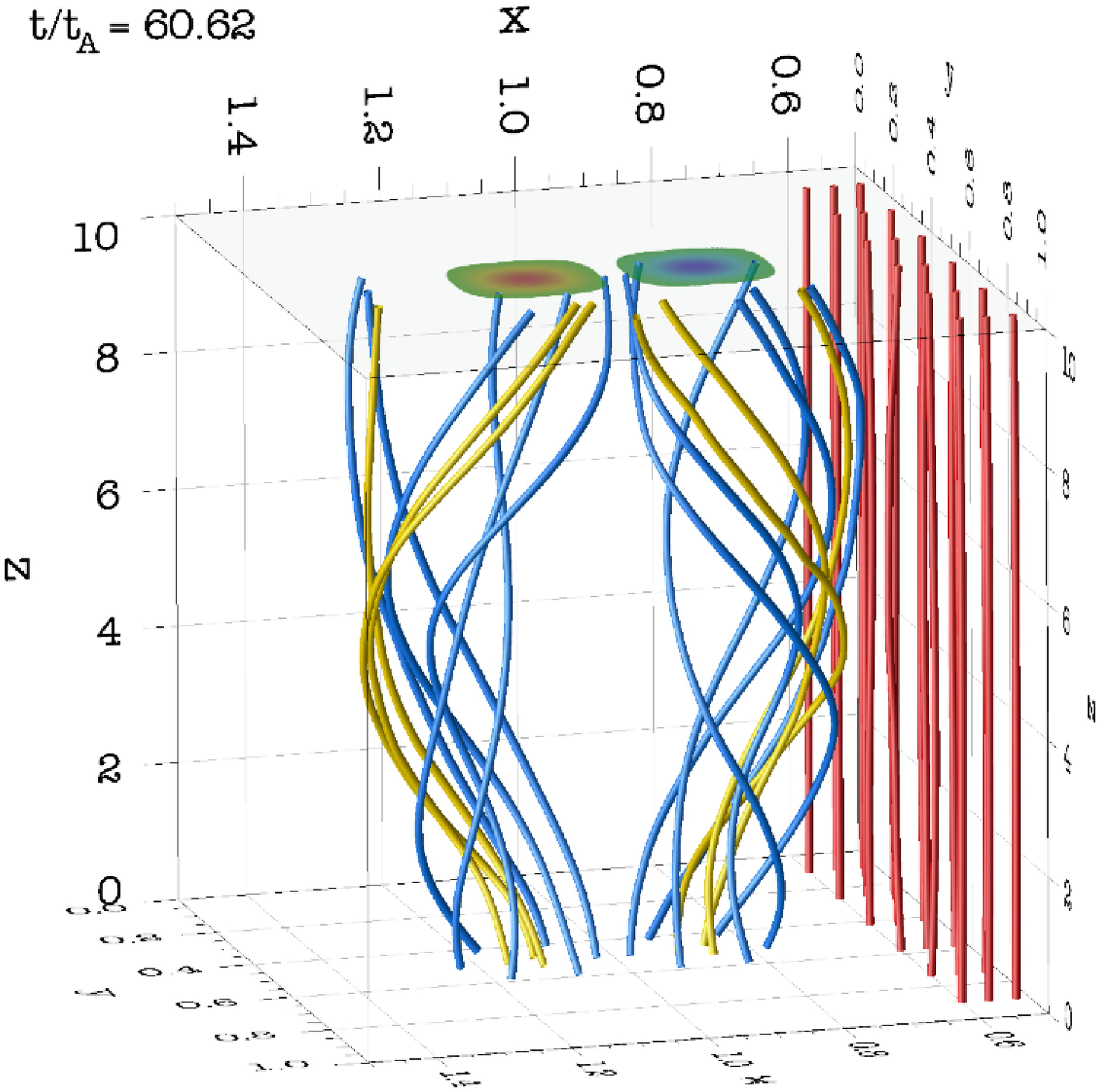}\hspace{1em}
		\includegraphics[scale=.35]{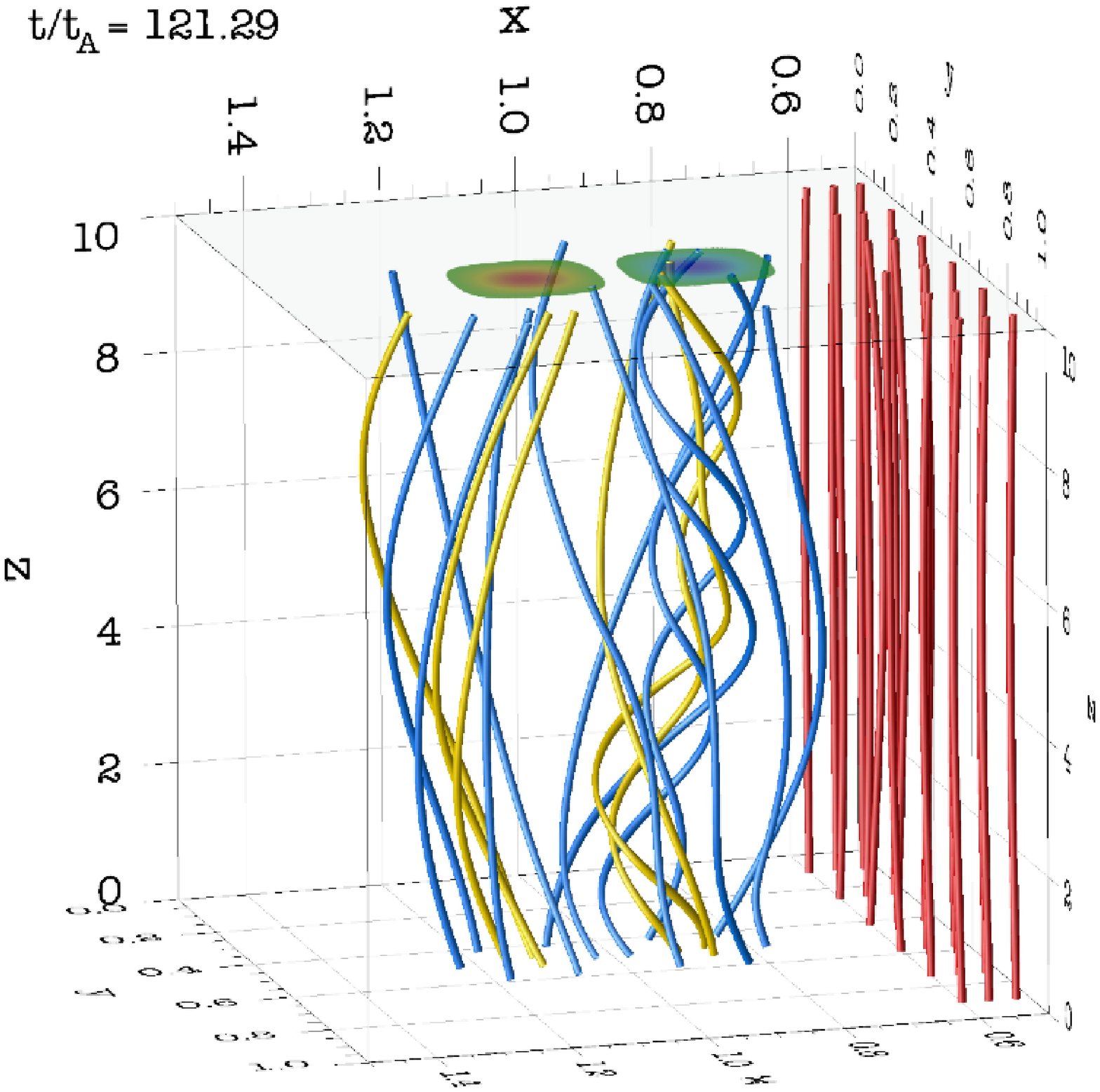}\\[.5em]
		\includegraphics[scale=.35]{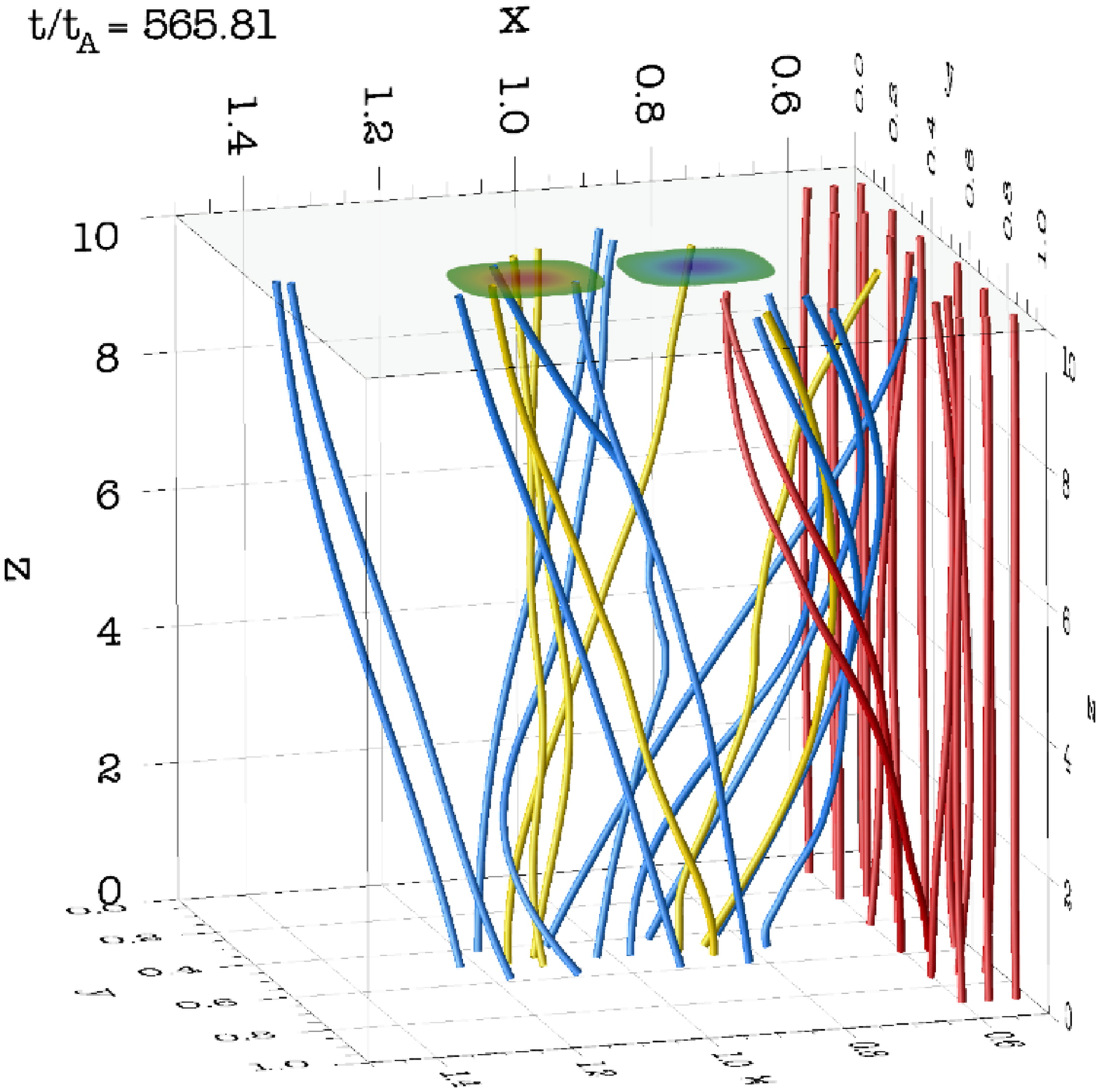}\hspace{1em}
		\includegraphics[scale=.35]{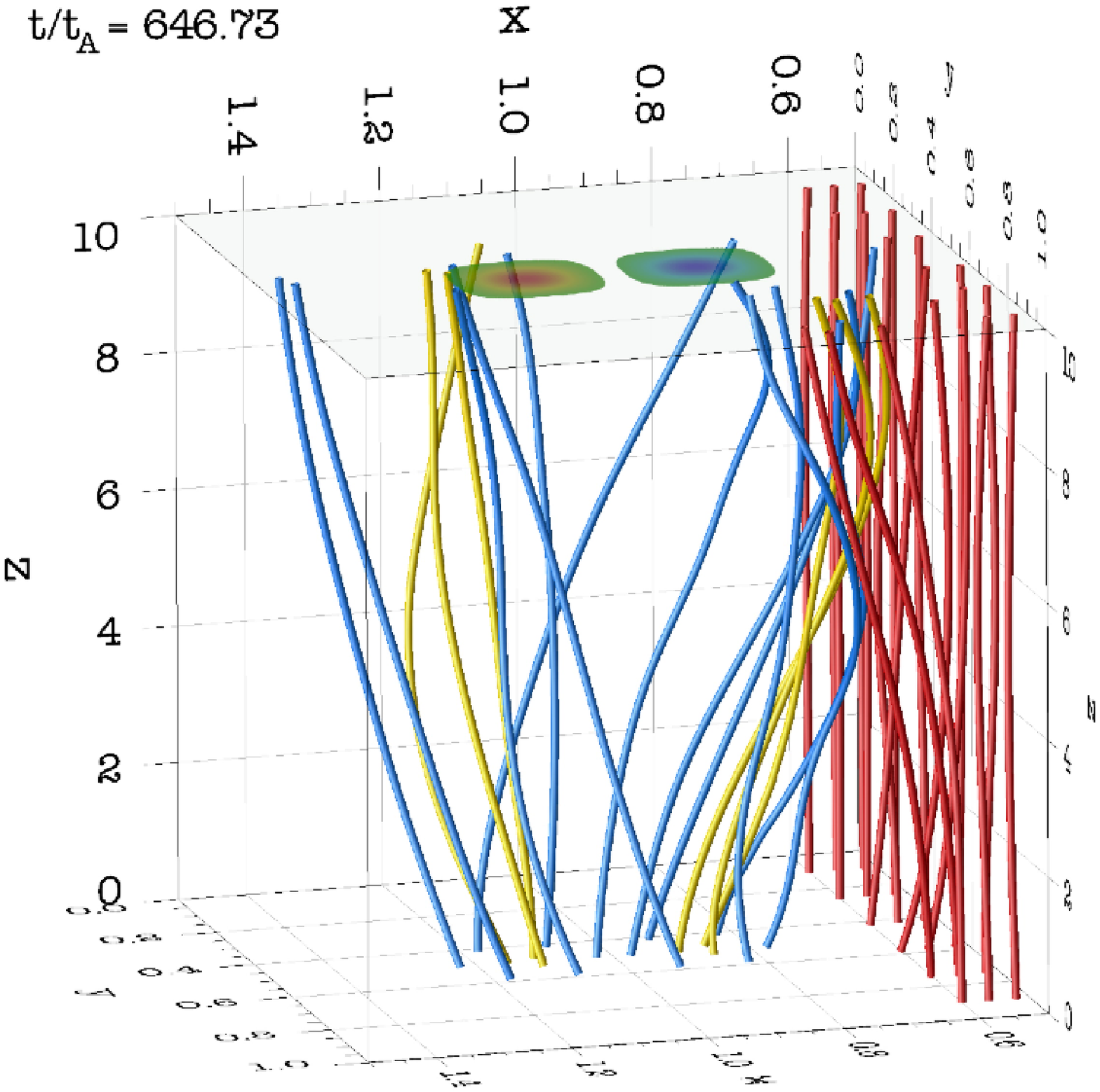}\\[.3em]
		\caption{\emph{Run B}: Three-dimensional view of magnetic field lines at selected times (always traced from the same locations in the motionless photospheric boundary z=0, as visible in the panels). In the linear stage ($t = 20.28\, \tau_A$) the counter rotating boundary vortices (shown in color in the plane $z=10$) twist into helices the magnetic field lines in the corresponding regions underneath them. Those outside this region remain straight, a sample of which is shown in red. Kink instability releases magnetic energy and untwists the field lines ($t = 40.46\, \tau_A$). In subsequent dynamics field lines between the two flux tubes do not interchange connectivity among them, but a moderate inverse cascade occurs slightly enlarging both flux tubes ($t$=60.62 and 121.29 $\tau_A$), until it saturates ($t$=565.81 and 646.73 $\tau_A$) as a balance between Poynting flux and energy dissipation is achieved (Figures~\ref{fig:fig8} and \ref{fig:fig9}). 
The box has been rescaled for an improved visualization, the axial length (along $z$) is ten times the length of the  orthogonal cross section (along $x$-$y$). \label{fig:fig7}}
	\end{centering}
\end{figure*}

The simulation discussed in this section, run~B, has almost all parameters as run~A discussed in the previous section, except that the two photospheric vortices in the top plate z=L are counter-rotating, i.e., they rotate in opposite directions. A minor difference is a slightly higher value for the hyperdiffusion coefficient that is now $Re_4 = 5\times10^{19}$. The Alfv\'en velocity is still $B_0 = 200$, corresponding to 200~km/s in conventional units. The total duration is  slightly longer at $\sim$~650 axial Alfv\'en crossing times $\tau_A = L/B_0$, where L=10 is the axial box length. 

Figure~\ref{fig:fig6} shows the field lines of the orthogonal magnetic field component $\mathbf{b}$ and the current density $j$ (in color) in the mid-plane z=5 at selected times. As seen for run~A the fields initially evolve linearly according to Equation~(\ref{eq:lin1}). At time t=0.61~t$_A$ the magnetic field and current density are then a mapping of the boundary velocity and vorticity fields, and they display oppositely directed currents in the flux tube centers (and consequently oppositely directed return current at their edges so than the integrated current vanishes in each flux tube). Again the field lines have  a slight departure from a circular shape toward the flux tube edges, and at later times the field line tension straightens them out into a circular shape (e.g., see t=33.63~t$_A$). 

In contrast to run~A now the magnetic field lines of \textbf{b} are parallel along the boundary between the two flux tubes around the plane x=1, and therefore they \emph{cannot reconnect}. We then expect the dynamics to be very different because the flux tubes cannot merge as they do in run~A. In fact now the internal kink mode develops to some extent in similar fashion to the case with a single vortex. As shown in Figure~\ref{fig:fig6} at t=38.26~t$_A$ kink instability transitions to the nonlinear stage. For a single vortex the transition to the nonlinear stage occurs at a later time (t$\sim$84~t$_A$) with much more twisted field lines (Equation~(\ref{eq:lin1})). The quicker development of kink instability is probably due to the proximity of the two flux tubes that in the process of straightening into a circular shape their orthogonal magnetic field lines produce, due to their proximity, an additional perturbation to the magnetic field. 

Once the system transitions to the fully nonlinear stage (t~$\gtrsim$~40~t$_A$) an inverse cascade appears to occur for the individual flux tubes, but in a limited manner respect to the single vortex case (t=47.32, and 347.46~t$_A$). In fact even at later stages it never proceeds beyond what can be observed at time t=347.46~t$_A$. A three-dimensional view of selected field lines at various stages of the dynamics is shown in Figure~\ref{fig:fig7}. At times t=565.81 and 646.73~t$_A$ the region where field lines are twisted has not increased its transverse scale beyond what already reached at t=347.46~t$_A$, i.e., the \emph{inverse cascade appears to have stopped} transferring energy toward the large scales.

\begin{figure}
	\begin{centering}
		\includegraphics[scale=.57]{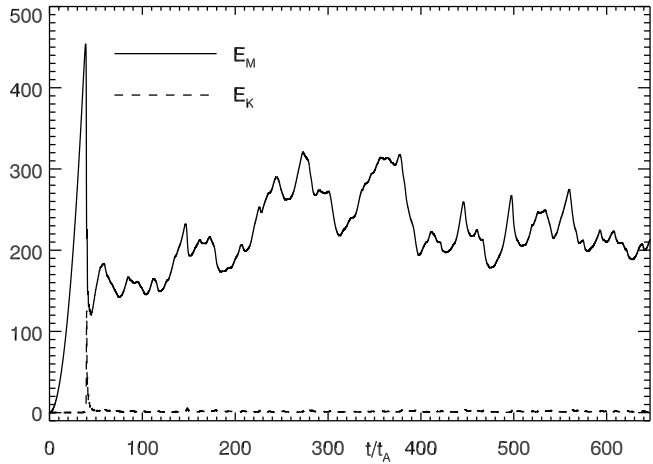}
		\caption{\emph{Run B}: Magnetic ($E_M$) and kinetic ($E_K$) 
			for counter rotating vortices.
			Magnetic ($E_M$) and kinetic ($E_K$) energies
			as a function of time for counter-rotating boundary vortices. $\tau_A = L/B_0$ is the axial Alfv\'en crossing time.
			\label{fig:fig8}}
	\end{centering}
\end{figure}

\begin{figure}
	\begin{centering}
		\includegraphics[scale=.57]{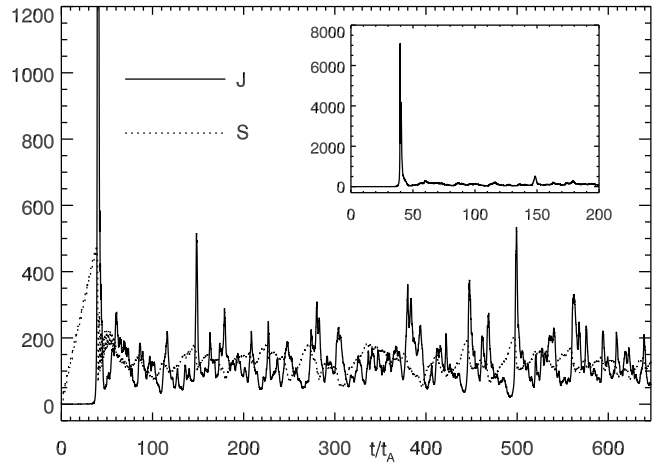}
		\caption{\emph{Run B}: Ohmic ($J$) dissipation rate and
			the integrated Poynting flux $S$ (the injected power) versus 
			time for counter-rotating boundary vortices. Inset shows the ohmic dissipative peak of kink instability around time t~$\sim$~40~t$_A$.
			\label{fig:fig9}}
	\end{centering}
\end{figure}

The temporal evolution of the total magnetic ($E_M$) and kinetic ($E_K$) energies in Figure~\ref{fig:fig8}, and of the ohmic dissipation rate $J$ and integrated Poynting flux $S$ in Figure~\ref{fig:fig9} gives us further insight into the dynamics. We can notice around t~$\sim$~40~t$_A$ the big dissipative peak due to the transition of kink instability to the nonlinear stage (see also inset in Figure~\ref{fig:fig9}), and the corresponding energy drop in Figure~\ref{fig:fig8}. After kink instability a statistically steady state is reached for all quantities, including magnetic energy that now fluctuates around its mean value (Figure~\ref{fig:fig8}) instead of growing steadily in time as in the corotating vortices case (Figure~\ref{fig:fig4}). Additionally it can be seen that respect to run~A (Figure~\ref{fig:fig5}) now the integrated Poynting flux $S$ has the same average as the ohmic dissipation rate $J$, while for run~A it was distinctly higher thus originating the energy surplus to sustain an inverse cascade of energy.

\emph{The reason for the lack of an inverse cascade originates from the fact that this would make the single flux tubes larger in the transverse direction.} Because the field line topology does not allow the two tubes to reconnect and merge they would then displace each other in opposite directions thus de-centering each flux tube respect to the corresponding boundary vortex that originates them. On the other hand the power injected from the boundary is given by the integrated \emph{Poynting flux (Equation~(\ref{eq:poy})) that  depends crucially on the scalar product between the magnetic field \textbf{b} and the boundary velocity \textbf{u}$^L$.} 

Defining the \emph{correlation} between the two fields as $C(\mathbf{b}, \mathbf{u}^L) = \mathbf {b} \cdot \mathbf{u}^L / ( \langle b^2 \rangle \langle (u^L)^2 \rangle )^{1/2}$, where $\langle \ldots \rangle$ indicates the r.m.s. value in the plane z=L, and taking into account that magnetic energy is approximately constant along~z in this type of simulations \citep[e.g., see Figure~5 in][]{2008ApJ...677.1348R}, without writing explicitly all constant terms we obtain from Equation~(\ref{eq:poy}) that $S \propto E_M^{1/2} \langle C(\mathbf{b}, \mathbf{u}^L) \rangle$. \emph{The correlation and therefore~$S$ is clearly maximized when the flux tube is centered below the respective forcing vortex (with $\mathbf{b}$ approximately proportional to $\mathbf{u}^L$), while decreases when it gets de-centered.} Thus in the present simulation \emph{a balance must be reached} between the integrated Poynting flux~$S$ and the ohmic dissipation rate~$J$. In fact if $S$ were higher than $J$ as in the corotating vortices case (run~A), the surplus energy would sustain \emph{an inverse cascade, but that would de-center the flux tubes and decrease the Poynting flux} to the point that it matches the ohmic dissipation rate, thus \emph{stopping the inverse cascade}.

In run~A the \emph{two flux tubes merge} forming a single flux tube that continues to expand in the transverse direction through the inverse cascade, but \emph{the new flux tube remains approximately centered under the two photospheric vortices}. A closer inspection of the data shows that the field lines of the orthogonal component of the magnetic field $\mathbf{b}$ in the plane z=L  form a single magnetic island that in the fully nonlinear stage is alternatively centered under one of the two vortices or in between them, in any of these cases the correlation between $\mathbf{u}$ and $\mathbf{b}$ remains then positive at the boundary and the Poynting flux fluctuates around its average value that remains higher than the average dissipation rate (Figure~\ref{fig:fig5}).

\subsection{Run~C} \label{sec:runc}

\begin{figure*}
	\begin{centering}
		\includegraphics[scale=.31]{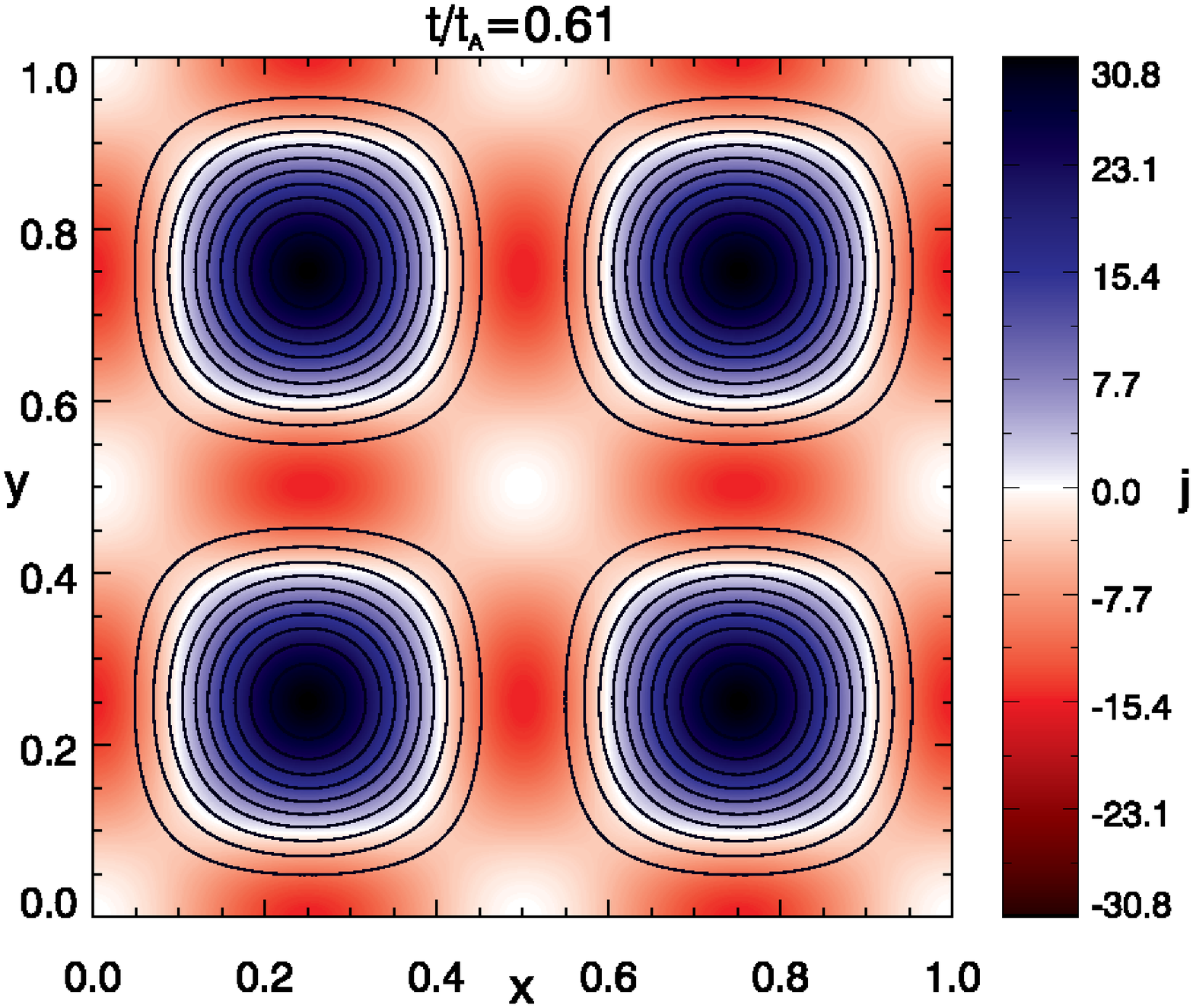}
		\includegraphics[scale=.31]{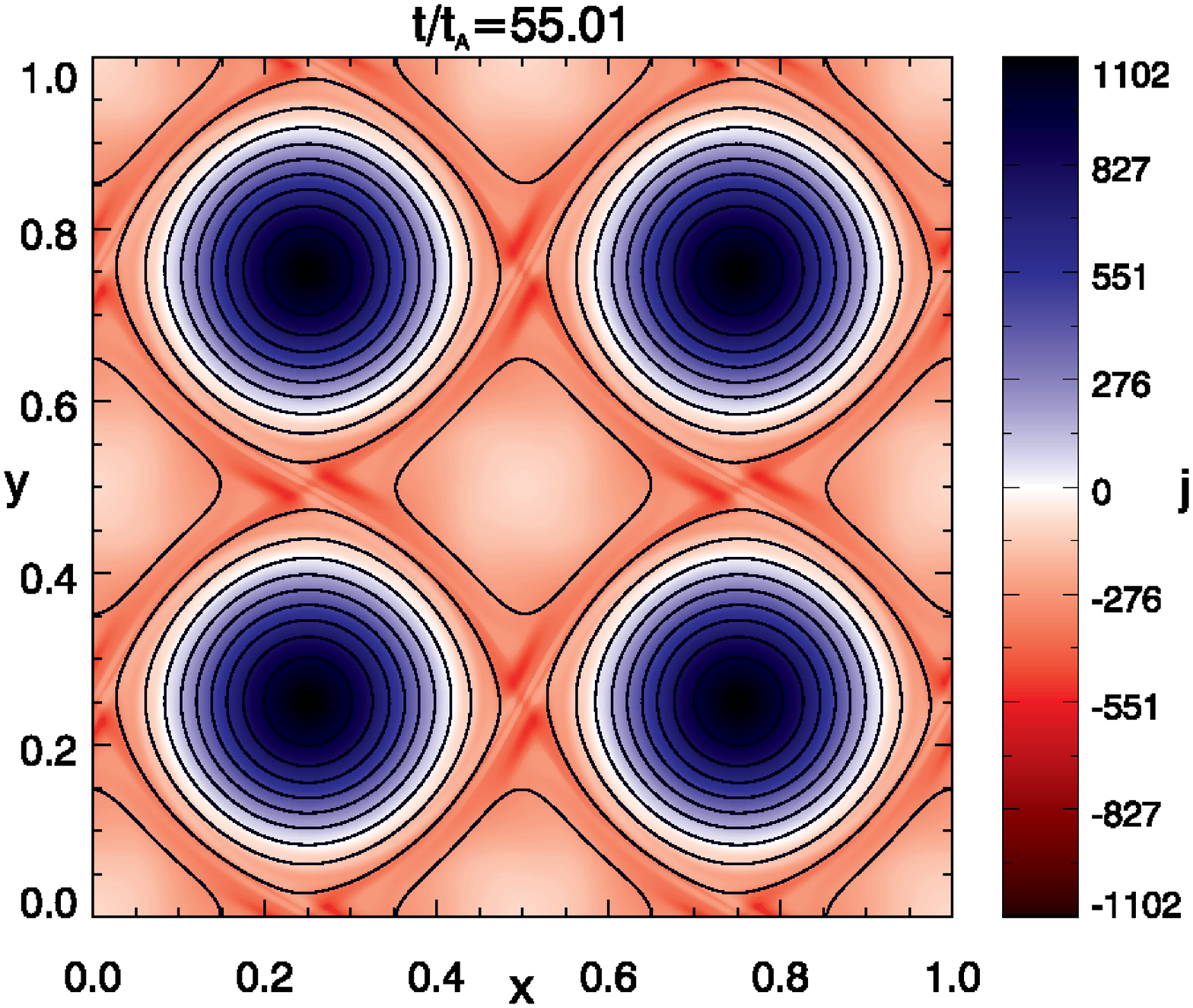}\\[1em]
		\includegraphics[scale=.31]{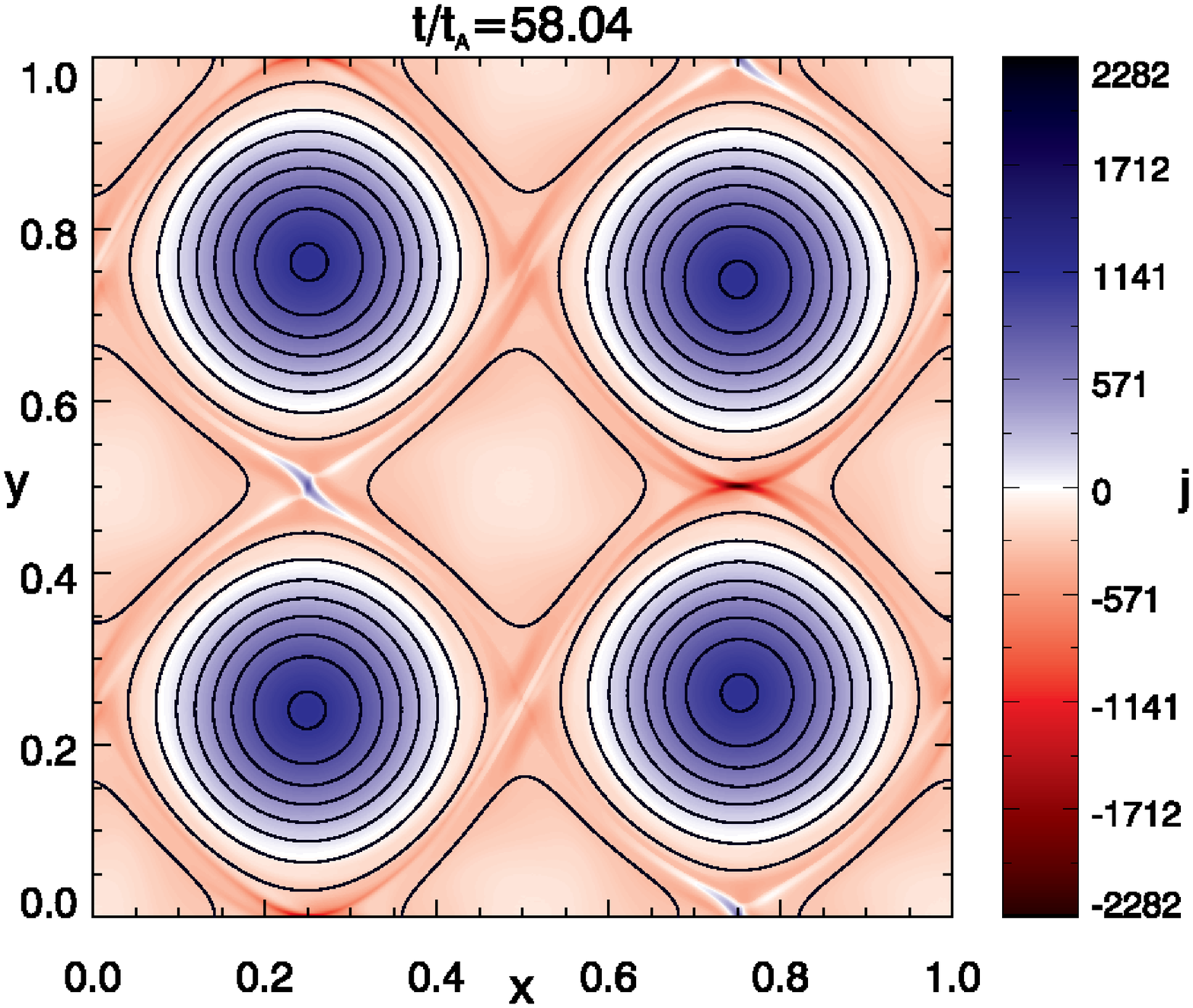}
		\includegraphics[scale=.31]{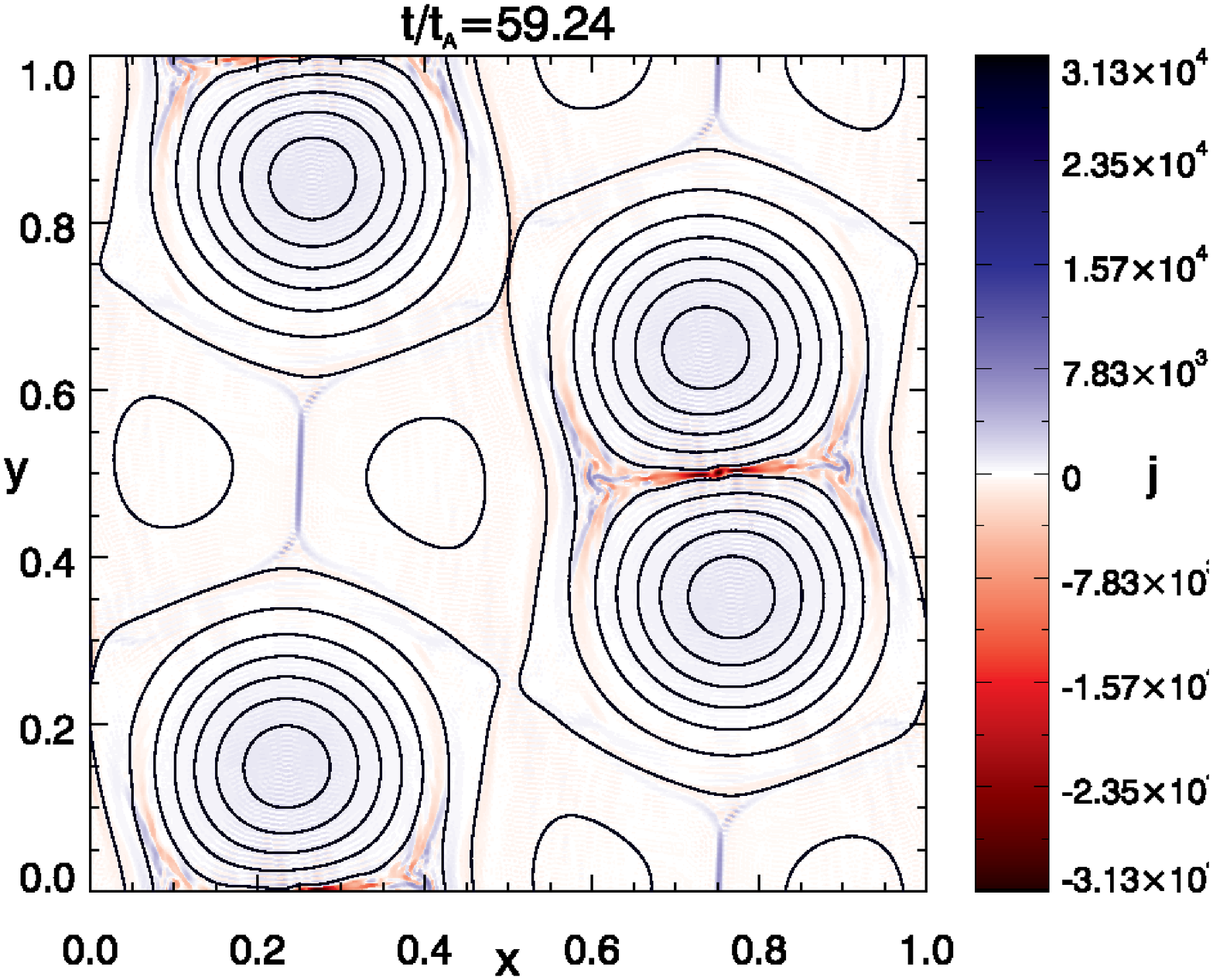}\\[1em]
		\includegraphics[scale=.31]{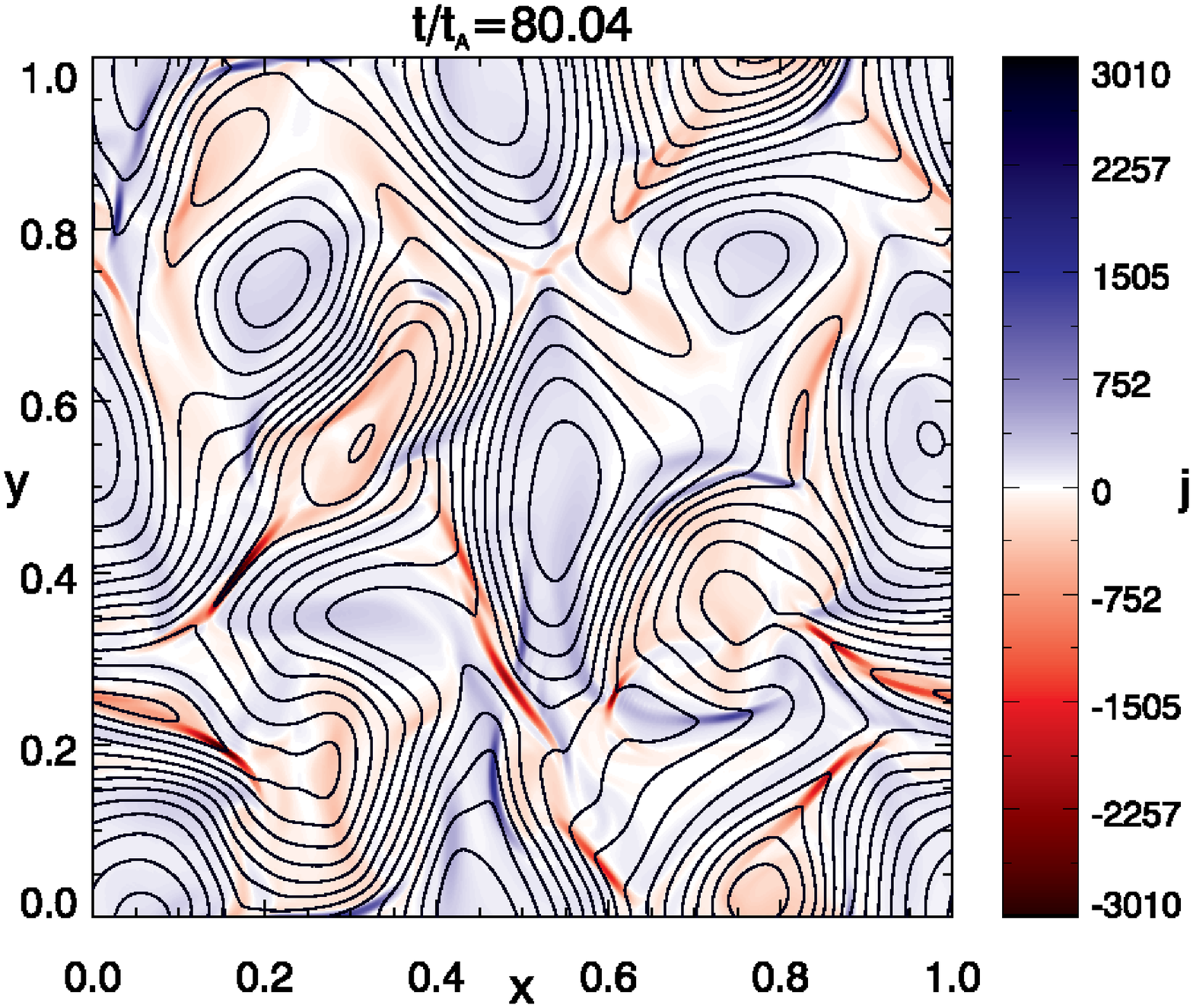}
		\includegraphics[scale=.31]{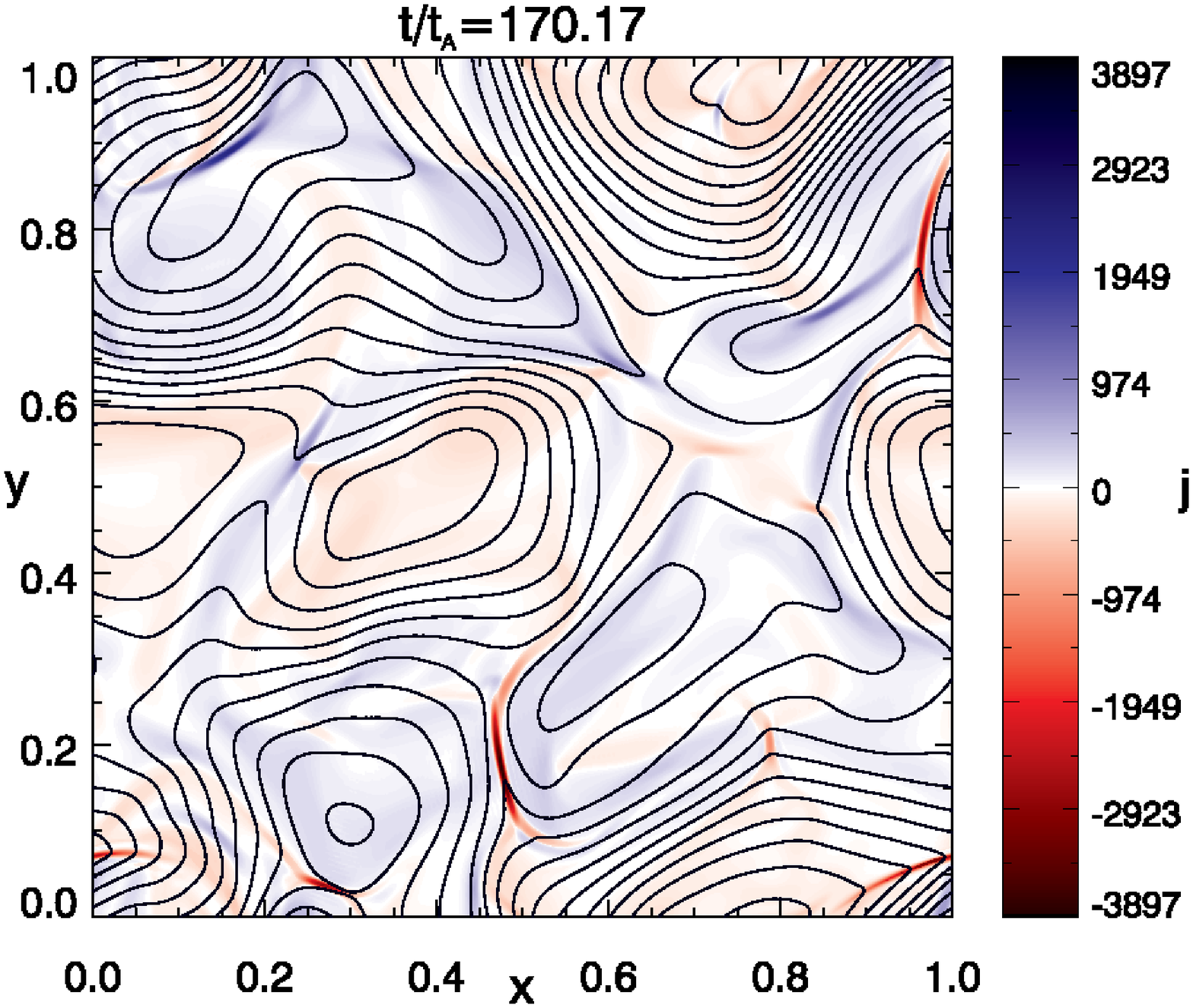}
		\caption{\emph{Run~C -- Four corotating vortices }: Axial component of the current $j$ (in color) and field lines of the orthogonal magnetic field in the midplane ($z=5$) at selected times. At the beginning of the linear stage ($t = 0.61\, \tau_A$) the orthogonal magnetic field is a mapping of the boundary vortices [see linear analysis, Equation~(\ref{eq:lin1})]. Still in the linear stage but at later times ($t = 55.01\, \tau_A$) the field line tension straightens out in a circular shape the vortices mapping. A collective internal kink mode develops transitioning to the nonlinear stage around $t \sim 58.04\, \tau_A$, with the four flux tubes merging into two. Subsequently in the fully nonlinear stage (t=80.04, and 170.17~t$_A$) nonlinearity does not allow the system to return to a linear-like mapping as at times $t \lesssim 58\, t_A$, with dynamics overall similar to those obtained with disordered photospheric motions \citep{2007ApJ...657L..47R, 2008ApJ...677.1348R}. \\
	(An animation of this figure is available.) \label{fig:fig10}}
	\end{centering}
\end{figure*}

Furthermore we perform a numerical simulation that uses the boundary vortex described in Equations~(\ref{eq:f0})-(\ref{eq:f02}) as a building block for the boundary forcing, but we now use four corotating vortices of linear extension $\Delta$=1/2 (double respect to $\Delta$=1/4 used in Runs~A and B) in a computational box spanning $0\le x,y\le 1$. The vortices do fill completely the boundary plate z=L, and because of periodicity at the x--y boundaries there is no empty space around them ({\bf since they are repeated in all directions}) unlike in runs~A and B. Because in the linear stage the magnetic field maps the boundary velocity (Equation~(\ref{eq:lin1})) this can be visualized from the magnetic field lines of $\mathbf{b}$ in Figure~\ref{fig:fig10} at time t=0.61~t$_A$. This numerical simulation has a grid with n$_x \times$ n$_y \times$ n$_z$ = 512$^2 \times$~208, employs standard diffusion with n=1 and Re=800 (because of the larger vortex size we can avoid to use hyperdiffusion). We use same boundary forcing and initial conditions as \cite{Klimchuk2009, Klimchuk2010}, but those early simulations had much lower resolution so that the higher diffusion prevented the system from transitioning to the nonlinear stage by reaching a diffusive equilibrium as discussed in \cite{2013ApJ...771...76R}. This simulation allows us to show the difference between the dynamics developing with a space filling boundary forcing in the plane z=10 versus a localized forcing as implemented in runs~A and B, and to explore dynamics relevant to the inverse cascade of twist in the solar corona \citep{Antiochos2013}, and MHD avalanches for coronal heating \citep{Hood2016}.

\begin{figure}
	\begin{centering}
		\includegraphics[width=\columnwidth]{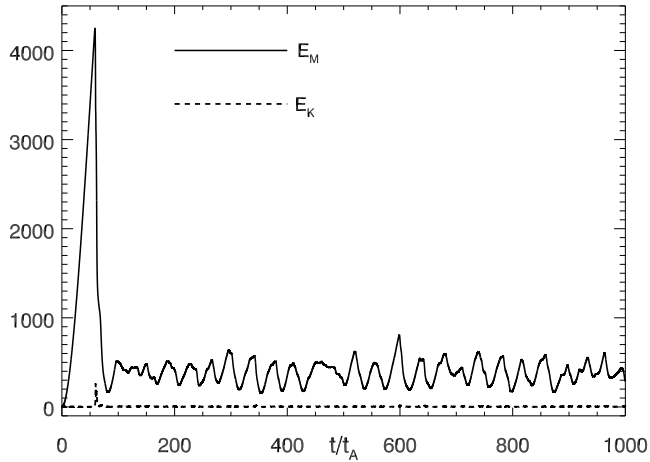}
		\caption{\emph{Run C}: Magnetic ($E_M$) and kinetic ($E_K$) as a function of time. \label{fig:fig11}}
	\end{centering}
\end{figure}

As shown in Figure~\ref{fig:fig10} after the linear stage (t=0.61~t$_A$) a collective internal kink mode develops transitioning to the nonlinear stage around t=58.04~t$_A$, with the original four flux tubes merging two by two thus forming two flux tubes (t=59.24~t$_A$). At this point, once in the fully nonlinear stage, the boundary velocity vortices continue to twist and inject magnetic energy at the large scale $\Delta$=1/2, and also in this case the resulting nonlinear magnetic field that is generated never return to the laminar fields of the linear stage (Equations~(\ref{eq:lin1})-(\ref{eq:lin2})). Rather, the dynamics and magnetic field strongly resemble those that occur when the top and bottom boundaries velocity fields are made of distorted velocity vortices as implemented in our previous work \citep{2007ApJ...657L..47R, 2008ApJ...677.1348R}, with the orthogonal magnetic field component~$\mathbf{b}$ structured in many distorted magnetic islands and current sheets (t=80.04, and 170.17~t$_A$).

\begin{figure}
	\begin{centering}
		\includegraphics[width=\columnwidth]{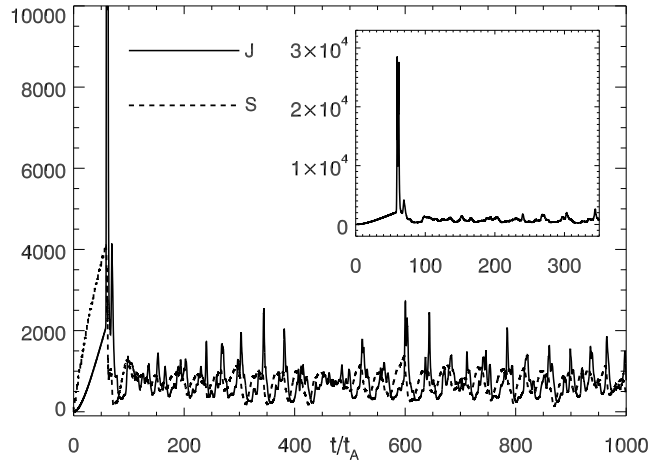}
		\caption{\emph{Run C}: Ohmic dissipation rate ($J$) and integrated Poynting flux ($S$) versus time. \label{fig:fig12}}
	\end{centering}
\end{figure}

Additional insight into the dynamics is given by the r.m.s.\ of magnetic ($E_M$) and kinetic ($E_K$) energies shown in Figure~\ref{fig:fig11} and of Ohmic dissipation rate $J$ and integrated Poynting flux shown in Figure~\ref{fig:fig12}. They display dynamics very similar to those exhibited by the simulations with distorted photospheric vortices \citep{2007ApJ...657L..47R, 2008ApJ...677.1348R} with some distinctive features.

In particular for run~C a big dissipative peak around time t=60~t$_A$ is present due to the kink-like instability developing around that time. On the opposite no similar large dissipative events occur in our previous simulations with distorted vortices where no kink instability develops. However, in common with those early simulations a statistically steady state is reached in the subsequent fully nonlinear stage, when all the r.m.s.\  quantities fluctuate around their mean values, with kinetic energy much smaller than magnetic energy and the integrated Poynting flux balancing on average ohmic dissipation. The main distinctive difference during the nonlinear stage is that in our previous simulations \citep{2007ApJ...657L..47R, 2008ApJ...677.1348R} the ohmic dissipation rate was uniformly fluctuating around its mean value, while here it exhibits a distinctive skewness with bursty dissipative peaks above its average value (Figure~\ref{fig:fig12}), a feature held in common with all previous simulation with ordered vortices at the boundary, including runs~A and B (see Figures~\ref{fig:fig5} and \ref{fig:fig9}), and also for the simulation with a single vortex at the top boundary \citep[see Figure~3 in][]{2013ApJ...771...76R}. The distorted vortices we have used in \citep{2007ApJ...657L..47R, 2008ApJ...677.1348R} have alternate sense of twist among them, therefore besides magnetic reconnection field lines can also get untwisted by the photospheric vortices as their connectivity changes. In the simulations presented here only run~B has vortices with alternate twist, but as we have seen in Section~\ref{sec:runb}, their isolation from nearby vortices does not allow that from happening.

The reason for which in the nonlinear stage a statistically steady state is reached quickly for the simulations with four space-filling vortices (run~C) while energy continues to grow for an extended time for the simulations with two (run~A) or just one \citep{2013ApJ...771...76R} corotating vortices is simple from the energetic point of view. As shown in \cite{2008ApJ...677.1348R} the integrated Poynting flux entering from the photospheric planes, that we indicate with $\epsilon_{in}$, and that injects energy at the large scale $\ell_c$ (the convective length-scale, with $\ell_c = \Delta$ in the runs discussed here) is proportional to $\delta b_{\ell_c}$ (the magnetic field intensity of the component at the scale $\ell_c$). I.e., considering all other parameters fixed, including loop length and Alfv\'en velocity associated to the axial field $B_0$, then $\epsilon_{in} \propto \delta b_{\ell_c}$ \citep[see Eq.~(64) in][]{2008ApJ...677.1348R}. 

On the other hand the energy flowing from the large to the small scales in the turbulent cascade, the so-called spectral energy flux $\epsilon$ scales proportionally to a higher power, in fact $\epsilon \propto \delta b_{\ell_c}^{\alpha+3}$, where $\alpha \geq 0$ is determined by the magnetic energy spectral index \citep[see Eq.~(53) in][]{2008ApJ...677.1348R}. The point here is that when a statistically steady state is reached as for run~C the injection and spectral fluxes are equal on average, i.e., $\epsilon_{in} = \epsilon$, and $\delta b_{\ell_c}$ has a critical value for which both fluxes are the same, that we indicate with $\delta b_{\ell_c}^\ast$. Interestingly when $\delta b_{\ell_c} < \delta b_{\ell_c}^\ast$ the Poynting flux is larger than the spectral flux $\epsilon_{in} > \epsilon$ making the magnetic field intensity grow, but when it grows beyond the critical value $\delta b_{\ell_c} > \delta b_{\ell_c}^\ast$ then the spectral flux transports energy away from the large scales faster then the Poynting flux injects it $\epsilon > \epsilon_{in}$ and the magnetic field intensity decreases toward its critical value. This mechanism clearly leads the magnetic field intensity to fluctuate around its critical value in a statistically steady way as observed.

For two isolated corotating vortices and a single one the magnetic field intensity does not reach initially its critical value because the inverse energy cascade that enlarges the flux tube leads to a weaker magnetic field intensity respect to the case with many nearby flux tubes where such expansion cannot occur. Nevertheless in run~A after the two flux tube merge the dynamics are similar to those of the single vortex, and for that case we have carried out the simulation for much longer times finding that when the flux tube expands enough that it interacts with the nearby identical flux tubes (via the periodic boundary conditions) a statistically steady state is then reached \citep{2013ApJ...771...76R}, as expected from the energetics of the four space-filling vortices in run~C. 

Regarding the possibility of an inverse cascade of energy and twist toward the largest possible scale allowed by the system \citep{Antiochos2013} the most interesting result of run~C is that no inverse cascade occurs, as clearly visible from Figure~\ref{fig:fig10}. This aspect is further investigated and discussed in the next section.

\subsection{On Inverse Cascades in a Line-tied Corona: Runs~D, E and F.} \label{sec:c16}

Inverse cascades in MHD turbulence have been studied extensively \citep[e.g.,][]{Politano1989, 1992PhFlB...4.3070M, 2003matu.book.....B}, and early two dimensional works modeling line-tied loops found that an inverse cascade occurred \citep{1996ApJ...457L.113E, 1998ApJ...497..957G, 1999PhPl....6.4146E}. Later, three-dimensional reduced MHD simulations found that line-tying inhibits the development of an inverse cascade for hot coronal loops. Keeping fixed as parameters the loop length $L$, the orthogonal scale of convective-mimicking photospheric velocity $\ell_c$ and its r.m.s. $u_{ph}$, and the Alfv\'en velocity associated to the strong guide magnetic field $B_0$, and keeping the photospheric velocity constant in time, the solutions of the reduced MHD equations are a family depending on the single parameter:
\begin{equation}
f=\frac{\ell_c B_0}{L u_{ph}}.  \label{eq:rpar}
\end{equation}
For typical hot solar coronal loops we can estimate $B_0 \sim 2 \times 10^3$~km/s, $L\sim 4 \times 10^4$~km, while for the photospheric velocity $\ell_c \sim 10^3$~km and $u_{ph} \sim 1$~km/s, thus yielding $f \sim 50$. For this reason \cite{2007ApJ...657L..47R, 2008ApJ...677.1348R} have performed a parametric study of coronal loop dynamics, with different values of $f$ (specifically $f\sim$~1.8, 7, 14, and 35), finding only for the lower value of $f\sim$~1.8 the presence of an inverse cascade, namely that energy injected at the forcing scale with wavenumber n=4, besides giving rise to a direct energy cascade toward smaller scales, flows also toward larger scale with energy at wavenumber n=1 (the largest accessible scale) growing higher then n=4. But for all higher values no inverse cascade develops, i.e., energy modes with $n < 4$ are always smaller than n=4.

The simulations carried out recently by \cite{Zhao2015, Knizhnik2015, Knizhnik2017} with circular photospheric vortices appear to show the development of an inverse cascade. For them we can estimate that in dimensionless form $\ell_c \sim 1/4$, $L=1$, $B_0=1$, and $u_{ph} \sim 0.2$, so that $f\sim 1.25$. Compared to hot coronal loops, these are longer loops or loops with weaker magnetic fields. 

\begin{figure}
	\begin{centering}
		\includegraphics[width=\columnwidth]{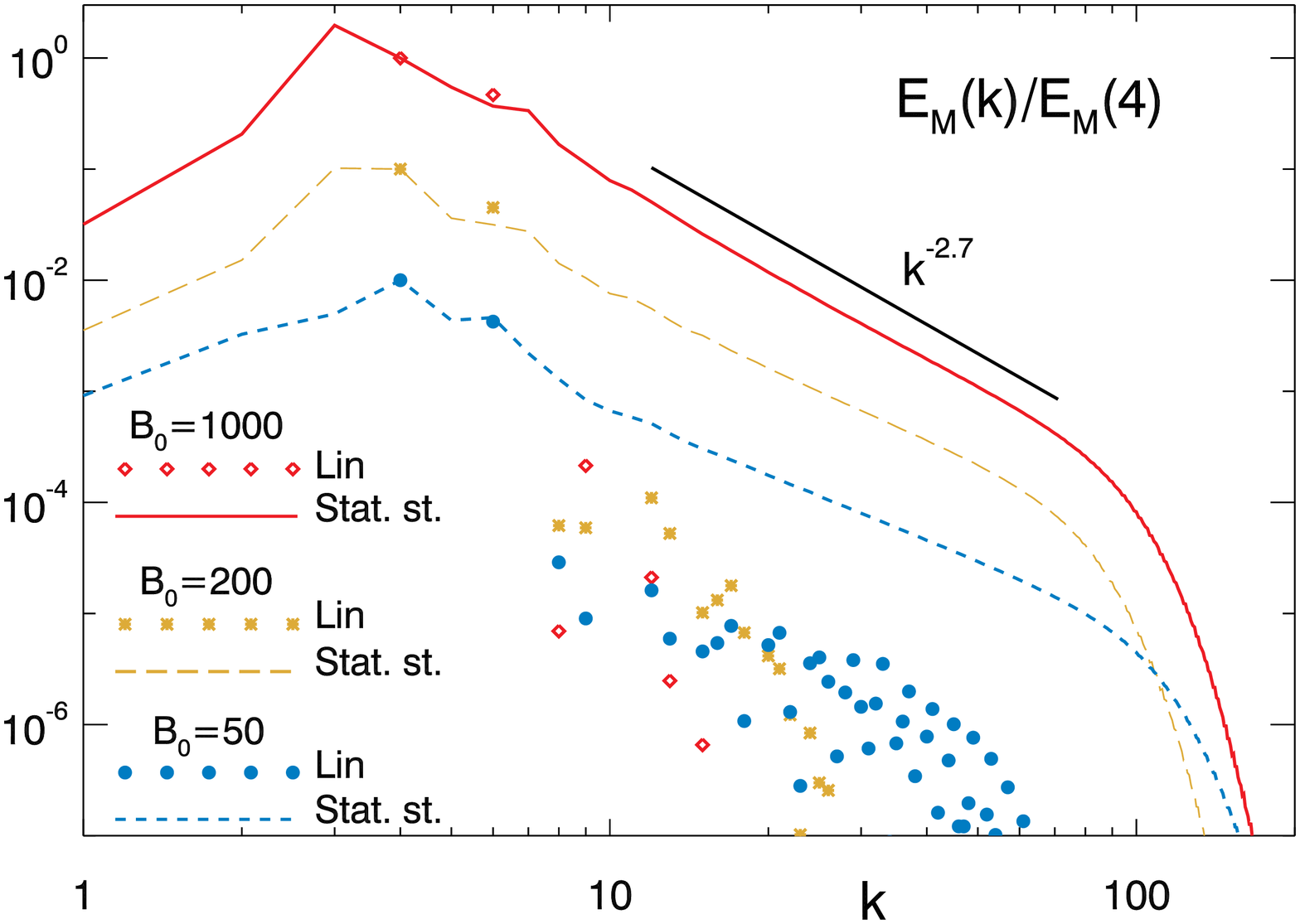}
		\caption{\emph{Runs~D, E, and F}: Magnetic energy spectra, during the \emph{linear} stage (symbols) and in the fully nonlinear \emph{statistically steady} stages (lines) for simulations D, E and F with respectively $B_0$=1000, 200, and 50. No inverse cascade develops, although slightly more energy is present at wavenumber n=1 respect to n=4 for lower values of $B_0$, corresponding to lower values of the parameter $f$ (Eq.~\ref{eq:rpar}). For an improved visualization the spectra are normalized to the value of the energy at wavenumber n=4, and for $B_0$=200 and 50 they are multiplied by a scaling factor not to overlap with each other.
			\label{fig:fig13}}
	\end{centering}
\end{figure}

The parametric study carried out by \cite{2008ApJ...677.1348R} was not using co-rotating photospheric vortices, but a more complex forcing with distorted vortices and zero total vorticity. The simulation described in the previous section (run~C) with four co-rotating vortices has $L=10$, $B_0=200$, $u_{ph} \sim 1/\sqrt{2}$, and $\ell_c = \Delta =1/2$, for which $f=14$, and as can be seen from Figures~10 no significant inverse cascade occurs, but the initial wavenumber is about $n=2$, and this might limit the room available for the development of an inverse cascade. For this reason we have carried out three simulations with sixteen photospheric vortices each (see Eq.~(\ref{eq:co16})), so that the energy injection wavenumber is about $n=4$ (corresponding to a vortex length of about $\ell_c \sim 1/4$). The loop length $L=10$ and photospheric velocity r.m.s. $u_{ph}=1/\sqrt{2}$ are same for all simulations, while 
the Alfv\'en velocity is respectively $B_0$= 1000, 200, and 50 for runs~D, E, and F,
thus yielding for the parameter $f$ (Eq.~(\ref{eq:rpar})) the values $f$= 35, 7, 1.8. 

The magnetic energy spectra are shown in Figure~\ref{fig:fig13}. During the linear stage the magnetic field is a map of the photospheric velocity field (Equation~\ref{eq:lin1}) so that most of the energy is predominantly at the injection scales in modes n=4 and 5 (marked with symbols in the Figure). In the fully nonlinear stage when turbulent dynamics sets in and all relevant quantities fluctuate around their mean value, a direct energy cascade develops leading to the formation of steep spectra similarly to our previous simulations with distorted vortices, and in a similar fashion no inverse cascade develops, but a tendency to a higher value of the ratio $E_M(1)/E_M(4)$ for lower values of $B_0$ and therefore $f$ is present, with $E_M(1)/E_M(4)$=3.2\% and 9.1\% for respectively $B_0$=1000 and 50.

These results therefore suggest that there is a hierarchy of loops for which an inverse cascade can occur or be inhibited. An inverse cascade does not develop for loops with $f \gtrsim 1.8$. Although for all simulations that we have carried out a direct energy cascade that forms small scales current sheets does develop \cite{2016ApJ...817...47D, Dahlburg2018} have shown that it is only for high values of $f$ that a significant radiative emission occurs ($f \gtrsim 30$). 

Regarding the inverse energy cascade model for closed loops envisioned by \cite{Antiochos2013}, we can conclude that it does not seem to occur for loops with $f \gtrsim 1.8$, and in particular for those active regions loops that shine bright in X-rays and EUV. Additionally caution has to be taken also for lower values of $f$. In fact it is to be noticed that indicating with $\tau_A = L/B_0$ the Alfv\'en crossing time, and with $\tau_p=\ell_c/u_{ph}$ the photospheric timescale the parameter $f$ (Eq.~(\ref{eq:rpar})) can be written as:
\begin{equation}
f=\frac{\tau_p}{\tau_A}.  \label{eq:rpar2}
\end{equation}
Since the photospheric timescale $\tau_p$ represents also the lifetime of convective cells, and therefore also the characteristic timescale of the photospheric velocity, while it is acceptable to use as boundary forcing a velocity constant in time for loops with high values of $f$ for which the photospheric timescale is larger than the Alfv\'en crossing time, for lower values of this ratio the impact of a velocity pattern changing in time at the boundary is significant. In fact as shown recently in \cite{Rappazzo2018} for $\tau_p/\tau_A \lesssim 3$ the higher decorrelation at the boundary between velocity and magnetic fields caused by the rapidly changing velocity leads do a strong decrease of the Poynting flux and radiative emission, and to overall vanishing dynamics. 

Magnetic helicity injection depends critically by the scalar product of the magnetic vector potential and the velocity field at the boundary, therefore its behavior with higher frequency boundary forcings might have a similarly strong decrease.

\section{Conclusions and Discussion} \label{sec:cd}

In this paper we have investigated the dynamics of a cartesian reduced MHD model of a closed coronal region forced at the photosphere by two or more vortices. Isolated photospheric vortices have been observed on the Sun \citep{Brandt1988, Bonet2008, Bonet2010} particularly in intergranular lanes and between supergranular cells, and they can have important effects on the corona and heliosphere \citep{Velli1999, 2012Natur.486..505W, Panasenco2014}. 

We have previously investigated the dynamics induced by a single isolated photospheric vortex with our reduced MHD model \citep{2013ApJ...771...76R}. Those results were in strong agreement with previous simulations and analysis \citep{1996A&A...308..935B, 1997SoPh..172..257V, 1998ApJ...494..840L, 1990ApJ...361..690M, 2002A&A...387..687G, 2008A&A...485..837B, 2009A&A...506..913H}, namely that an isolated vortex at the photosphere will twist the field lines into a helix, and once the twist is beyond the instability threshold the system will undergo internal kink instability (assuming a small perturbation is present). 

We then continued the simulation into the fully nonlinear stage to understand the dynamics when the photospheric vortex is not applied to originally straight magnetic field lines, but rather when twisting is applied to a magnetic field with finite amplitude broadband magnetic  fluctuations. In the fully nonlinear stage the presence of an already structured field does not allow the forced configuration to return to a laminar state with orderly and smoothly twisted field lines (laminar helical equilibria). Nevertheless field lines do get twisted by the photospheric vortex and exhibit an approximately constant r.m.s.\ twist of about 180$^\circ$ from top to bottom (for the specific conditions and parameters of that simulation). Additionally an inverse cascade of magnetic energy develops in time as also field lines not connected at their footpoints with the photospheric vortex get progressively twisted, so that in physical space the region where field lines are twisted expands in time in the orthogonal direction (see Section~\ref{sec:sv}). The inverse cascade can store a significant amount of magnetic energy that becomes available when it interacts with other magnetic structures, as seen in \cite{2013ApJ...771...76R} where we verified that interactions with  neighboring structures occur when the transverse scale reaches the computational box size, i.e., when  interactions with copies of itself occur because of the periodic boundary conditions, giving rise to dissipative events in the micro-flare range with $\Delta E \sim 2 \times 10^{25}$~erg. 

To further understand the development of inverse cascades of energy and twist in the solar corona, we have discussed in this paper simulations with two velocity vortices (corotating or counter-rotating, Sections~\ref{sec:runa} and \ref{sec:runb}) applied at the center of the top photospheric boundary. Additionally we have carried out two numerical simulation with respectively four (Section~\ref{sec:runc}) and sixteen (Section~\ref{sec:c16}) corotating vortices that fill the photospheric plate entirely.

The results of the two simulations with two photospheric vortices either corotating or counter-rotating are in full agreement with previous studies of this kind \citep{Lau1996, 1999ApJ...519..884K, Linton2001}: two corotating photospheric vortices bring about two flux tubes where field lines of the orthogonal magnetic field component \textbf{b} are anti-parallel and therefore reconnect. In fact we find that magnetic reconnection occurs and leads to the merging of the originally distinct two flux tubes. In our specific case with the two vortices one next to the other, kink instability does not have time to develop because magnetic reconnection and merging start to occur earlier than the internal kink mode requires to develop. Nevertheless as we discussed in Section~\ref{sec:runa}, if we increased the distance between the two photospheric vortices, while we do expect that the further they are from each other the more fully the internal kink mode can develop and kink instability transition to the nonlinear stage, the overall dynamics would not be that different because we have already seen in \cite{2013ApJ...771...76R} that the nonlinear stage of kink instability for a single vortex leads to an inverse cascade of magnetic energy, i.e., field lines are twisted in a disorderly manner and the linear extent of the region where they are twisted increases in time. We expect that setting the two vortices a certain distance apart will still lead to the two regions of twisted field lines to merge (this hypothesis and the specific dynamics at varying distance between the vortices should be investigated in future work).

Once merging between the two regions of twisted field lines occurs then the dynamics are very similar to that of a single photospheric vortex, i.e., an inverse cascade of magnetic energy develops, sustained by the Poynting flux injected from the boundary, and driven by the Lorentz force resulting from the small scale current sheets created by the direct cascade of energy and the orthogonal magnetic field, so that field lines will get twisted outside the region with field lines directly connected to the photospheric vortices, thus expanding the flux tubes in the orthogonal direction.

For two counter-rotating photospheric vortices (run~B, Section~\ref{sec:runb}), the topology with parallel field lines in the boundary region between the two flux tubes prevents them from reconnecting and merging. Kink instability will then develop in both flux tubes, but in the fully non linear stage the inverse cascade will be very limited because, as described in Section~\ref{sec:runb}, the inverse cascade makes the field lines bundle region expand, and pushing against each other without being able to reconnect and merge the two bundles push each other away and become de-centered from the respective vortices that originate them. As discussed here for the first time, this causes a strong decrease of the value of the integrated Poynting flux (Equation~(\ref{eq:poy})), that depends critically on the correlation between the vortex velocity field \textbf{u}$^L$ and the coronal magnetic field \textbf{b}. The field line bundles then find a dynamic balance between energy injection and ohmic dissipation that prevents them from fully developing an inverse cascade of magnetic energy.

We have then considered four corotating vortices set side by side and filling completely the photospheric plane z=L (run~C, Section~\ref{sec:runc}). In this case in the linear regime a  collective kink mode develops and subsequently the system transitions to the non-linear stage by merging the original four flux tubes into two nonlinear field lines bundles. Subsequently the system never returns to the linear laminar configuration with four smooth and ordered flux tubes, but they exhibit the formation and dissipation of disordered magnetic islands and current sheets in similar fashion to what we have already observed for our simulations with distorted photospheric vortices mimicking the effective orthogonal velocity field of solar convective cells \citep{2007ApJ...657L..47R, 2008ApJ...677.1348R}. 

Recently \citep{Hood2016} have proposed that the solar corona is made by marginally stable flux tubes prone to release energy via kink instability in the fashion of MHD avalanches (giving rise to self-organized criticality or SOC), thus bringing about an X-ray and EUV emitting hot corona. Nevertheless as shown by the simulations discussed so far, even with four corotating vortices in the non-linear stage the orthogonal magnetic field lines are very distorted and never return to the laminar condition in which they might be kink unstable. In our view this conclusion is also supported by the simulation with three adjacent vortices carried out by \cite{Reid2018}). This suggests that a mechanism to posit a corona structured in flux tubes marginally unstable to kink is lacking.

Additionally, no significant inverse cascade develops in the simulation with four corotating vortices (run~C). To better understand the conditions under which inverse cascades develop in closed coronal structures we have carried out three additional simulations with sixteen vortices.

As recently proposed by \cite{Antiochos2013} the inverse cascade of helicity may have an impact on the boundary structure between open and closed regions and the dynamics developing there. The simulations carried in this framework by \cite{Zhao2015, Knizhnik2015, Knizhnik2017} indeed appear to give always rise to an inverse cascade of magnetic energy and helicity when the system is shuffled at its footpoints by corotating vortices. Nevertheless they always use the same characteristic parameters for the considered closed region, in particular the same Alfv\'en velocity $B_0$ and axial length $L$.

As discussed in Section~\ref{sec:c16} the reduced MHD solutions depend on the single parameter $f = \ell_c B_0/L u_{ph}$, where $\ell_c$ is the convective length-scale and $u_{ph}$ the r.m.s.\ of the photospheric velocity. While $\ell_c$ and $u_{ph}$ have characteristic fixed values for solar convection, loop length $L$ and Alfv\'en velocity $B_0$ have substantial variations in the magnetically closed corona. Therefore while for typically hot loops $f \sim 50$, longer loops or loops with weaker magnetic fields will exhibit lower values for $f$. 

As discussed in more detail in Section~\ref{sec:c16} we have carried out three simulations with $f = 35, 7, 1.8$. In all of them no inverse cascade develops, although the energy at the largest scales increases respect to the energy at the injection scale for lower values of $f$, similarly to what observed in our previous simulations with distorted vortices and zero net vorticity \citep{2008ApJ...677.1348R}. Nevertheless for values of $f$ smaller than about 3 ($f \lesssim 3$) the time variation of photospheric convective cells cannot be neglected, and could lead to a substantial decrease of the helicity injection, analogously to the decrease of Poynting flux observed by \cite{Rappazzo2018} (see Section~\ref{sec:c16} for a more thorough discussion). There is therefore a hierarchy of loops that depending on their parameter $f$ can develop an inverse cascade of energy and twist at different rates or it can be entirely inhibited.

We conclude then that the dynamics induced in the closed corona by photospheric vortices can be very different depending on how they are arranged. This has consequences for coronal heating, namely higher energy releases can occur when more energy can be stored, e.g., for a single or two corotating vortices, but also for the initial stage of a coronal mass ejection. Indeed photospheric rotations have been detected within the region of interest to CMEs and they are thought to play a role in their initiation process \citep{Toeroek2013}. Although typically twisted field lines are found from the force-free extrapolation of the photospheric magnetic field at various stages of the initiation of a CME \citep[e.g.,][]{Amari2015,Amari2018}, the technique used specifically looks for equilibria configurations so that it is natural to think of kink instability as playing a key role in the initiation of CMEs. Our simulations suggests though the situation can be much more complicated, that twisted field lines in the presence of photospheric rotations or vortices do not need to be in equilibrium and might not be unstable to kink instability at all. Rather a complex inverse cascade process able to expand the field line bundles could occur that could in turn destabilize the CME region, bypassing in this way the need for kink instabilities \citep{Toeroek2013}. We have here neglected any effects due to curvature, that already by itself can lead, when not confined by overarching closed field lines, to flux tube expansion  \citep{Amari1996}.

\acknowledgments
This research was supported in part by the National Aeronautics and Space Administration Grant No.~NNH17AE96I issued through the Heliophysics Grand Challenge Research Program, NRL contract No.~N00173-17-P-3044, and by the NASA Parker Solar Probe Observatory Scientist grant No.~NNX15AF34G. RBD also was supported in part by the Naval Research Laboratory 6.1~program. Computing resources supporting this work were provided by the NASA High-End Computing (HEC) Program through the NASA Advanced Supercomputing (NAS) Division at the Ames Research Center and by NRL LCP\&FD.


\begin{thebibliography}{}
	\expandafter\ifx\csname natexlab\endcsname\relax\def\natexlab#1{#1}\fi
	
	\bibitem[{Amari {et~al.}(2014)Amari, Canou, \& Aly}]{Amari2014}
	Amari, T., Canou, A., \& Aly, J.-J. 2014, \nat, 514, 465
	
	\bibitem[{Amari {et~al.}(2018)Amari, Canou, Aly, Delyon, \&
		Alauzet}]{Amari2018}
	Amari, T., Canou, A., Aly, J.-J., Delyon, F., \& Alauzet, F. 2018, \nat, 554,
	211
	
	\bibitem[{Amari {et~al.}(2015)Amari, Luciani, \& Aly}]{Amari2015}
	Amari, T., Luciani, J.-F., \& Aly, J.-J. 2015, \nat, 522, 188
	
	\bibitem[{Amari {et~al.}(1996)Amari, Luciani, Aly, \& Tagger}]{Amari1996}
	Amari, T., Luciani, J.~F., Aly, J.~J., \& Tagger, M. 1996, \apjl, 466, L39
	
	\bibitem[{Antiochos(2013)}]{Antiochos2013}
	Antiochos, S.~K. 2013, \apj, 772, 72
	
	\bibitem[{Bareford {et~al.}(2016)Bareford, Gordovskyy, Browning, \&
		Hood}]{Bareford2016}
	Bareford, M.~R., Gordovskyy, M., Browning, P.~K., \& Hood, A.~W. 2016,
	\solphys, 291, 187
	
	\bibitem[{{Baty} \& {Heyvaerts}(1996)}]{1996A&A...308..935B}
	{Baty}, H., \& {Heyvaerts}, J. 1996, \aap, 308, 935
	
	\bibitem[{{Biskamp}(2003)}]{2003matu.book.....B}
	{Biskamp}, D. 2003, {Magnetohydrodynamic Turbulence} (Cambridge: Cambridge
	University Press)
	
	\bibitem[{Bonet {et~al.}(2008)Bonet, M{\'a}rquez, S{\'a}nchez~Almeida, Cabello,
		\& Domingo}]{Bonet2008}
	Bonet, J.~A., M{\'a}rquez, I., S{\'a}nchez~Almeida, J., Cabello, I., \&
	Domingo, V. 2008, \apjl, 687, L131
	
	\bibitem[{Bonet {et~al.}(2010)Bonet, M{\'a}rquez, S{\'a}nchez~Almeida,
		Palacios, Mart{\'{\i}}nez~Pillet, Solanki, del Toro~Iniesta, Domingo,
		Berkefeld, Schmidt, Gandorfer, Barthol, \& Kn{\"o}lker}]{Bonet2010}
	Bonet, J.~A., M{\'a}rquez, I., S{\'a}nchez~Almeida, J., {et~al.} 2010, \apjl,
	723, L139
	
	\bibitem[{Brandt {et~al.}(1988)Brandt, Scharmer, Ferguson, Shine, \&
		Tarbell}]{Brandt1988}
	Brandt, P.~N., Scharmer, G.~B., Ferguson, S., Shine, R.~A., \& Tarbell, T.~D.
	1988, \nat, 335, 238
	
	\bibitem[{{Browning} {et~al.}(2008){Browning}, {Gerrard}, {Hood}, {Kevis}, \&
		{van der Linden}}]{2008A&A...485..837B}
	{Browning}, P.~K., {Gerrard}, C., {Hood}, A.~W., {Kevis}, R., \& {van der
		Linden}, R.~A.~M. 2008, \aap, 485, 837
	
	\bibitem[{{Dahlburg} {et~al.}(2012){Dahlburg}, {Einaudi}, {Rappazzo}, \&
		{Velli}}]{2012A&A...544L..20D}
	{Dahlburg}, R.~B., {Einaudi}, G., {Rappazzo}, A.~F., \& {Velli}, M. 2012, \aap,
	544, L20
	
	\bibitem[{{Dahlburg} {et~al.}(2016){Dahlburg}, {Einaudi}, {Taylor},
		{Ugarte-Urra}, {Warren}, {Rappazzo}, \& {Velli}}]{2016ApJ...817...47D}
	{Dahlburg}, R.~B., {Einaudi}, G., {Taylor}, B.~D., {et~al.} 2016, \apj, 817, 47
	
	\bibitem[{Dahlburg {et~al.}(2018)Dahlburg, Einaudi, Ugarte-Urra, Rappazzo, \&
		Velli}]{Dahlburg2018}
	Dahlburg, R.~B., Einaudi, G., Ugarte-Urra, I., Rappazzo, A.~F., \& Velli, M.
	2018, \apj, 868, 116
	
	\bibitem[{{Dmitruk} \& {G{\'o}mez}(1997)}]{1997ApJ...484L..83D}
	{Dmitruk}, P., \& {G{\'o}mez}, D.~O. 1997, \apjl, 484, L83
	
	\bibitem[{{Dmitruk} \& {G{\'o}mez}(1999)}]{1999ApJ...527L..63D}
	---. 1999, \apjl, 527, L63
	
	\bibitem[{{Dmitruk} {et~al.}(2003){Dmitruk}, {G{\'o}mez}, \&
		{Matthaeus}}]{2003PhPl...10.3584D}
	{Dmitruk}, P., {G{\'o}mez}, D.~O., \& {Matthaeus}, W.~H. 2003, Physics of
	Plasmas, 10, 3584
	
	\bibitem[{{Einaudi} \& {Velli}(1999)}]{1999PhPl....6.4146E}
	{Einaudi}, G., \& {Velli}, M. 1999, Physics of Plasmas, 6, 4146
	
	\bibitem[{{Einaudi} {et~al.}(1996){Einaudi}, {Velli}, {Politano}, \&
		{Pouquet}}]{1996ApJ...457L.113E}
	{Einaudi}, G., {Velli}, M., {Politano}, H., \& {Pouquet}, A. 1996, \apjl, 457,
	L113
	
	\bibitem[{{Georgoulis} {et~al.}(1998){Georgoulis}, {Velli}, \&
		{Einaudi}}]{1998ApJ...497..957G}
	{Georgoulis}, M.~K., {Velli}, M., \& {Einaudi}, G. 1998, \apj, 497, 957
	
	\bibitem[{{Gerrard} {et~al.}(2002){Gerrard}, {Arber}, \&
		{Hood}}]{2002A&A...387..687G}
	{Gerrard}, C.~L., {Arber}, T.~D., \& {Hood}, A.~W. 2002, \aap, 387, 687
	
	\bibitem[{{Gold} \& {Hoyle}(1960)}]{1960MNRAS.120...89G}
	{Gold}, T., \& {Hoyle}, F. 1960, \mnras, 120, 89
	
	\bibitem[{{Hood} {et~al.}(2009){Hood}, {Browning}, \& {van der
			Linden}}]{2009A&A...506..913H}
	{Hood}, A.~W., {Browning}, P.~K., \& {van der Linden}, R.~A.~M. 2009, \aap,
	506, 913
	
	\bibitem[{Hood {et~al.}(2016)Hood, Cargill, Browning, \& Tam}]{Hood2016}
	Hood, A.~W., Cargill, P.~J., Browning, P.~K., \& Tam, K.~V. 2016, \apj, 817, 5
	
	\bibitem[{{Kadomtsev} \& {Pogutse}(1974)}]{1974JETP...38..283K}
	{Kadomtsev}, B.~B., \& {Pogutse}, O.~P. 1974, Soviet Journal of Experimental
	and Theoretical Physics, 38, 283
	
	\bibitem[{Klimchuk {et~al.}(2009)Klimchuk, Nigro, Dahlburg, \&
		Antiochos}]{Klimchuk2009}
	Klimchuk, J.~A., Nigro, G., Dahlburg, R.~B., \& Antiochos, S.~K. 2009, AGU Fall
	Meeting Abstracts, SM42B
	
	\bibitem[{Klimchuk {et~al.}(2010)Klimchuk, Nigro, Dahlburg, \&
		Antiochos}]{Klimchuk2010}
	Klimchuk, J.~A., Nigro, G., Dahlburg, R.~B., \& Antiochos, S.~K. 2010, in
	American Astronomical Society Meeting Abstracts, Vol. 216, American
	Astronomical Society Meeting Abstracts \#216, 847
	
	\bibitem[{{Knizhnik} {et~al.}(2015){Knizhnik}, {Antiochos}, \&
		{DeVore}}]{Knizhnik2015}
	{Knizhnik}, K.~J., {Antiochos}, S.~K., \& {DeVore}, C.~R. 2015, \apj, 809, 137
	
	\bibitem[{Knizhnik {et~al.}(2017)Knizhnik, Antiochos, \& DeVore}]{Knizhnik2017}
	Knizhnik, K.~J., Antiochos, S.~K., \& DeVore, C.~R. 2017, \apj, 835, 85
	
	\bibitem[{{Kondrashov} {et~al.}(1999){Kondrashov}, {Feynman}, {Liewer}, \&
		{Ruzmaikin}}]{1999ApJ...519..884K}
	{Kondrashov}, D., {Feynman}, J., {Liewer}, P.~C., \& {Ruzmaikin}, A. 1999,
	\apj, 519, 884
	
	\bibitem[{Lau \& Finn(1996)}]{Lau1996}
	Lau, Y.-T., \& Finn, J.~M. 1996, Physics of Plasmas, 3, 3983
	
	\bibitem[{Linton {et~al.}(2001)Linton, Dahlburg, \& Antiochos}]{Linton2001}
	Linton, M.~G., Dahlburg, R.~B., \& Antiochos, S.~K. 2001, \apj, 553, 905
	
	\bibitem[{{Lionello} {et~al.}(1998){Lionello}, {Velli}, {Einaudi}, \&
		{Miki{\'c}}}]{1998ApJ...494..840L}
	{Lionello}, R., {Velli}, M., {Einaudi}, G., \& {Miki{\'c}}, Z. 1998, \apj, 494,
	840
	
	\bibitem[{{Malara} {et~al.}(1992){Malara}, {Veltri}, \&
		{Carbone}}]{1992PhFlB...4.3070M}
	{Malara}, F., {Veltri}, P., \& {Carbone}, V. 1992, Physics of Fluids B, 4, 3070
	
	\bibitem[{{Mikic} {et~al.}(1990){Mikic}, {Schnack}, \& {van
			Hoven}}]{1990ApJ...361..690M}
	{Mikic}, Z., {Schnack}, D.~D., \& {van Hoven}, G. 1990, \apj, 361, 690
	
	\bibitem[{{Montgomery}(1982)}]{1982PhST....2...83M}
	{Montgomery}, D. 1982, Physica Scripta Volume T, 2, 83
	
	\bibitem[{Panasenco {et~al.}(2014)Panasenco, Martin, \& Velli}]{Panasenco2014}
	Panasenco, O., Martin, S.~F., \& Velli, M. 2014, \solphys, 289, 603
	
	\bibitem[{{Parker}(1972)}]{1972ApJ...174..499P}
	{Parker}, E.~N. 1972, \apj, 174, 499
	
	\bibitem[{{Parker}(1988)}]{1988ApJ...330..474P}
	---. 1988, \apj, 330, 474
	
	\bibitem[{{Parker}(1994)}]{1994ISAA....1.....P}
	---. 1994, {Spontaneous current sheets in magnetic fields} (New York: Oxford
	University Press)
	
	\bibitem[{Politano {et~al.}(1989)Politano, Pouquet, \& Sulem}]{Politano1989}
	Politano, H., Pouquet, A., \& Sulem, P.~L. 1989, Physics of Fluids B, 1, 2330
	
	\bibitem[{{Rappazzo}(2006)}]{2010PhDT.......193R}
	{Rappazzo}, A.~F. 2006, PhD thesis, Universit{\`a} di Pisa
	
	\bibitem[{Rappazzo {et~al.}(2018)Rappazzo, Dahlburg, Einaudi, \&
		Velli}]{Rappazzo2018}
	Rappazzo, A.~F., Dahlburg, R.~B., Einaudi, G., \& Velli, M. 2018, \mnras, 478,
	2257
	
	\bibitem[{{Rappazzo} \& {Velli}(2011)}]{2011PhRvE..83f5401R}
	{Rappazzo}, A.~F., \& {Velli}, M. 2011, \pre, 83, 065401
	
	\bibitem[{{Rappazzo} {et~al.}(2010){Rappazzo}, {Velli}, \&
		{Einaudi}}]{2010ApJ...722...65R}
	{Rappazzo}, A.~F., {Velli}, M., \& {Einaudi}, G. 2010, \apj, 722, 65
	
	\bibitem[{{Rappazzo} {et~al.}(2013){Rappazzo}, {Velli}, \&
		{Einaudi}}]{2013ApJ...771...76R}
	---. 2013, \apj, 771, 76
	
	\bibitem[{{Rappazzo} {et~al.}(2007){Rappazzo}, {Velli}, {Einaudi}, \&
		{Dahlburg}}]{2007ApJ...657L..47R}
	{Rappazzo}, A.~F., {Velli}, M., {Einaudi}, G., \& {Dahlburg}, R.~B. 2007,
	\apjl, 657, L47
	
	\bibitem[{{Rappazzo} {et~al.}(2008){Rappazzo}, {Velli}, {Einaudi}, \&
		{Dahlburg}}]{2008ApJ...677.1348R}
	---. 2008, \apj, 677, 1348
	
	\bibitem[{Reid {et~al.}(2018)Reid, Hood, Parnell, Browning, \&
		Cargill}]{Reid2018}
	Reid, J., Hood, A.~W., Parnell, C.~E., Browning, P.~K., \& Cargill, P.~J. 2018,
	\aap, 615, A84
	
	\bibitem[{{Strauss}(1976)}]{1976PhFl...19..134S}
	{Strauss}, H.~R. 1976, Physics of Fluids, 19, 134
	
	\bibitem[{T{\"o}r{\"o}k {et~al.}(2013)T{\"o}r{\"o}k, Temmer, Valori, Veronig,
		van Driel-Gesztelyi, \& Vr{\v s}nak}]{Toeroek2013}
	T{\"o}r{\"o}k, T., Temmer, M., Valori, G., {et~al.} 2013, \solphys, 286, 453
	
	\bibitem[{Velli \& Liewer(1999)}]{Velli1999}
	Velli, M., \& Liewer, P. 1999, \ssr, 87, 339
	
	\bibitem[{{Velli} {et~al.}(1997){Velli}, {Lionello}, \&
		{Einaudi}}]{1997SoPh..172..257V}
	{Velli}, M., {Lionello}, R., \& {Einaudi}, G. 1997, \solphys, 172, 257
	
	\bibitem[{{Wedemeyer-B{\"o}hm} {et~al.}(2012){Wedemeyer-B{\"o}hm}, {Scullion},
		{Steiner}, {van der Voort}, {de La Cruz Rodriguez}, {Fedun}, \&
		{Erd{\'e}lyi}}]{2012Natur.486..505W}
	{Wedemeyer-B{\"o}hm}, S., {Scullion}, E., {Steiner}, O., {et~al.} 2012, \nat,
	486, 505
	
	\bibitem[{{Zank} \& {Matthaeus}(1992)}]{1992JPlPh..48...85Z}
	{Zank}, G.~P., \& {Matthaeus}, W.~H. 1992, Journal of Plasma Physics, 48, 85
	
	{Zhao, L. and DeVore, C. R. and Antiochos, S. K. and Zurbuchen, T. H.}
	\bibitem[{{Zhao} {et~al.}(2015){Zhao}, {DeVore}, {Kosovichev}, {Antiochos}, \&
		{Zurbuchen}}]{Zhao2015}
	{Zhao}, J., {DeVore}, C.~R., {Antiochos}, S.~K., \&
	{Zurbuchen}, T.~H. 2015, \apj, 805, 61
	
\end{thebibliography}
\end{document}